\documentclass[journal]{IEEEtran}
\usepackage{algorithm}
\usepackage{algorithmic}
\usepackage{helvet}
\usepackage{courier}
\usepackage{subfigure}
\usepackage{graphicx}
\usepackage{multirow}
\usepackage{bm}
\usepackage{cite}

\ifCLASSINFOpdf
\else
\fi
\begin{document}
%
\title{Personalized Next Point-of-Interest Recommendation via Latent Behavior Patterns Inference}
%
%
%

\author{Jing~He,~\IEEEmembership{Member,~IEEE,}
        Xin~Li,~\IEEEmembership{Member,~IEEE,}
        Lejian~Liao,~\IEEEmembership{Member,~IEEE}
        and~William~K.Cheung,~\IEEEmembership{Member,~IEEE}
\thanks{(Corresponding author: Xin Li).}
\thanks{Jing He, Xin Li and Lejian Liao are with the Beijing Engineering Research Center of High Volume Language 

Information Processing and Cloud Computing Application,School
of Computer Science and Technology, Beijing Institute of Technology, 10081 Beijing, China (e-mail: skyhejing@bit.edu.cn; 

xinli@bit.edu.cn; liaolj@bit.edu.cn).}
}

%
%

\markboth{IEEE Transactions on Cybernetics,~Vol.~, No.~, August~}%
{Shell \MakeLowercase{\textit{et al.}}: Bare Demo of IEEEtran.cls for IEEE Journals}
%



\maketitle

\begin{abstract}
In this paper, we address the problem of personalized next Point-of-interest (POI) recommendation which has become an important and very challenging task for location-based social networks (LBSNs), but not well studied yet. With the conjecture that, under different contextual scenarios, human exhibits distinct mobility pattern, we attempt here to jointly model the next POI recommendation under the influence of user's latent behavior pattern. We propose to adopt a third-rank tensor to model the successive check-in behaviors. By integrating categorical influence into mobility patterns and aggregating user's spatial preference on a POI, the proposed model deal with the next new POI recommendation problem by nature. By incorporating softmax function to fuse the personalized Markov chain with latent pattern, we furnish a Bayesian Personalized Ranking (BPR) approach and derive the optimization criterion accordingly. Expectation Maximization (EM) is then used to estimate the model parameters. We further develop a personalized model by taking into account personalized mobility patterns under the contextual scenario to improve the recommendation performance. Extensive experiments on two large-scale LBSNs datasets demonstrate the significant improvements of our model over several state-of-the-art methods. 
\end{abstract}

\begin{IEEEkeywords}
IEEE, IEEEtran, journal, \LaTeX, paper, template.
\end{IEEEkeywords}

%
\IEEEpeerreviewmaketitle

\section{Introduction}
\IEEEPARstart{O}{nline} social networks allow hundreds of millions of Internet users world wide to access to vast amount of information on an unprecedented scale. In recent years, there have been an increased emphasis on developing the location-based social networks (LBSNs), such as Foursquare\footnote{https://foursquare.com}, Gowalla, Facebook Place\footnote{https://www.facebook.com/places}, and GeoLife, etc., where users can check-in at venues online and share their experiences towards point-of-interest (POIs) in the physical world via their mobile devices. For example, as of February 2017, Foursquare recorded 10 billion check-ins at 93 million point-of-interests (POIs), which are contributed by more than 50 million Foursquare users world wide\footnote{https://foursquare.com/about}. This so called check-in behavior has become a new culture of a modern life and can be used to study life patterns of millions of LBSN users. POI recommendation is one of the most important tasks in LBSN, which is to provide recommendations of places to users, and has attracted much attention as it is not only able to improve user viscosity to LBSN service provider but also to benefit for advertising agency to provide an effective way of launching advertisement to target the potential clients. In the broader context, an accurate POI recommendation is essential for urban computing \cite{zheng2014urban}, behavior informatics \cite{cao2012behavior} and control of the spread of infectious diseases \cite{eubank2004modelling}, etc. 

POI recommendation has become a popular research issue and attracted much effort from both academia and industry \cite{gao2015content,DBLP:conf/aaai/YuanCZQL14}. Yet achieving accurate personalized POI recommendation is challenging as the data available for each user is highly sparse. The sparsity is due to the fact that the check-in interactions are conducted by the users on a voluntary basis. ``Diligent'' users who keep checking-in on LBSN for every venue they visited in physical world are in fact rare. Collaborative filtering (CF) technique is widely adopted for recommender system and many CF models have been proposed to learn users' preferences on the POIs from the check-in data in the literature. The CF methods can be divided into two categories, namely memory-based CF and model-based CF. Memory-based CF models can be further divided into two sub-categories, namely user-based CF and item-based CF. Memory-based CF methods suffer much from the data sparsity problem, since the user-user or item-item similarities need to be calculated based on the common check-ins. 
Ye et al. \cite{ye2011exploiting} adopt linear interpolation to incorporate both social and geographical influences into the user-based CF framework for POI recommendation. Their experimental results show that user-based CF outperforms item-based CF for POI recommendation. Incorporating the geographical influence into the user-based CF model leads to a significant improvement in the recommendation performance, while the social influence has little impact on the performance. Model-based CF builds models using data mining techniques, such as matrix factorization (CF), on user ratings to perform the recommendations. Based on the observation that individual’s check-in locations are usually around several centers, Cheng et al. \cite{cheng2012fused} introduce a multi-center Gaussian model to infer the geographical influence and then combine it heuristically with MF for POI recommendation. However, aforementioned works overlooked the consecutive information of check-ins which is very important for POI recommendation, as human movements often exhibit sequential patterns. And a good POI recommendation should be able to provide a prompt recommendation with respect to users' current location.

Hence, different from aforementioned works, some researchers considered the task of next POI recommendation (also called successive POI recommendation) in LBSNs. This is an even harder task which is to be accurate on predicting user’s very next move amon tens of thousands of location candidates. The challenge of next POI recommendation results from the follow two reasons. First, the check-in can be considered as a type of implicit feedback, which is different from conventional 5-star rating data with explicitly denoting ``like" or ``dislike” to an item by different rating scores. The check-ins offer only positive examples that a user likes, and the POIs without check-ins are a mixture of real negative feedback (the location is unattractive for the user) and missing values (the user might visit the location in the future). Thus, the recommender system has to infer user preferences from the implicit feedback data, which makes the next POI recommendation very tough. Second, the check-in data for personalized successive interactions is very sparse. The users’ visiting history data is often transformed to user-POI check-in matrix whose sparsity is dramatically higher than that of user-item rating matrix in Netflix data \cite{yu2015survey}.Moreover, when considering the task of next POI recommendation, we propose to adapt a third-rank tensor to model the successive check-in behaviors, then the user-POI check-in data (matrix) needs to be separated and represented as the user-Current POI-Next POI tensor. This will make the data more sparse, and the density of the check-in tensor in experiments is $5.81\times 10^{-10}$ for Foursquare dataset within Los Angeles, $5.85\times 10^{-9}$ for Foursquare dataset within New York City and $1.01\times 10^{-6}$ for Gowalla dataset respectively, which makes the next POI recommendation task more difficult. 

\begin{figure}[t]
	\centering
	\subfigure[LA]{\includegraphics[clip=true,width=1.6in]{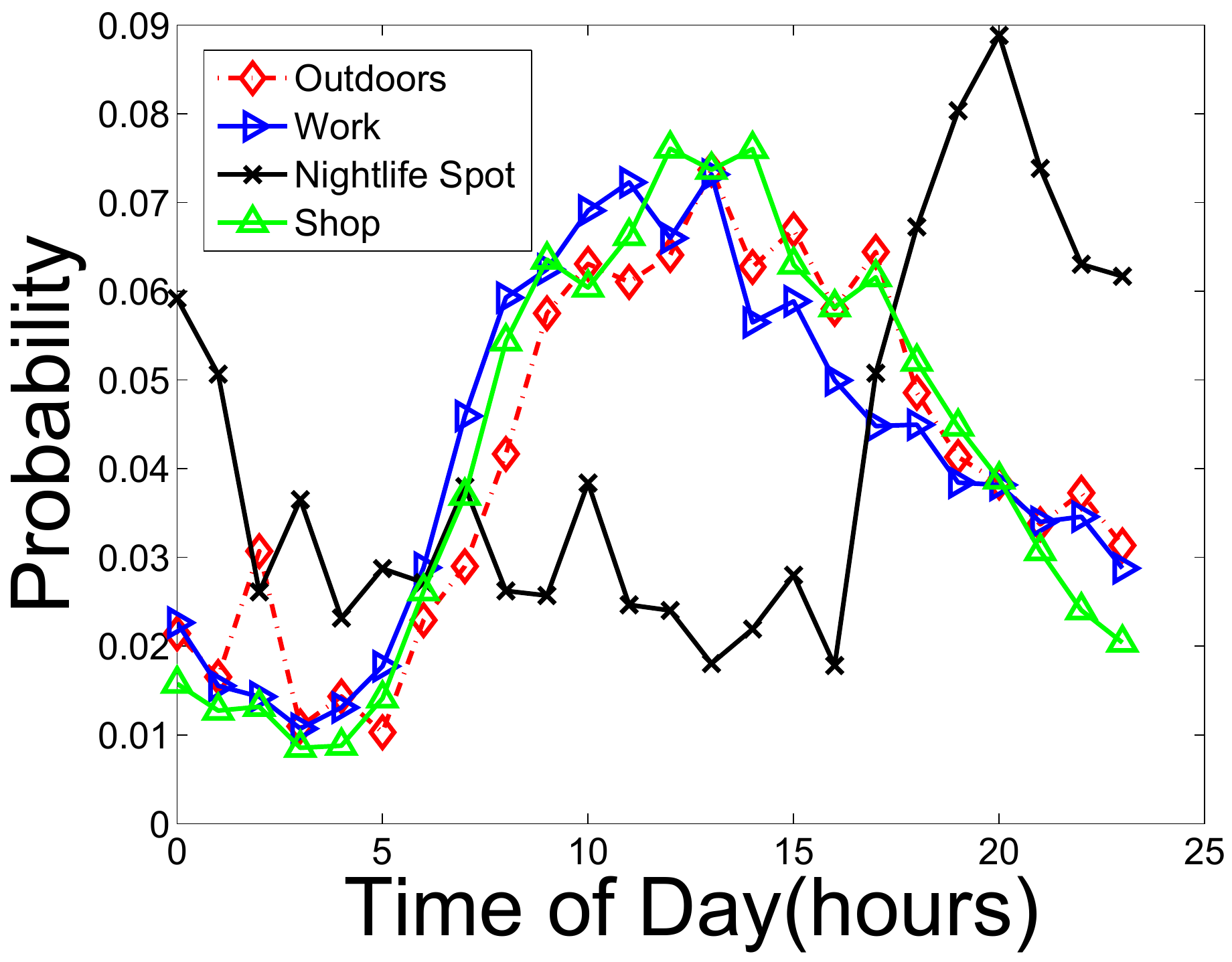}
		\label{fig:time_LA} }
	\subfigure[LA]{\includegraphics[clip=true,width=1.6in]{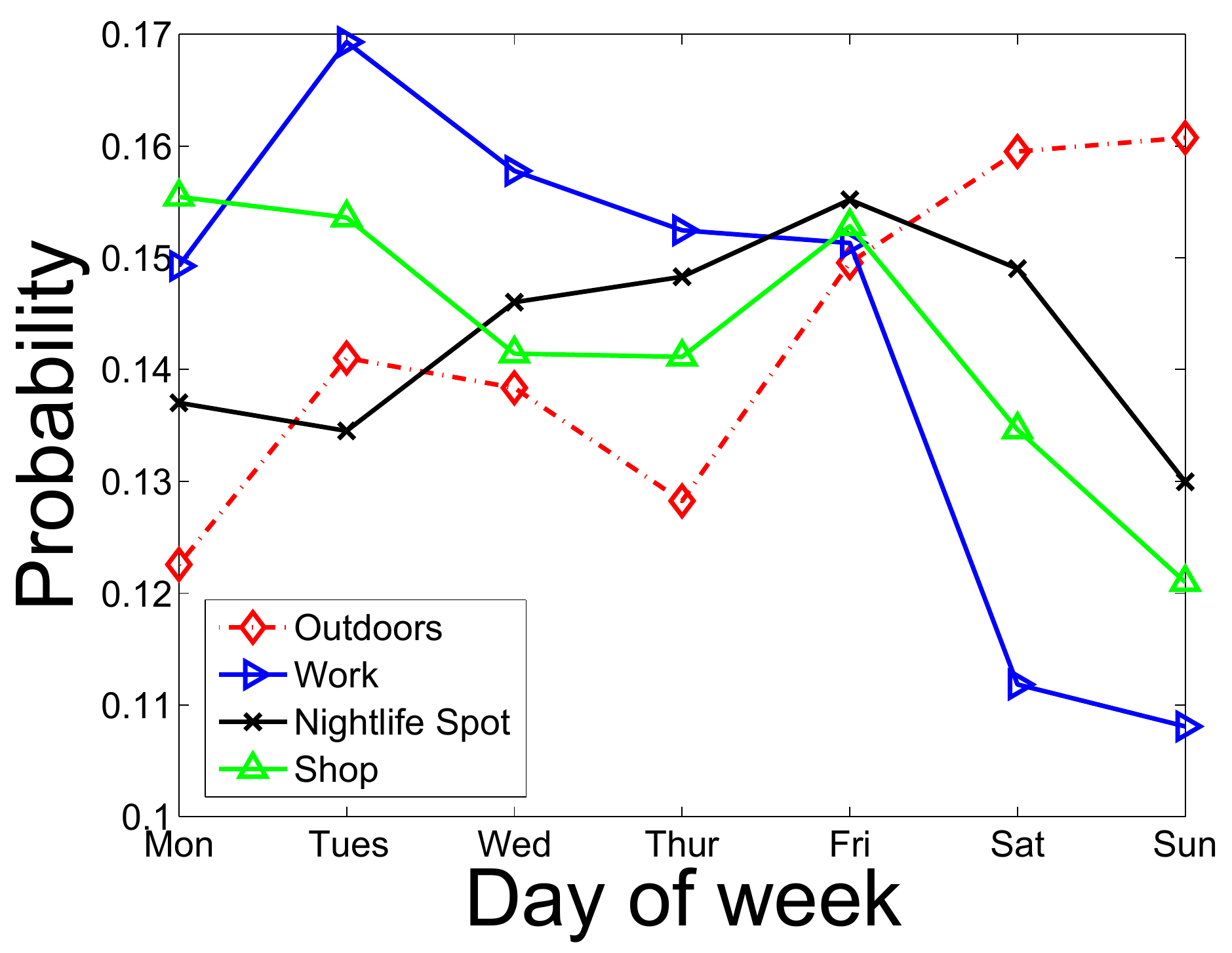}
		\label{fig:week_LA} }
	\subfigure[NYC]{\includegraphics[clip=true,width=1.6in]{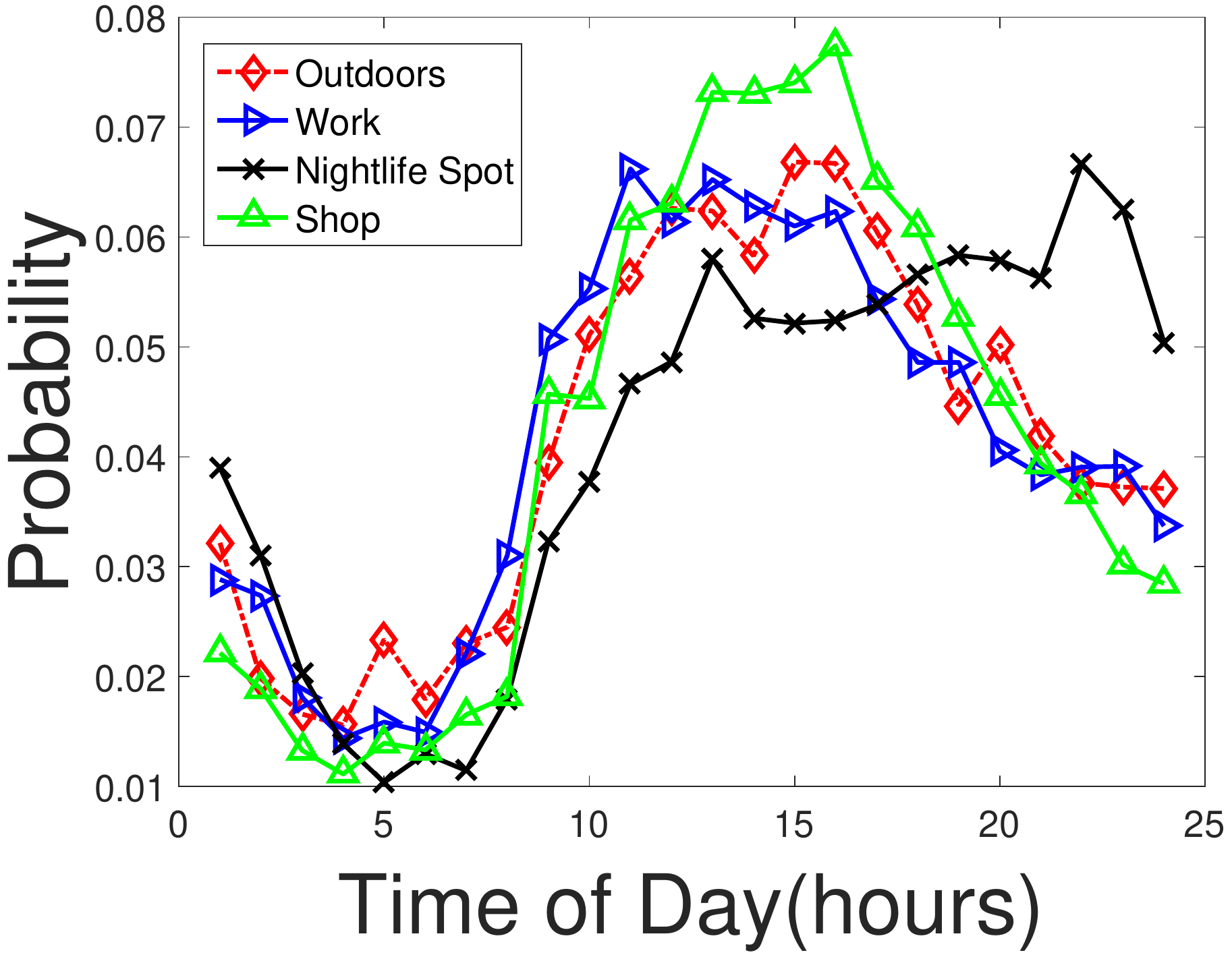}
		\label{fig:time_NY} }
	\subfigure[NYC]{\includegraphics[clip=true,width=1.6in]{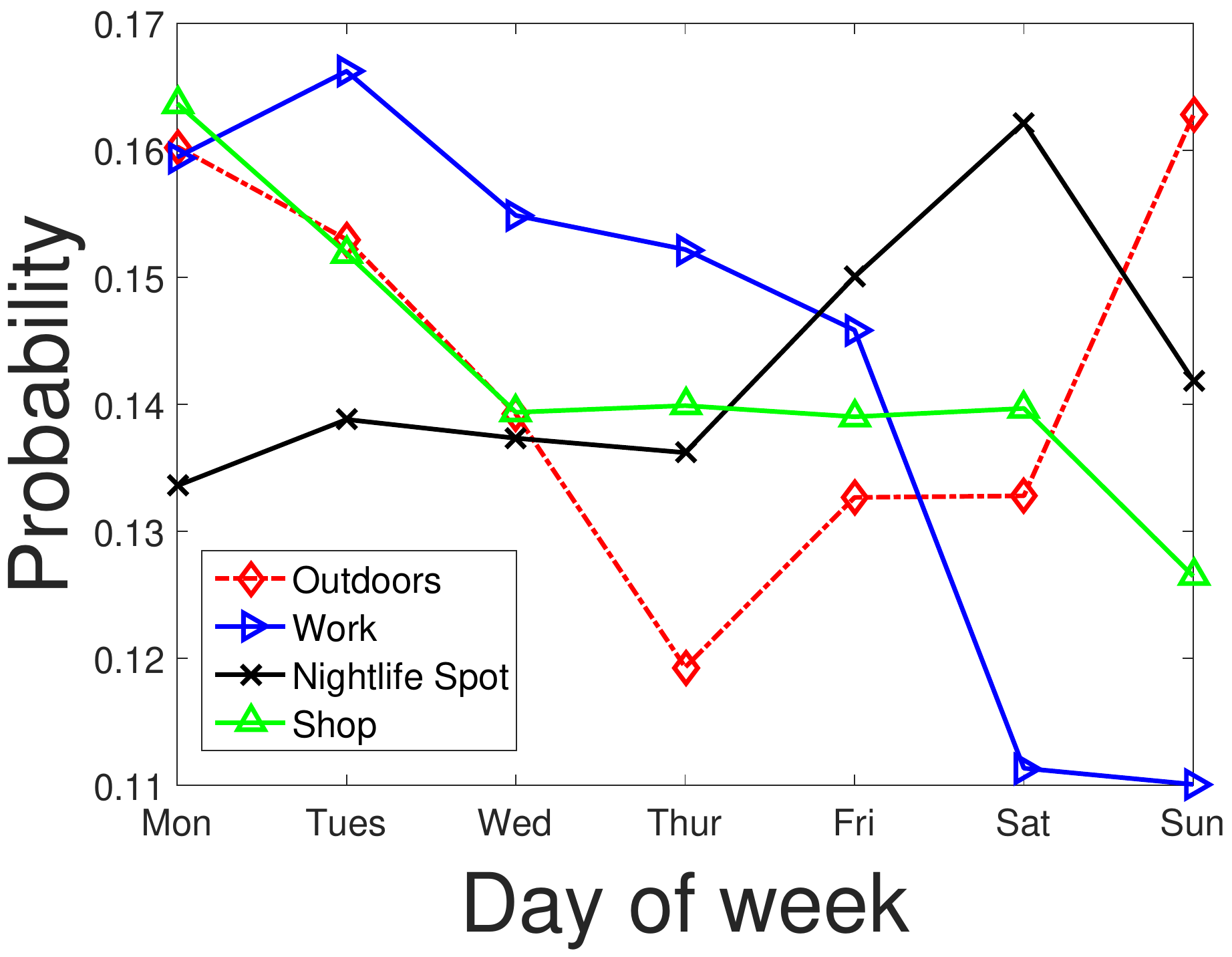}
		\label{fig:week_NY} }
	\caption{Check-in Periodicity Analysis}
\end{figure}

Human mobility has been well known for its periodic property \cite{Eagle09,Lizhenhui2010,Cho2011}. For next POI recommendation, we focus more on the transition periodicity of location categories. For example, people may regularly stop by coffee stalls, starbucks stores, to grab a cup of coffee on their way to work in the morning, which can be explained as a periodic transition pattern from coffee shop to workplace on weekday morning. The next POI is highly likely related to the previous POI. For example, after taking part in intense outdoor activities, e.g., hiking, running, some user may prefer to have high-protein meals in restaurants like Steakhouse rather than a Juicy Bar. Fig.\ref{fig:time_LA} and Fig.\ref{fig:week_LA} plot the check-in probabilities of the top-4 most popular location categories over time of day (hours) and day of week respectively, based on the check-in data of LA, collected from Foursquare (the detailed data description will be seen in Section \uppercase\expandafter{\romannumeral3}). The categorical mobility periodicity is very obvious, e.g., the work places are often checked on weekdays. Another interesting observation is that the check-ins of nightspots occur most often on Friday and least often on Sunday. Fig.2 plots the transition probabilities between categories along with the day of week. We observe that the transition preference shown in Fig.2(a) is significantly different from that of Fig.2(g) but somehow similar to that of Fig.2(b), which indicates that there exist several latent transition patterns and such patterns may play a key role for our next POI recommendation. However, such patterns are learned from all users' visit history (i.e., global patterns), thus suffering from lack of personalization as users may exhibit distinct latent transition patterns under the same contextual scenario. To support personalized latent behavior patterns, we extend the proposed global model to accommodate personalized pattern distribution. In addition, the personalized model provides much more flexibility and interpretability than global model through enough check-in history and rich contextual features. For example, as shown in Fig.2, the transition preference of LA users is different from that of NYC users.

In addition to providing personalized recommendations of next POIs to users, our proposed model also recommend new POIs that users may be interested in but have not visited before. More specifically, the next new POI recommendation problem is to recommend new POIs in terms of the historical check-ins of the user to be visited next give the user's current location. Recently, this task becomes increasingly popular and useful, since it not only helps users to explore interesting new places in the city, but also creates the opportunities for businesses to increase their revenues by attracting and discovering potential customers. Thus, we further consider the task of next new POI recommendation in LBSNs, which is a much harder task than standard successive POI recommendation, as it is challenging to infer user preference for potential new POIs from the unobserved transitions based on the sparse historical data.  

\begin{figure}[t]
	\centering
	\subfigure[LA]{\includegraphics[clip=true,width=1.6in]{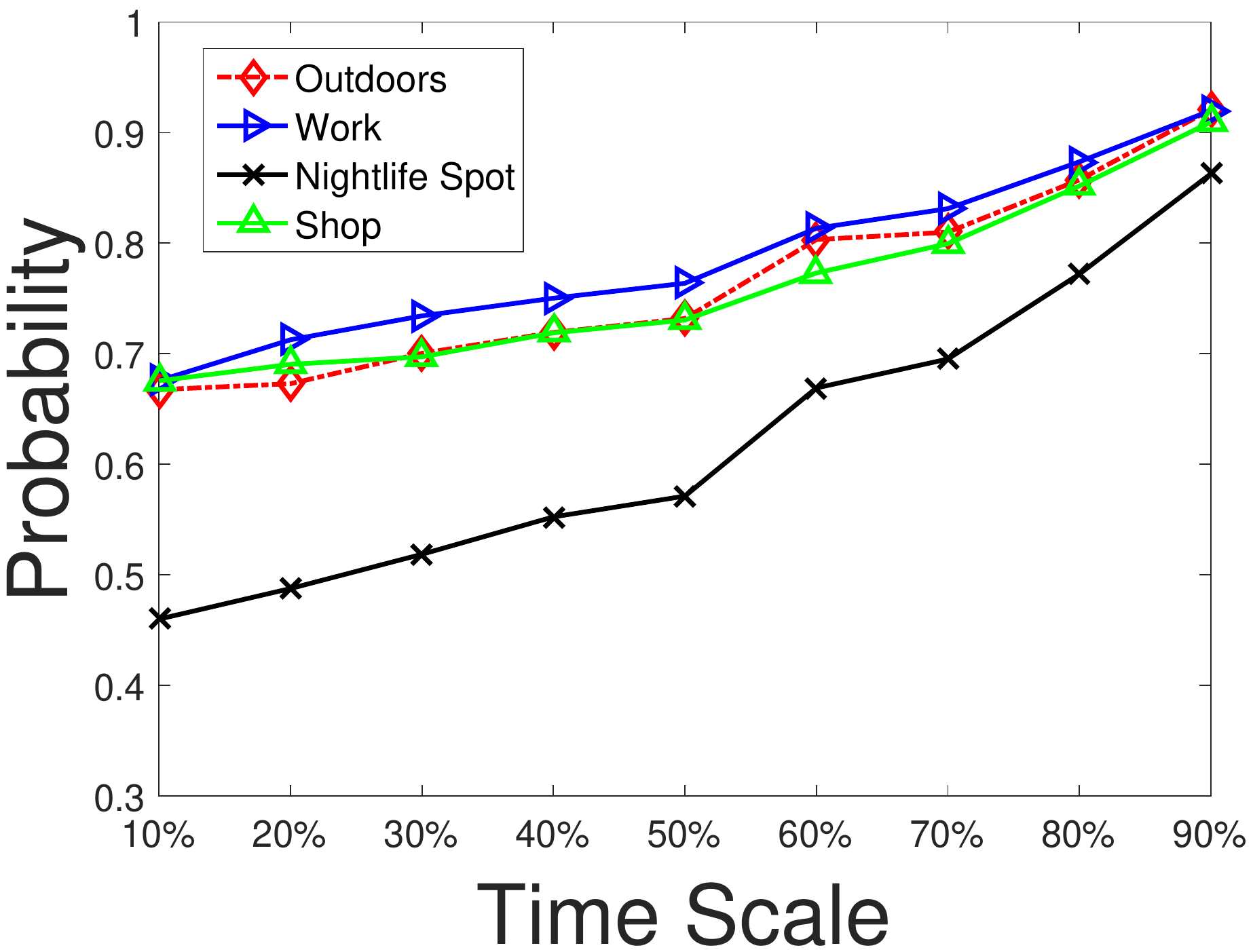}
		\label{fig:new_Category_LA} }
	\subfigure[NYC]{\includegraphics[clip=true,width=1.6in]{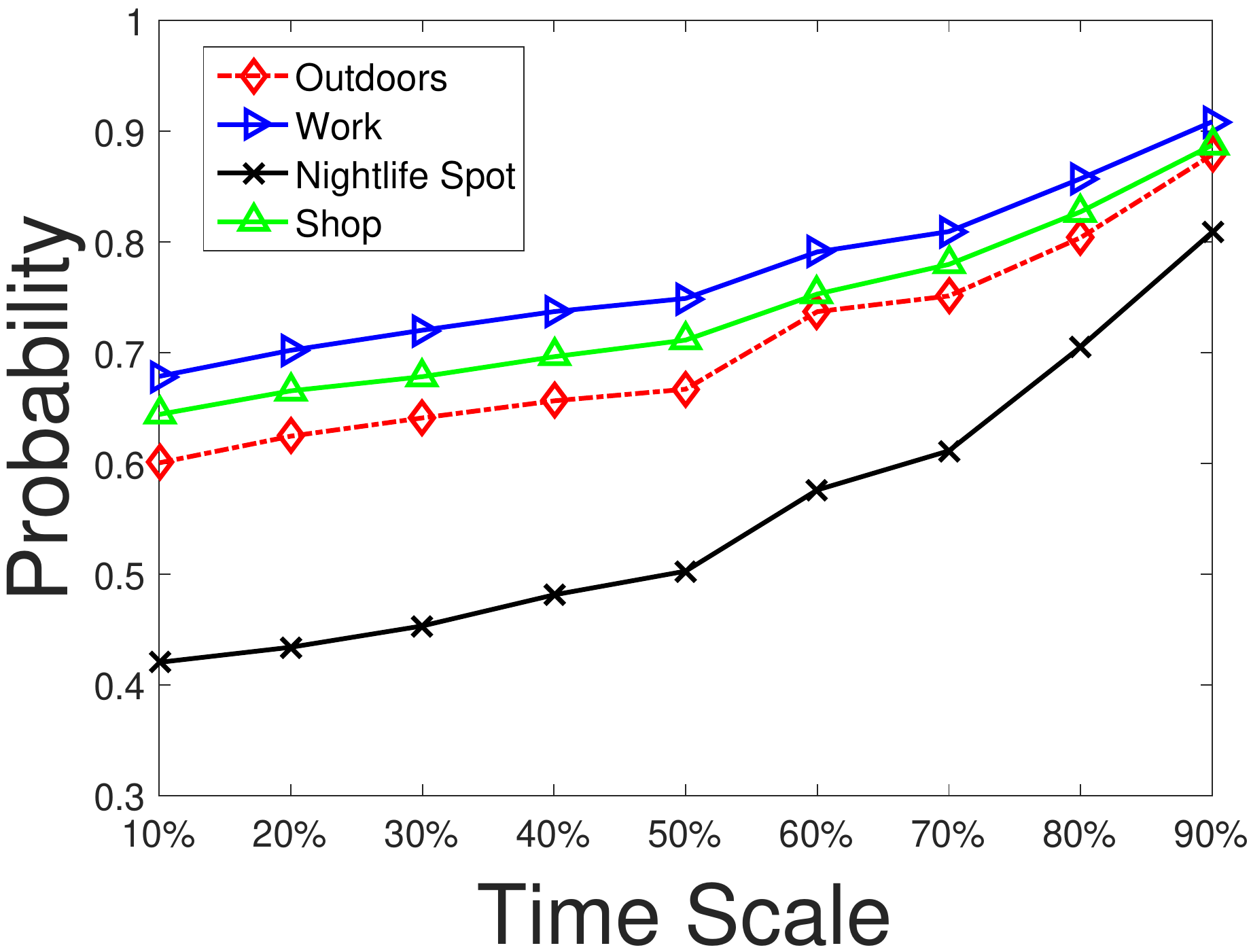}
		\label{fig:new_Category_NYC} }
	\caption{The ratio of new POIs on Location Category along with time scale} \label{fig:new_category}
\end{figure}

In fact, our observation is that there is a big fraction of new POIs to be visited in both datasets (see Section \uppercase\expandafter{\romannumeral3} for more details), which implies that the task of offering new POIs for users is important. Meanwhile, by exploring the proportion of new POIs for the top-4 most popular location categories, as the statistics shown in Fig.\ref{fig:new_category}, we find that users are more likely to visit new places at the early stage in both categories and the ratio of new POIs is distinct between each category. Again, we also observe that the category of Nightlife Spot has the lowest ratio for new POIs, which indicates that if a user visits a nightspot for the first time, there is a higher chance she might return and check-in again. Obviously, there are several latent behavior patterns for users and they are important for the next new POI recommendation. However, the traditional recommender systems cannot deal with the next new POI recommendation problem, because they only provide the routinely visited locations of the user for her next movement. In contrast, our proposed model can deal with the task of next new POI recommendation by nature, since it integrates categorical influence into pattern distribution and derives the spatial preference of users on new POIs to predict the transition probabilities of the users on the new POIs.

In this paper, we attempt to jointly model next POI recommendation under the influence of user's latent behavior pattern. Meanwhile, we observe that users often visit new POIs that they have not been visited before and the proposed model is able to recommend new POIs to be visited next given a user's current lication. We propose to adopt a third-rank tensor to model the successive check-in behaviors. By incorporating the softmax function to fuse the personalized Markov chain with the aforementioned latent pattern's influence, we furnish a Bayesian Personalized Ranking (BPR)\cite{Rendle2009BPR} approach and derive the optimization criterion accordingly. In the model learning phase, the Expectation Maximization (EM)\cite{Neal1998} is used to estimate the model parameters. 

\begin{figure*}[t]
	\centering
	\subfigure[Monday-LA]{\includegraphics[width=0.94in]{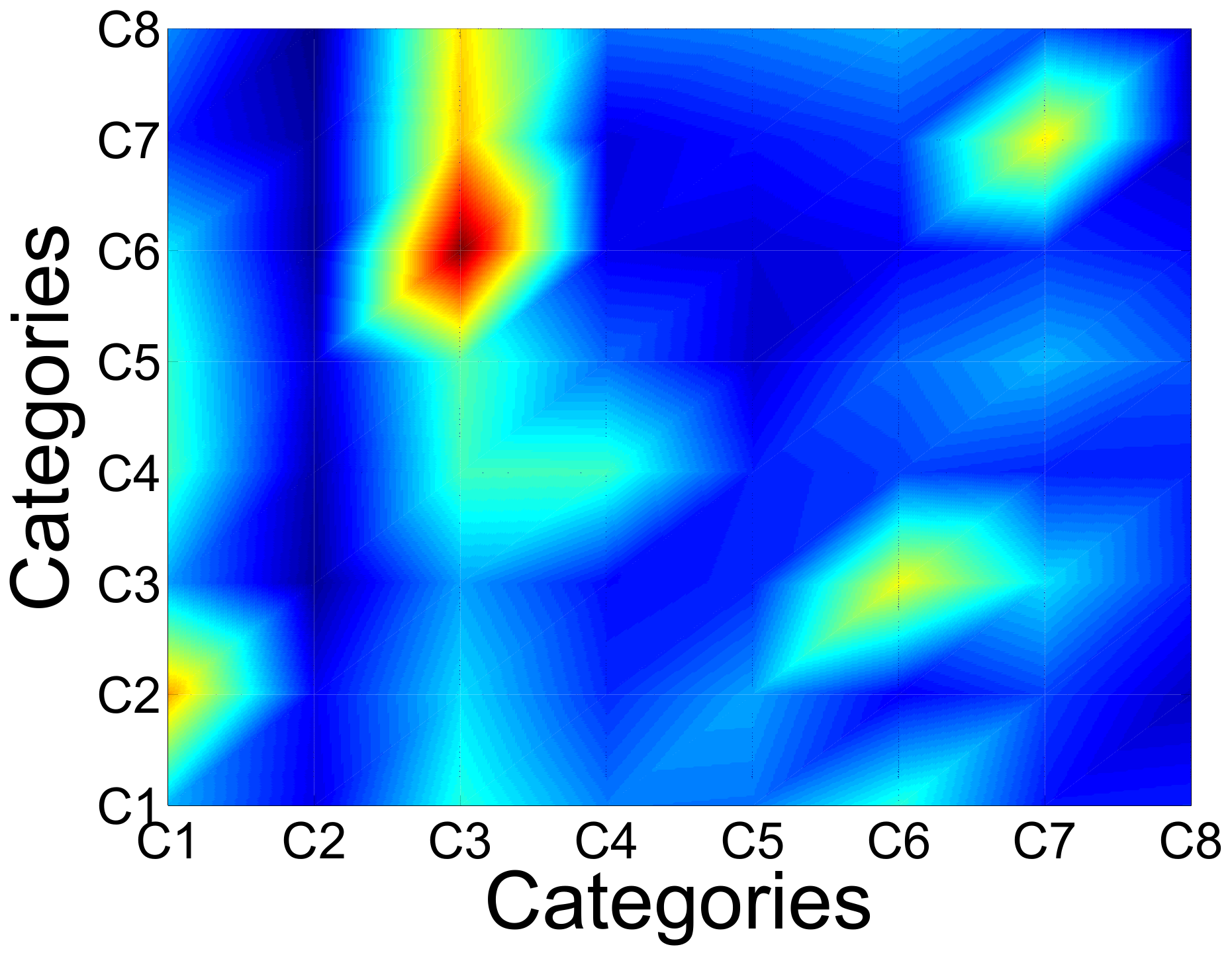}
		\label{fig:mon_LA} }
	\subfigure[Tuesday-LA]{\includegraphics[width=0.94in]{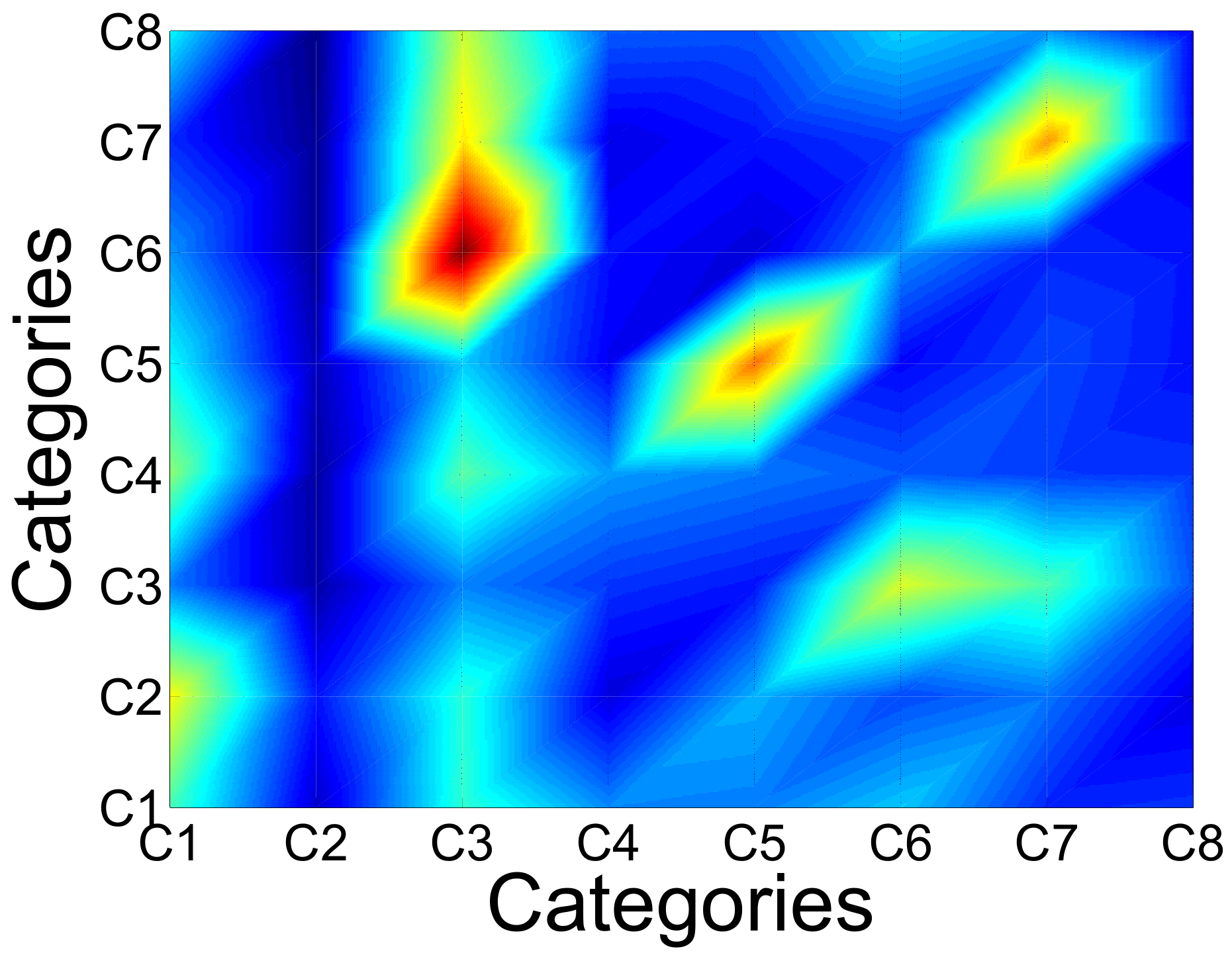}
		\label{fig:tues_LA} }
	\subfigure[Wednesday-LA]{\includegraphics[width=0.94in]{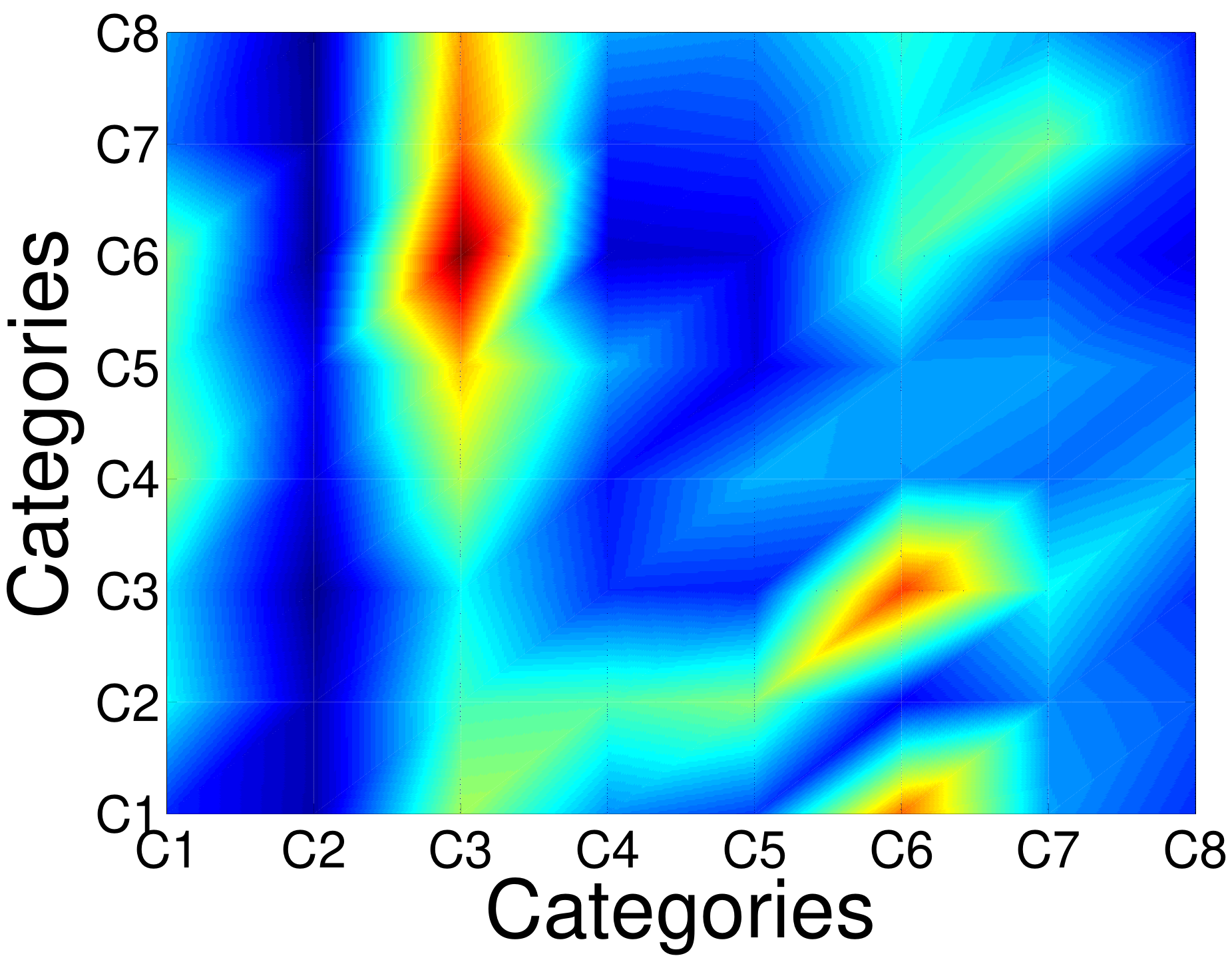}}
	\subfigure[Tuesday-LA]{\includegraphics[width=0.94in]{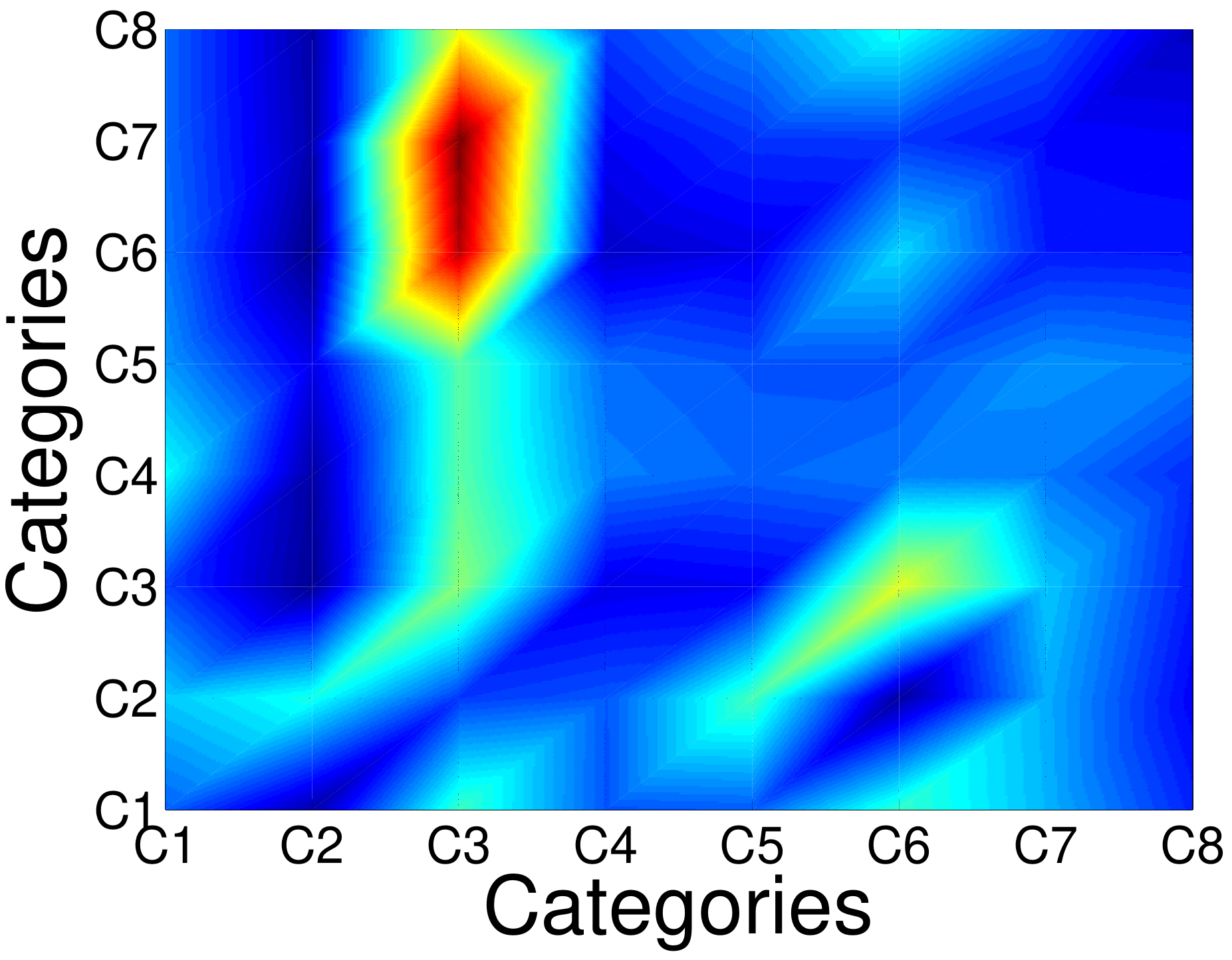}}
	\subfigure[Friday-LA]{\includegraphics[width=0.94in]{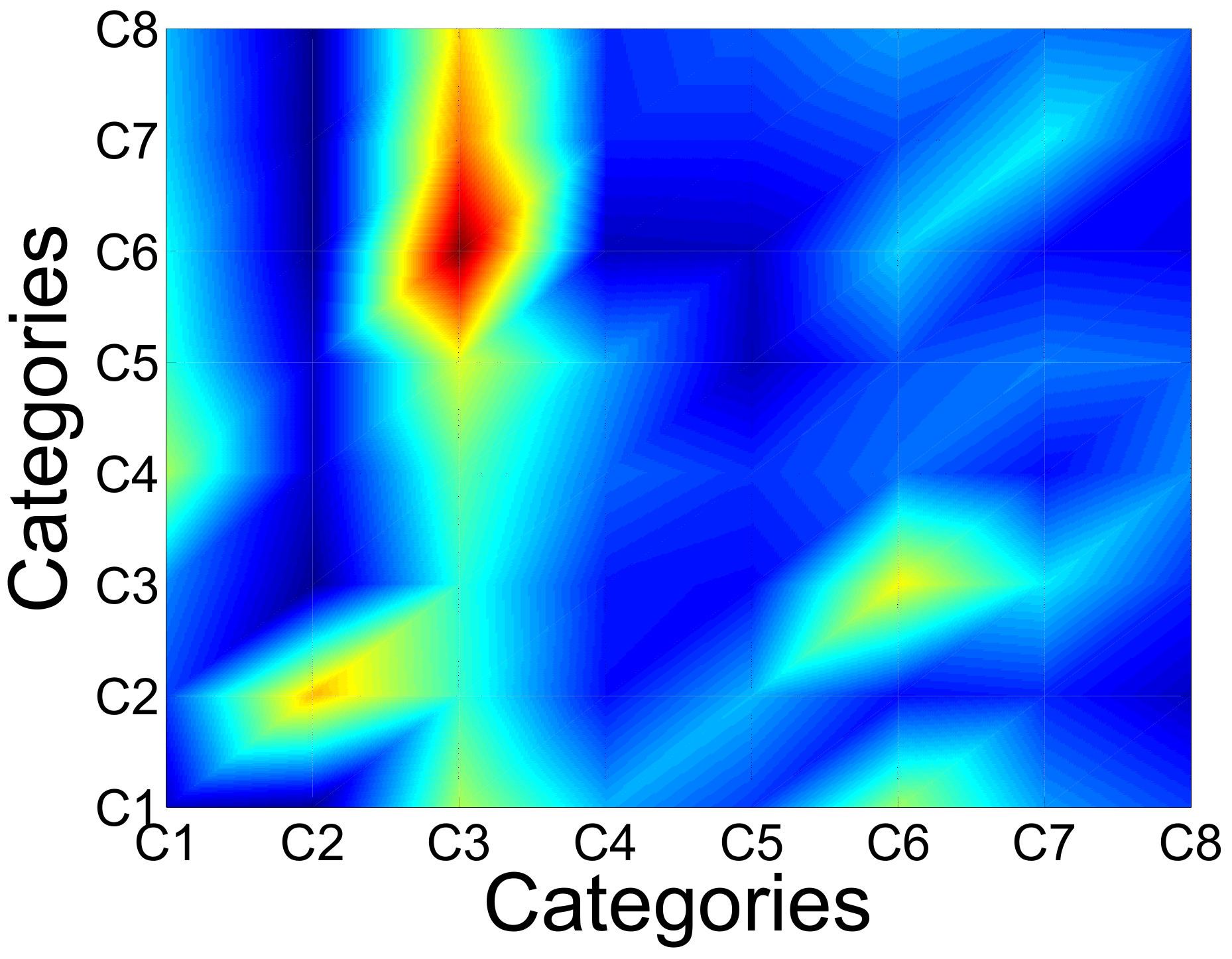}
		\label{fig:fri_LA} }
	\subfigure[Saturday-LA]{\includegraphics[width=0.94in]{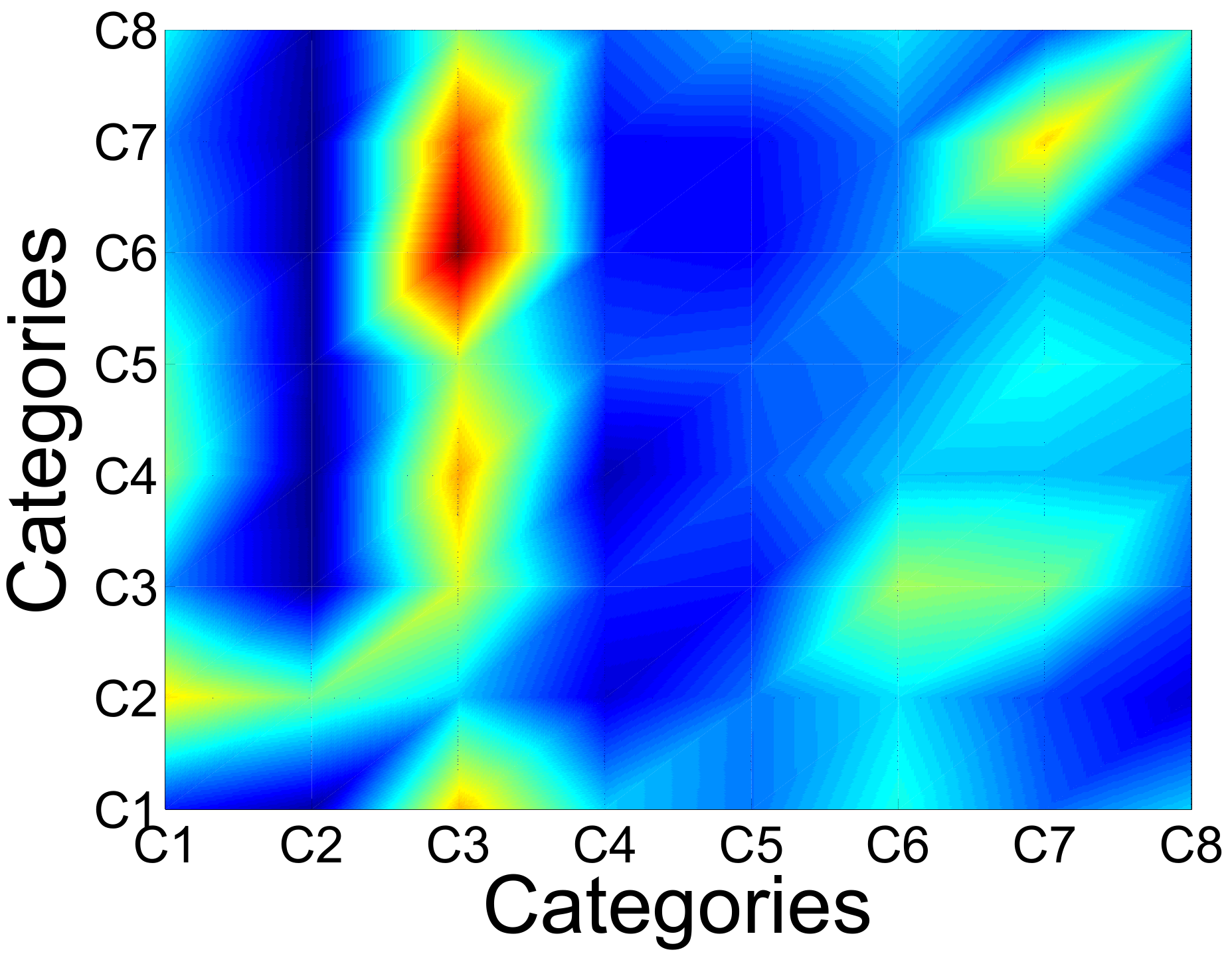}
		\label{fig:sat_LA} }
	\subfigure[Sunday-LA]{\includegraphics[width=0.96in]{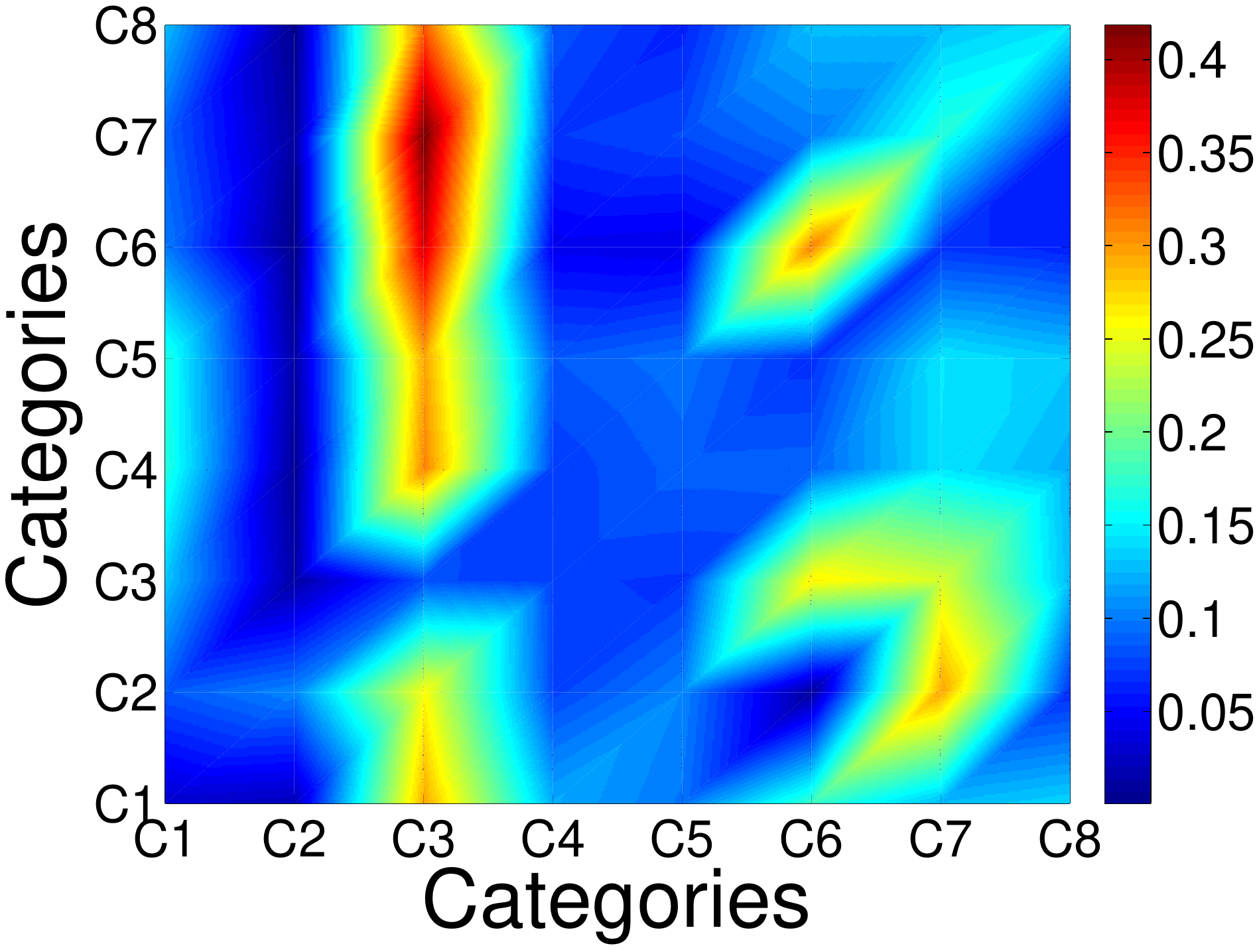} }
	\subfigure[Monday-NY]{\includegraphics[width=0.94in]{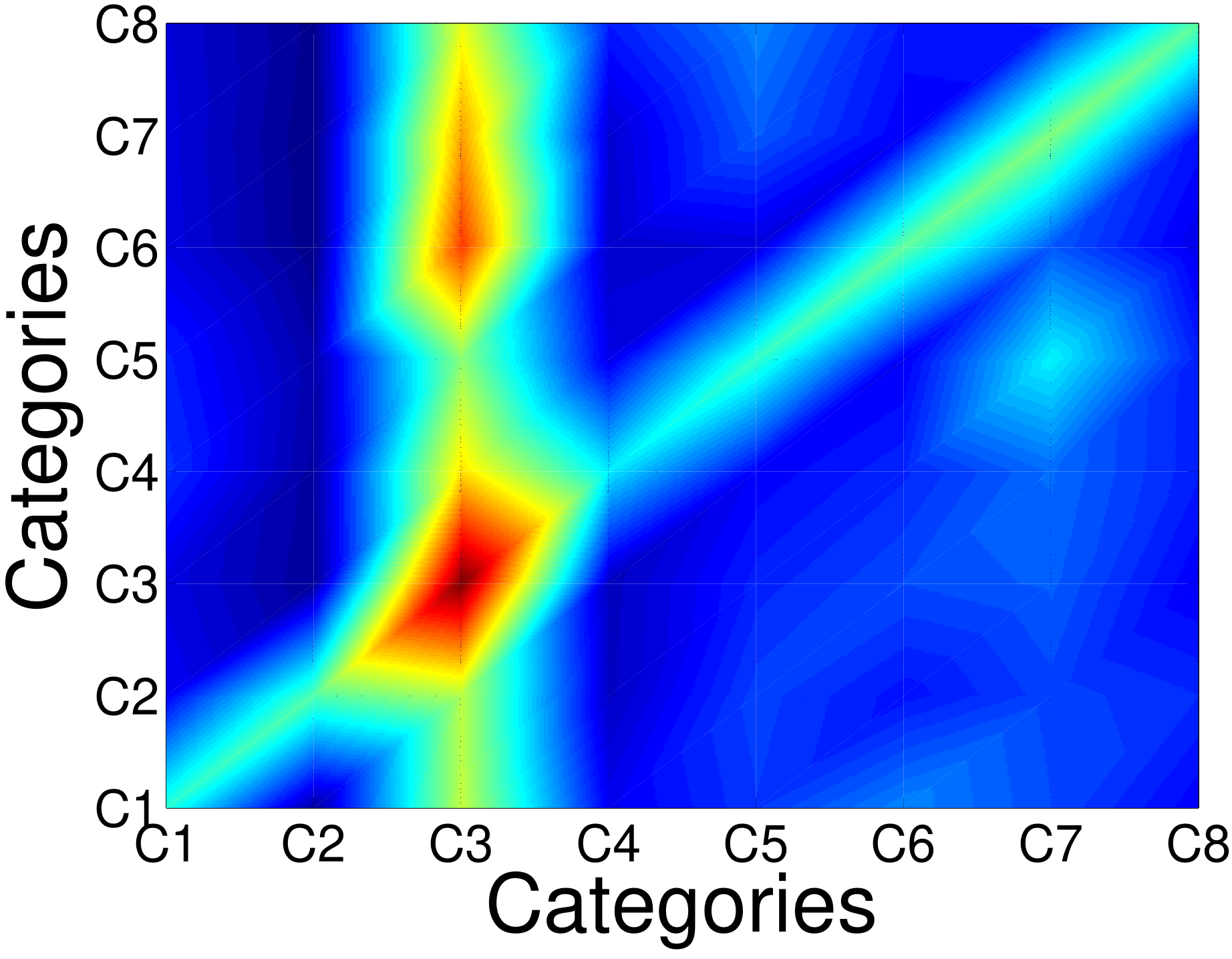}
		\label{fig:mon_LA} }
	\subfigure[Tuesday-NY]{\includegraphics[width=0.94in]{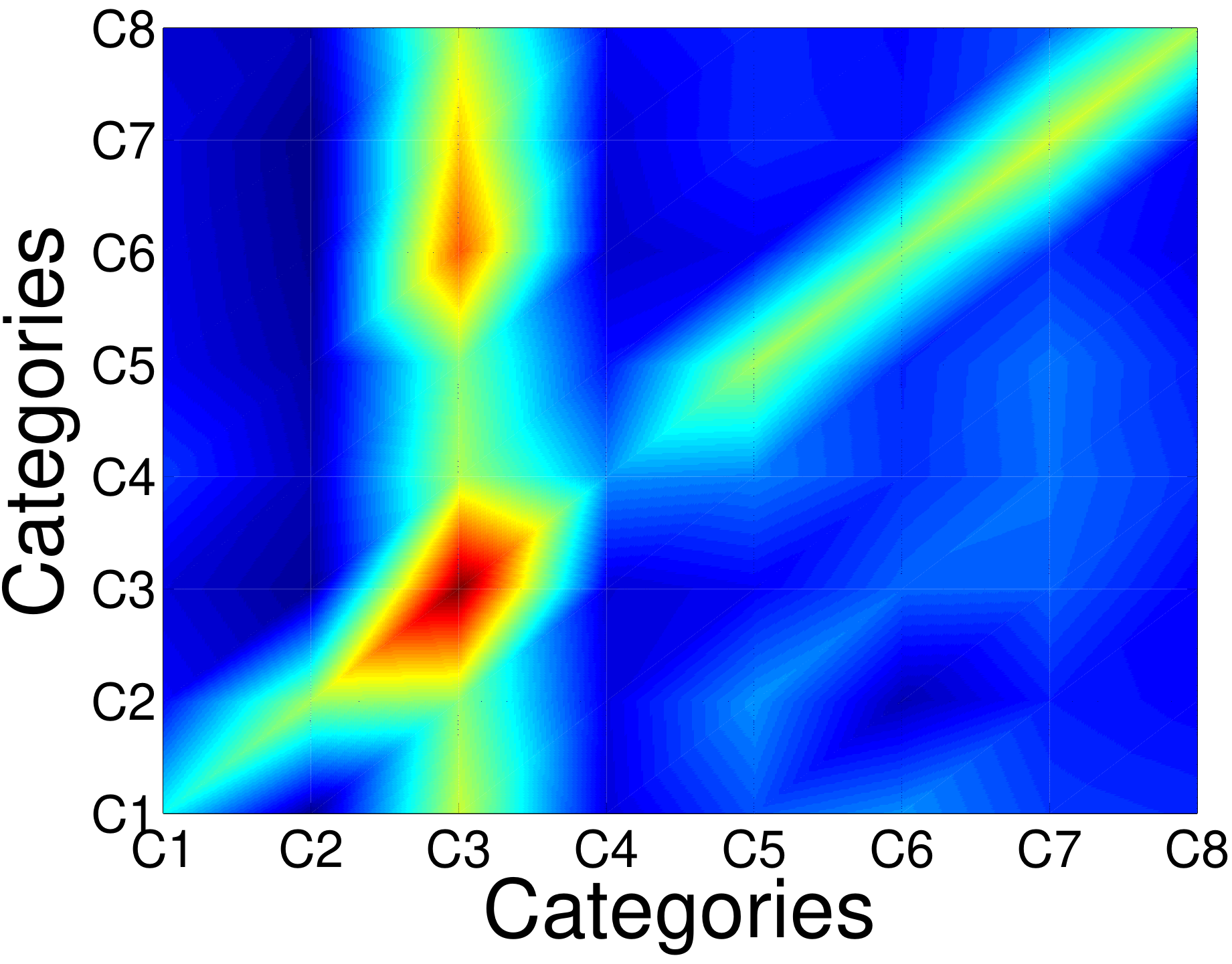}
		\label{fig:tues_LA} }
	\subfigure[Wednesday-NY]{\includegraphics[width=0.94in]{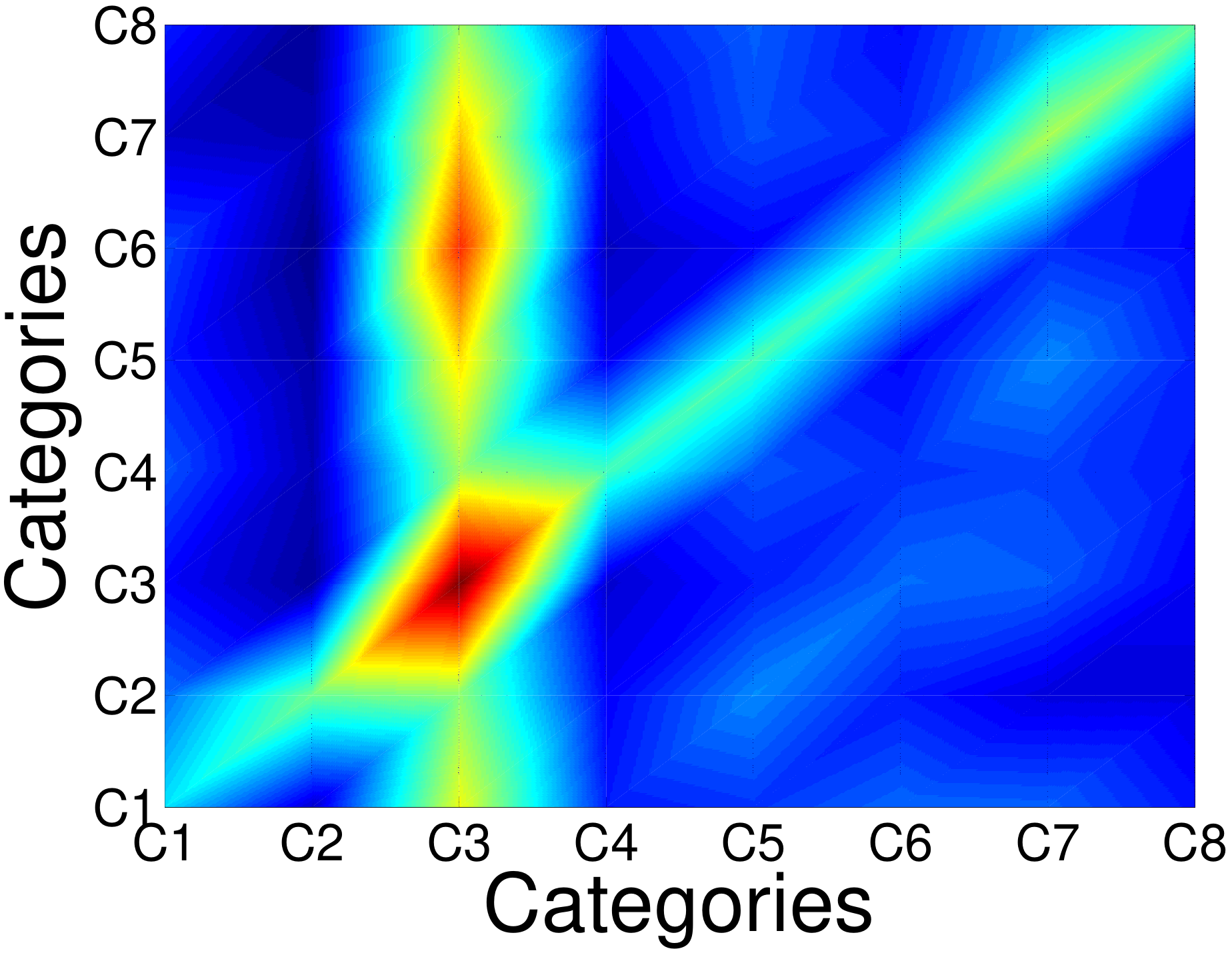}}
	\subfigure[Tuesday-NY]{\includegraphics[width=0.94in]{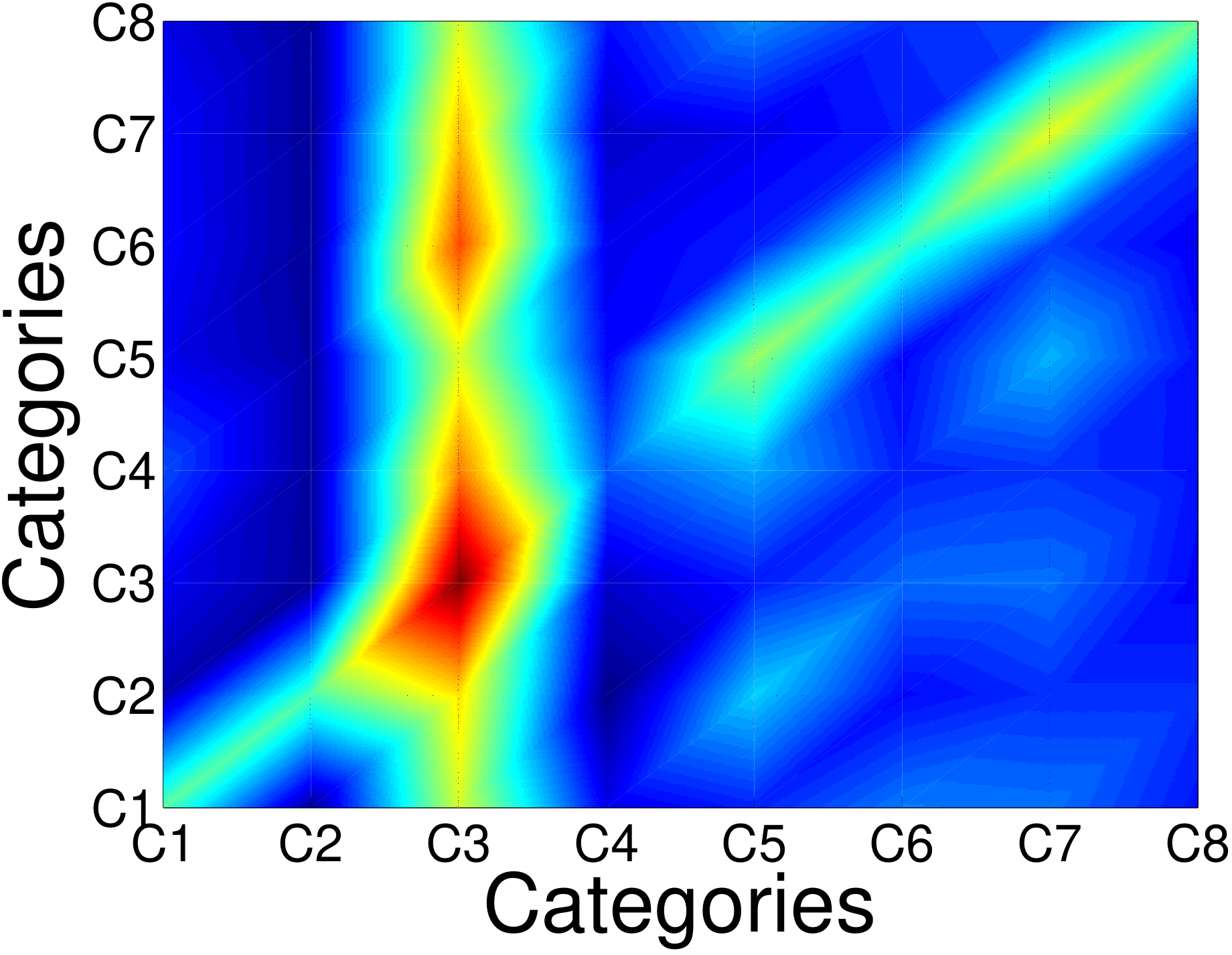}}
	\subfigure[Friday-NY]{\includegraphics[width=0.94in]{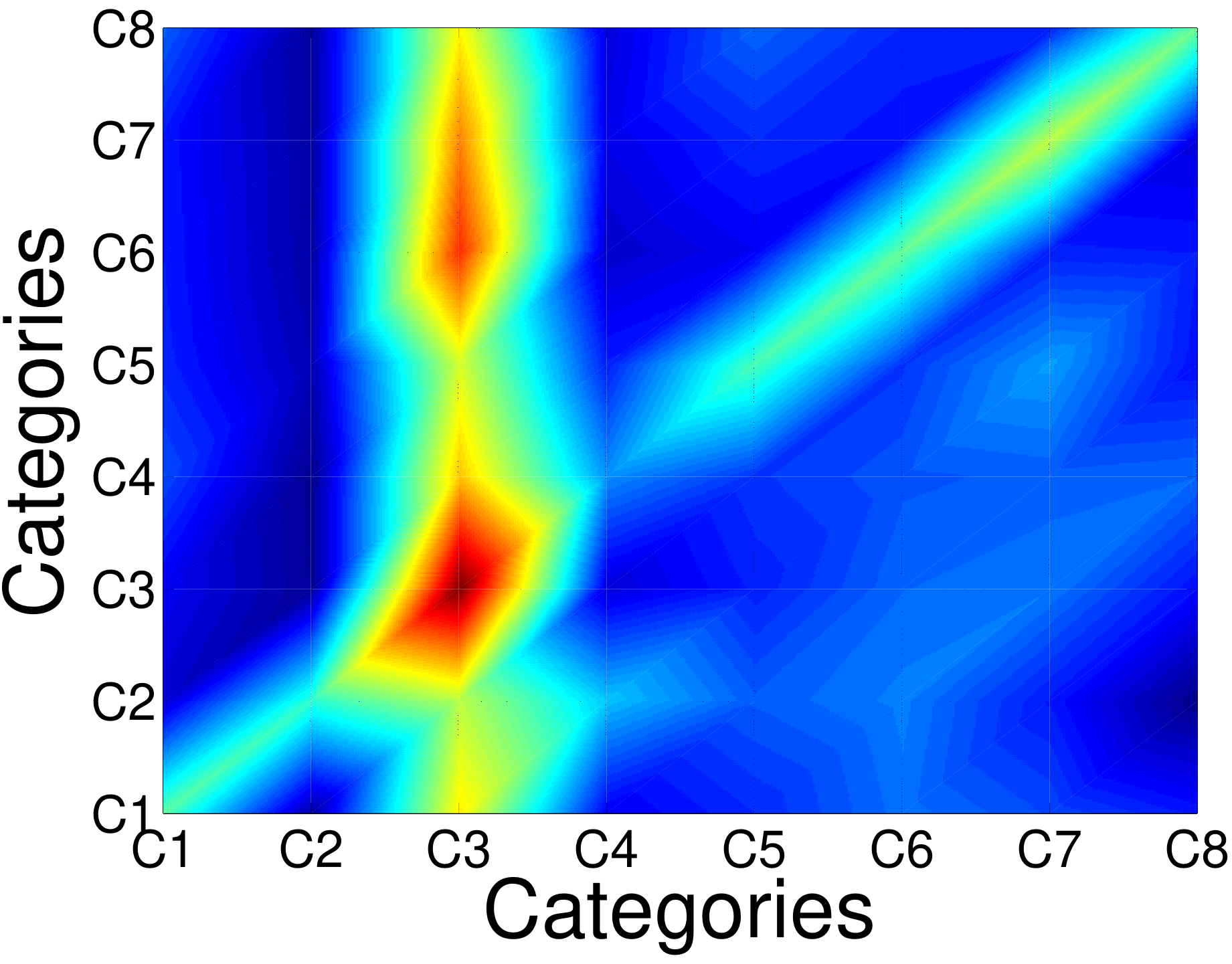}
		\label{fig:fri_LA} }
	\subfigure[Saturday-NY]{\includegraphics[width=0.94in]{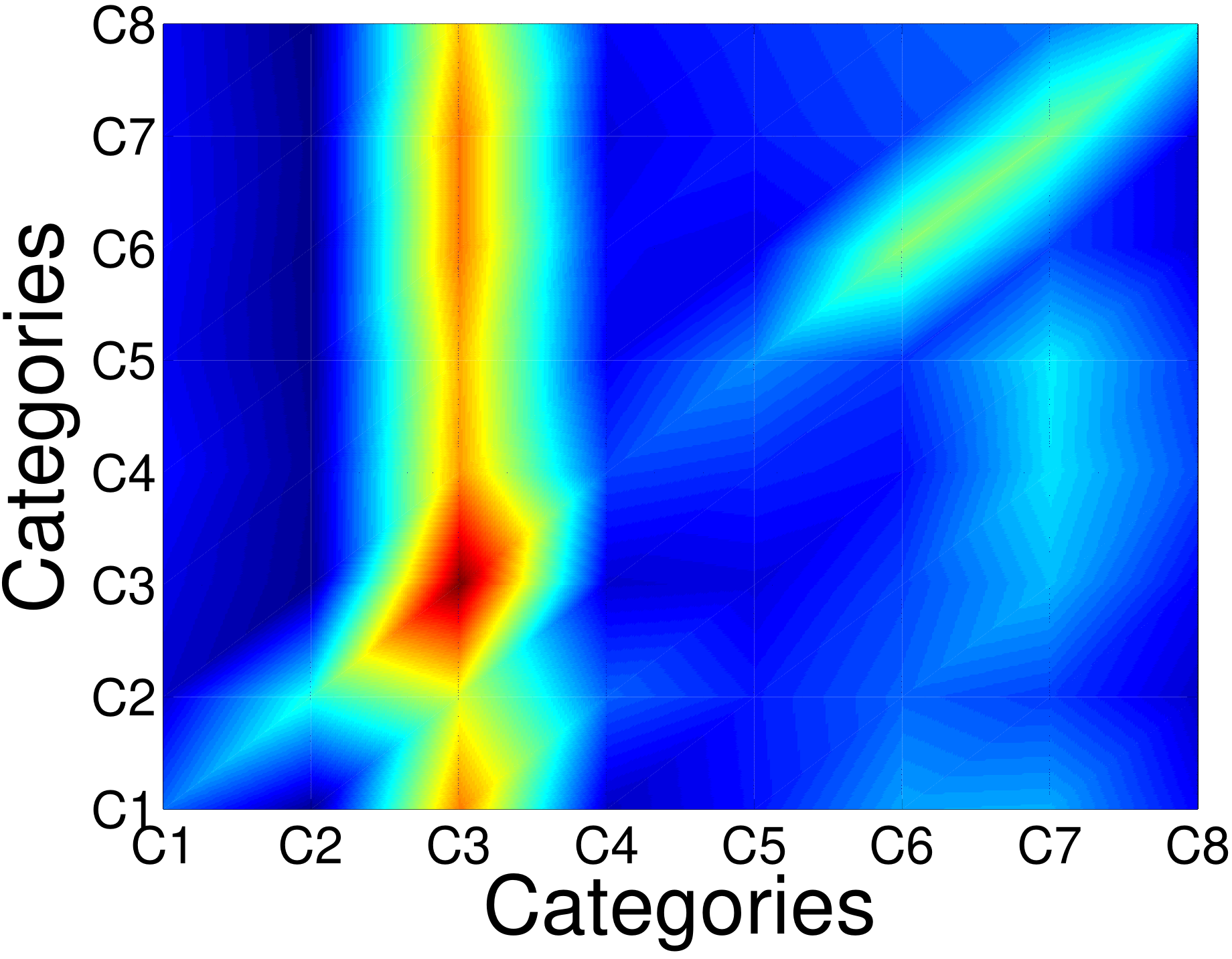}
		\label{fig:sat_LA} }
	\subfigure[Sunday-NY]{\includegraphics[width=0.96in]{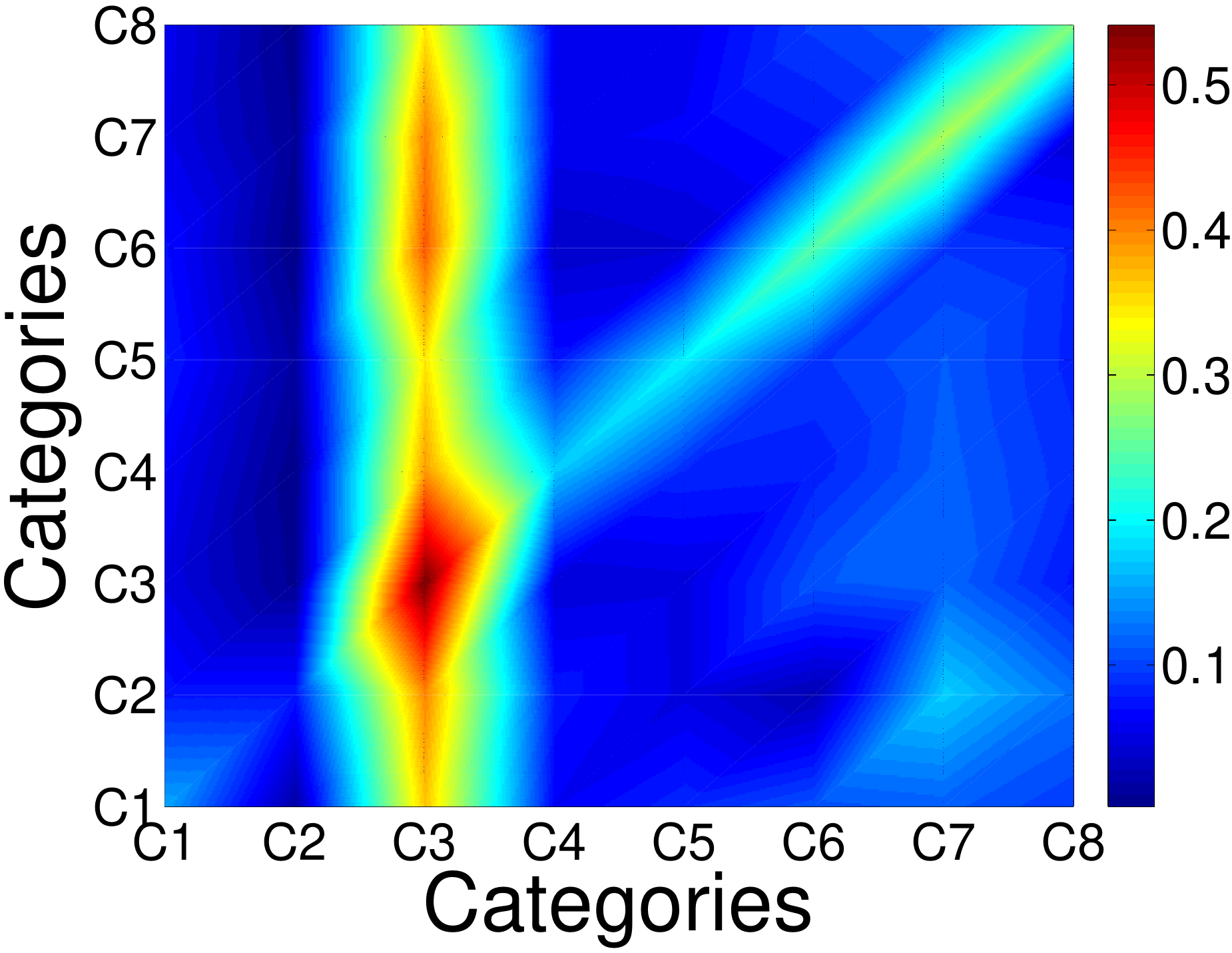} }
	\caption{Statistics on Location Category Transitions Along with The Day of Week. Categories=\{$c_1$ : Arts \& Entertainment, $c_2$: College \& University, $c_3$ : Food, $c_4$ : Outdoors, $c_5$ : Work, $c_6$ : Nightlife Spot, $c_7$ : Shop, $c_8$ : Travel Spot\}}
	\label{fig:category_transition}
\end{figure*}

The main contributions of this paper can be summarized as follows:
\begin{itemize}
	\item We propose a unified tensor-based latent model to fuse the observed successive check-in behavior with latent behavior preference for each user to address a personalized next POI recommendation problem. The corresponding optimization criterion and learning steps/tricks have been carefully studied.
	
	\item We evaluate the proposed model by detailed experiments on two large-scale LBSN datasets and demonstrate that our method outperforms other state-of-the-art POI recommendation approaches by a large margin.
\end{itemize}

\section{Related Work}
Location recommendation has received intensive attention recently due to a wide range of potential applications. It was studied on GPS trajectory logs of hundreds of monitored users \cite{zheng2009mining}. With the easy access of users' check-in data in LBSNs, many recent studies have been conducted for POI recommendation, which can be roughly classified into four categories:

1) \textbf{\textit{time-aware POI recommendation}} which mainly leverages the temporal influence on POIs to enhance the recommendation performance. Yuan et al. assume that users tent to visit different locations at different time and proposed time-aware POI recommendation algorithm. Specifically, they proposed approach extends the user-based POI recommendation by leveraging the time factor when computing the similarity between two users as well as considering the historical check-ins at time $t$, rather than at all time to make POI recommendation \cite{yuan2013time}.
Gao et al. investigated the temporal cyclic patterns of user check-ins in terms of temporal non-uniformness and temporal consecutiveness \cite{gao2013exploring}.
Yin et al. proposed a temporal recommender system and modeled the user behavior based on intrinsic interest as well as the temporal context \cite{chen2013terec}. Zhao et al. proposed a spatial-temporal latent ranking method to recommend users most possible successive POIs by designing a time indexing scheme to smoothly encode time stamps to particular time ids and then incorporating the time ids into the proposed model \cite{zhao2016stellar}.

2) \textbf{\textit{geographical influence enhanced POI recommendation}} which exploits the ``geographical clustering phenomenon'' of check-in activities to improve the POI recommendation system \cite{liu2013learning}. Liu et al. proposed a geographical probabilistic factor analysis framework for POI recommendation by combining geographical influence with Bayesian non-negative matrix factorization (BNMF). Specifically, they used a Gaussian distribution to represent a POI over a sampled region and BNMF is used to capture user preference from check-in data.
Ye et al. delved into POI recommendation by investigating the geographical influences among locations and proposed a system that combines user preferences, social influence
and geographical influence \cite{ye2011exploiting}.

3) \textbf{\textit{content-aware POI recommendation approaches}} which propose to detect users' current locations by analyzing their published tweets or to rank POIs by analyzing user's comments on them to alleviate the problem of data sparsity. 
Chen et al. build a detection model to mine user interest from short text and establish the mapping between location function and user interest \cite{chen2013interest}.
Gao et al. studied both POI-associated contents and user sentiment information (e.g., user comments) into POI recommendation and reported their good performance \cite{gao2015content}. 
However semantic analysis is a very challenging research issue as most of comments in LBSN are short and contextually ambiguous. 

4) \textbf{\textit{social influence enhanced POI recommendation}} which is inspired by the intuition that friends of LBSNs tend to have more common interests. By inferring the social relations, the quality of recommendation could be enhanced. However, there are other opinions of leveraging social influence in the literature, as previous studies also report a large number of friends share nothing in terms of POI \cite{ye2010location}. And E. Cho {\it et al.} report their findings that the long-distance travel is more influenced by social relations \cite{Cho2011}.

Some very recent works have incorporated group behaviors into recommender systems for enhancing performance. T. Yuan {\it et al.} proposed a Group-Sparse Matrix Factorization (GSMF) approach to factorize the rating matrices for multiple behaviors into a user and item latent factor space \cite{DBLP:conf/aaai/YuanCZQL14}. H. Wang {\it et al.} proposed a group-based algorithm for POI recommendation \cite{Grouppoi} by grouping users of similar interests based on their frequently visited locations' category hierarchy. Chen {\it et al.} proposed a novel two-step approach for personalized successive POI recommendation: First, group-based category recommendation by designing a group-based tensor model to predict the location category preference; then, category-based location recommendation by proposing a distance weighted-HITS algorithm to rank the locations under a selected location category \cite{chen2016effective}.

Recently, researches have started to pay attention to exploiting deep network for recommender systems. One line of research is to integrate visual signals into personalized recommendation, which is conducted by utilizing visual features extracted from images using (pre-trained) Deep Convolutional Neural Network (Deep CNN) \cite{he2016vbpr}. The other line of work is to employ Recurrent Neural Networks (RNN) for location recommendation, which models spatial temporal contexts in each layer with time-specific and distance-specific transition matrices \cite{liu2016predicting}.

The next POI recommendation is a newly emerging task and even challenging. In the literature, there  exist only few works in which the sequential influence between successive check-ins is not yet well-studied. S. Feng {\it et al.} proposed a personalized ranking metric embedding method (PRME) to model personalized check-in sequences for next new POI recommendation \cite{feng2015personalized}. C. Cheng {\it et al.} proposed a tensor-based FPMC-LR model by considering the order relationship between visitings \cite{cheng2013you}. However, the periodicity of check-in data and categorical influence are not well studied. Moreover, the candidate set of POIs is filtered by simply removing the venues far from the previous checked-in POI to deal with the data sparsity. The yielded smaller set leads to a lower computation cost at expense of neglecting the experience of users whose check-in behavior patterns are exclusive from the majority ones and a failure of predicting those far way POIs.

\section{Data Description and Characteristics of Check-ins}
Before introducing the proposed approach, in this section, we first introduce two real-world LBSN datasets used in this paper and then conduct some empirical analysis on them to explore the spatial influence, check-in counts, temporal influence and exploration for new locations of users' successive check-in behaviors.
\subsection{Datasets}
We choose three large-scale datasets from real-world LBSNs, Foursquare and Gowalla, to conduct the experiments. \textit{Foursquare} check-in data is within Los Angeles and New York City, provided by \cite{bao2012location}, while \textit{Gowalla} dataset is from \cite{cheng2012fused} with a complete snapshot. For both datasets, we removed the users who checked in LSBN less than 10 times (\textit{Note that the categorical information of POIs are not included in Gowalla dataset}). We split the two datasets into two non-overlapping sets: for each user, the earliest 80\% of check-ins as training sets and the remaining 20\% checkins as test sets to evaluate the performance of different algorithms. The densities of transition tensor $\chi$ are $5.81\times 10^{-10}$ for Foursquare-LA, $2.81\times 10^{-9}$ for Foursquare-NYC and $1.01\times 10^{-6}$ for Gowalla respectively, which is extremely sparse. The statistics of the two datasets are listed in Table \ref{tab:statistics}.

\begin{table}[h]
	\caption{\label{tab:statistics}Dataset Statistics}
	\center
	\newsavebox{\tableboxaaaa}
	\begin{lrbox}{\tableboxaaaa}
		\begin{tabular}{l|l|l|l|l|l}
			\hline
			\hline
			& \#User & \#POI & \#Check-in & \#Avg.check-in & \ Sparsity  \\
			\hline
			Fours.-LA & 2823 & 84937 & 130583 & 46.25 & $5.81\times 10^{-10}$  \\
			\hline
			Fours.-NYC & 2579 & 97013 & 157404 & 61.03 & $5.85\times 10^{-9}$  \\
			\hline
			Gowalla & 1388 & 11235 & 301678 & 217.35 & $1.01\times 10^{-6}$ \\
			\hline
		\end{tabular}
	\end{lrbox}
	\scalebox{0.93}{\usebox{\tableboxaaaa}}
\end{table}

\begin{figure}
	\centering
	\subfigure{\includegraphics[clip=true,width=1.6in]{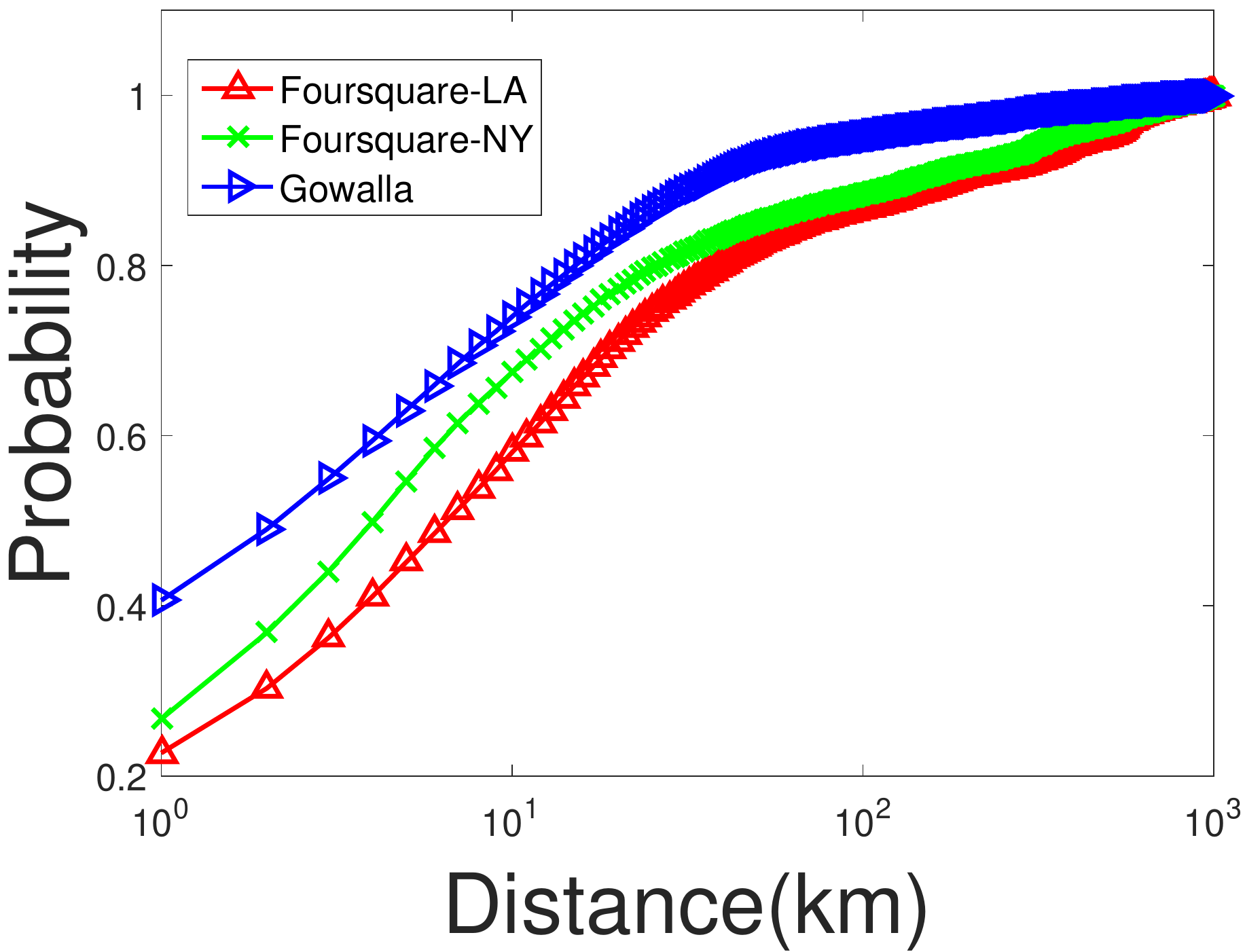}
		\label{fig:dataset_distance} }
	\subfigure{\includegraphics[clip=true,width=1.6in]{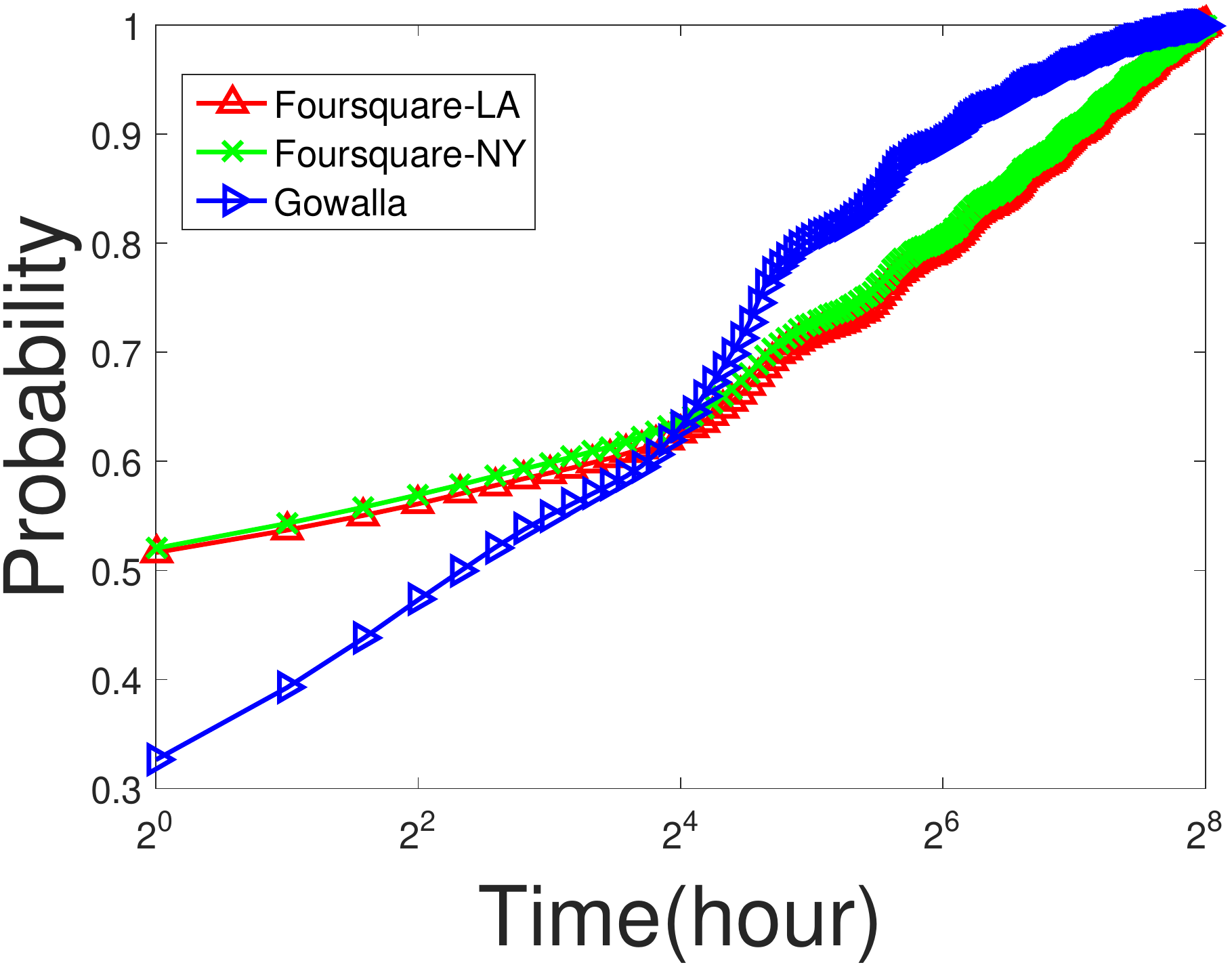}
		\label{fig:dataset_time} }
	\caption{CDF of time difference and geographical distance of two consecutive check-ins}
\end{figure}
\begin{figure}
	\centering
	\subfigure{\includegraphics[clip=true,width=1.6in]{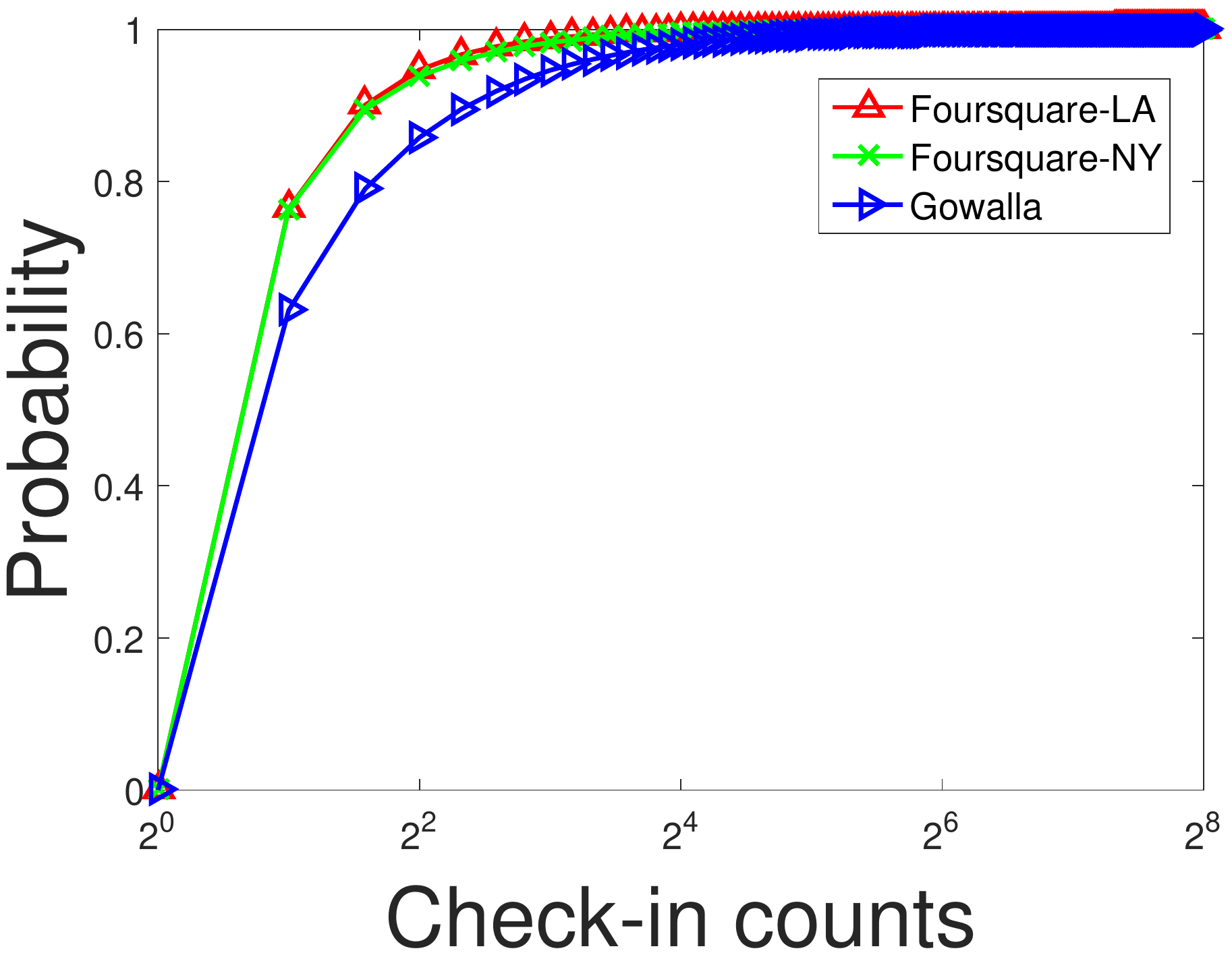}
		\label{fig:dataset_check_in_count} }
	\subfigure{\includegraphics[clip=true,width=1.6in]{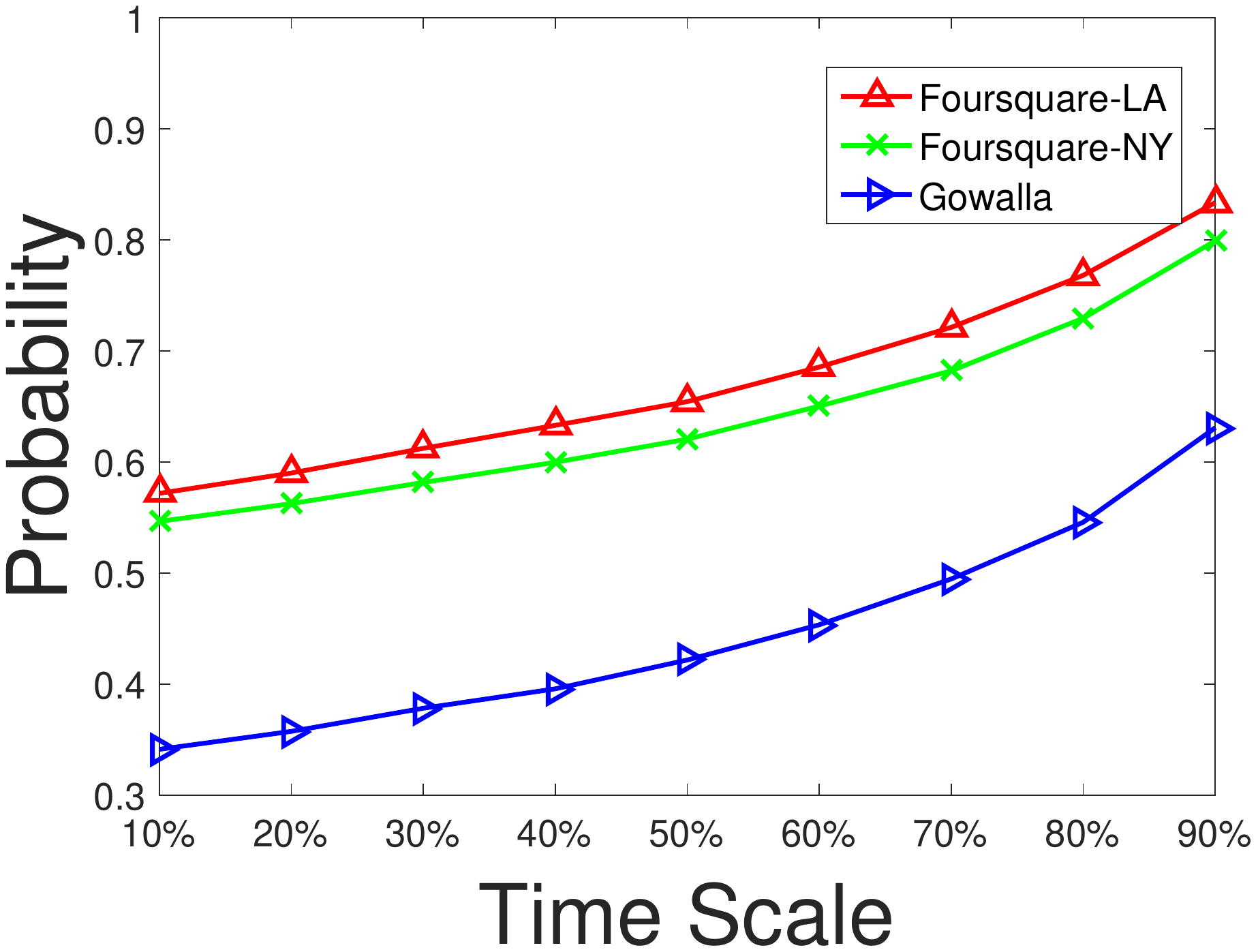}
		\label{fig:dataset_next_new} }
	\caption{(a) Check-ins probability vs. counts (b) The ratio of new POIs along with time scale }
\end{figure}

\subsection{Spatial Influence}
We compute the geographical distance of two consecutive check-ins and plot the cumulative distribution function (CDF) distribution in Fig.\ref{fig:dataset_distance}, which shows that users' movement is restricted by the spatial influence. More specifically, about 60\% of Foursquare-LA successive check-in behaviors, over 65\% of Foursquare-NYC and over 70\% of Gowalla happened within 10 km since last check-in, while when the distance increases to 100 km, the number of successive check-ins account to over 80\% for Foursquare-LA and Foursquare-NYC and over 90\% for Gowalla, respectively. The CDF curve increases fast when distance is small, which suggests that users' movements mostly occur within a localized region. This observation is reasonable since most user generally move periodically within a bounded region but occasionally travel long distance journey. That is, successive check-in POIs are generally spatially correlated and the close POIs have the stronger geographical correlations than the POIs that are far from each other. Thus, the preference of the user for the POI is inversely proportional to geographic distance. 

\subsection{Check-in Counts}
Fig.\ref{fig:dataset_check_in_count} shows the check-in counts for each POI on three datasets, which demonstrates that apart from a few frequently visiting POIs such as home and office, most POIs are visited less than 4 times, which account for 90\%, 90\% and 80\% of total visited POIs for Foursquare-LA, Foursquare-NYC and Gowalla respectively. Again, we also observe that over 23\% of Foursquare POIs and 35\% of Gowalla POIs is checked more than once, which suggests that users' check-in activity exhibits periodic pattern. This observation motivates us to exploit transition periodicity for POI recommendation.
From the observation in Fig.\ref{fig:dataset_distance} and Fig.\ref{fig:dataset_check_in_count}, we find that most POIs are visited occasionally within a short distance interval, which indicates that users' next movements are influenced by their current locations.
Hence, the proposed model takes into account time-critical for POI recommendation and recommend next POIs based on users' current location.

\subsection{Temporal Influence}
Fig.\ref{fig:dataset_time} shows the cumulative distribution function (CDF) of the time interval of two sequential check-ins, which demonstrates that more than 60\% successive check-ins occur in less than 16 hours in both datasets. Only less than 20\% of Foursquare successive check-in time and 10\% of Gowalla is more than 64 hours, which indicates that successive check-ins in shorter time interval contain stronger correlation.
By further studying the categories of successive POIs for the user in a short interval, we find that there is a strong correlation between them. 
As shown in Fig.\ref{fig:category_transition}, Food is always visited after Shop as users would like to dinner after shopping. So far we saw that successive check-ins contain a personalized Markov chain property, and intuitively we utilize the transition probability to solve the task of successive personalized POI recommendation.

\subsection{Exploration for New Locations}
Fig.\ref{fig:dataset_next_new} shows the ratio of new POIs over all users on three datasets along with time. For example, the ratio at 20\% time scale is the proportion of POIs visited after the latest 20\% of check-ins that have not been visited in the previois days. 
Obviously, the ratio of new POIs is pretty high (most of the ratios above 0.3) on both datasets, 
which means that there is over 30\% chance a user might explore new POIs.
More specifically, about 55\% of Foursquare check-ins and about 35\% of Gowalla are distributed among new POIs at 10\% time scale while when the time scale increases to 90\%, the number of new POIs accounts to about 80\% for Foursquare and about 60\% for Gowalla, respectively.
This obervation suggests that users in Foursquare prefer to explore new POIs than Gowalla users.
It is noted that users not only would like to explore new POIs, but also commute among a few routinely visited locations, as shown in Fig.\ref{fig:dataset_check_in_count}. Hence, a good POI recommender should be capable of predicting the periodicity of mobility and meet users' expectation to explore new POIs.

\section{Problem Definition}
Let $U =\{u_1,u_2,...,u_M\}$ be a set of LBSN users, and $L =\{l_1,_2,...,l_N\}$ be a set of locations, also called POIs, where each location is geocoded by \{longitude, latitude\}. The set of POIs visited by user $u$ before time $t$ is denoted by $L_u$, i.e. $L_u = \{L^1_u,..., L^{t-1}_u \}$.
The contextual feature vector is defined as $\textbf{g}(\textbf{c}) = \{g_1(c),...,g_F(c)\}$ which infers a specific contextual scenario $\textbf{c}$. The contextual features include previous location, time of day, day of week, previous location's category, etc. $F$ denotes the number of features.
Assuming there are $K$ latent behavior patterns determined by contextual scenarios, the pattern distribution can be represented as $\Pi = (\pi_1,...,\pi_K)$, {\it s.t.} $\sum_{k=1}^{K}\pi_k = 1$, where $\pi_k$ denotes the probability of the contextual scenario belonging to the $k_{th}$ latent pattern.
With the conjecture that the check-in behaviors are governed by the pattern-level preferences, the probability distribution over next POIs is then the mixture of each pattern-level preference towards those POIs. Our goal is to estimate the pattern distribution $\Pi$ and pattern-level preference, so as to recommend top-N venues to the user $u$ for his next move by combining the obtained pattern-level preferences.



\section{Proposed Method}
Our proposed model is to recommend next personalized POIs via the ranking of probabilities that user $u$ will move from location $i$ to next location $l$. Based on the first-order Markov chain property, the probabilities is given as:
\begin{eqnarray}
	x_{u,i,l} = p(L_{u,l}|\textbf{c})
\end{eqnarray}
where \textbf{c} denotes the contextual scenario. Thus, each user is associated with a specific transition matrix $\chi^u$ which in total generates a transition tensor $\chi \in [0,1]^{|U| \times |L| \times |L|}$ with each $\chi_{u,i,l}$ representing the observed transition record of user $u$ from location $i$ to location $l$.
To further boost the recommendation performance, here we study both personal preference and spatial preference.

\emph{Personal Preference}.
A general linear factorization model for estimating the transition tensor $\chi$ is the Tucker Decomposition (TD):
\begin{eqnarray}
		\hat{\chi} = C \times U \times I \times L
\end{eqnarray}
where $C$ is a core tensor and $U$ is the feature matrix for the users, $I$ is the feature matrix for the locations in the last transition and $L$ is the feature matrix for the next locations.
As the transitions of $\chi$ are partially observed, here we adopt the low-rank factorization model--- a special case of Canonical Decomposition which models the pairwise interaction between all three modes of the tensor (i.e. user $U$, location $I$, next location $L$), to fill up the missing information, given as:
\begin{eqnarray}\label{tensor}
	\hat{x}_{u,i,l} = u_{U,L} \cdot l_{L,U} + l_{L,I} \cdot i_{I,L} + u_{U,I} \cdot i_{I,U}
\end{eqnarray}
where $u_{U,L}$ and $l_{L,U}$ denote the latent factor vectors for users and next locations, respectively. Other notions are similarly defined. The term $u_{U,I} \cdot i_{I,U}$ can be removed since it is independent of location $l$ and does not affect the ranking result, as shown in \cite{rendle2010factorizing}, which generates a more compact expression for $\hat{x}_{u,i,l}$:
\begin{eqnarray}\label{tensor}
	\hat{x}_{u,i,l} = u_{U,L} \cdot l_{L,U} + l_{L,I} \cdot i_{I,L}.
\end{eqnarray}

 $ p(L^t_u \in i | L^t-1_u \in j) \times$An advantage of this model over TD is that the prediction and learning complexity is much lower than for TD. Furthermore even though TD subsume the pairwise interaction model, with standard regularization estimation procedures have problems identifying such a model \cite{rendle2010pairwise}.

\begin{figure*}[t]
	\centering
	\subfigure{\includegraphics[clip=true,width=2.3in]{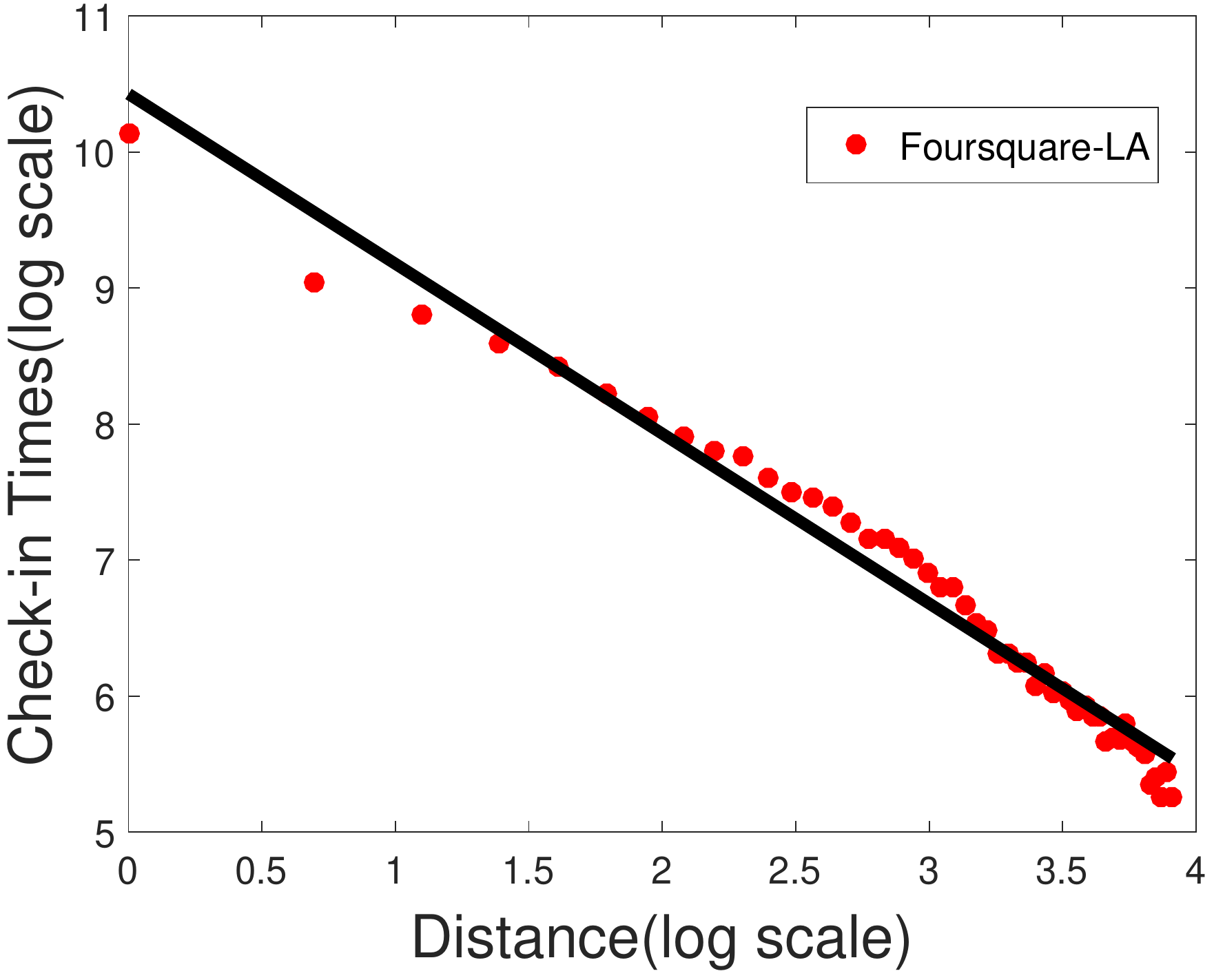} }
	\subfigure{\includegraphics[clip=true,width=2.3in]{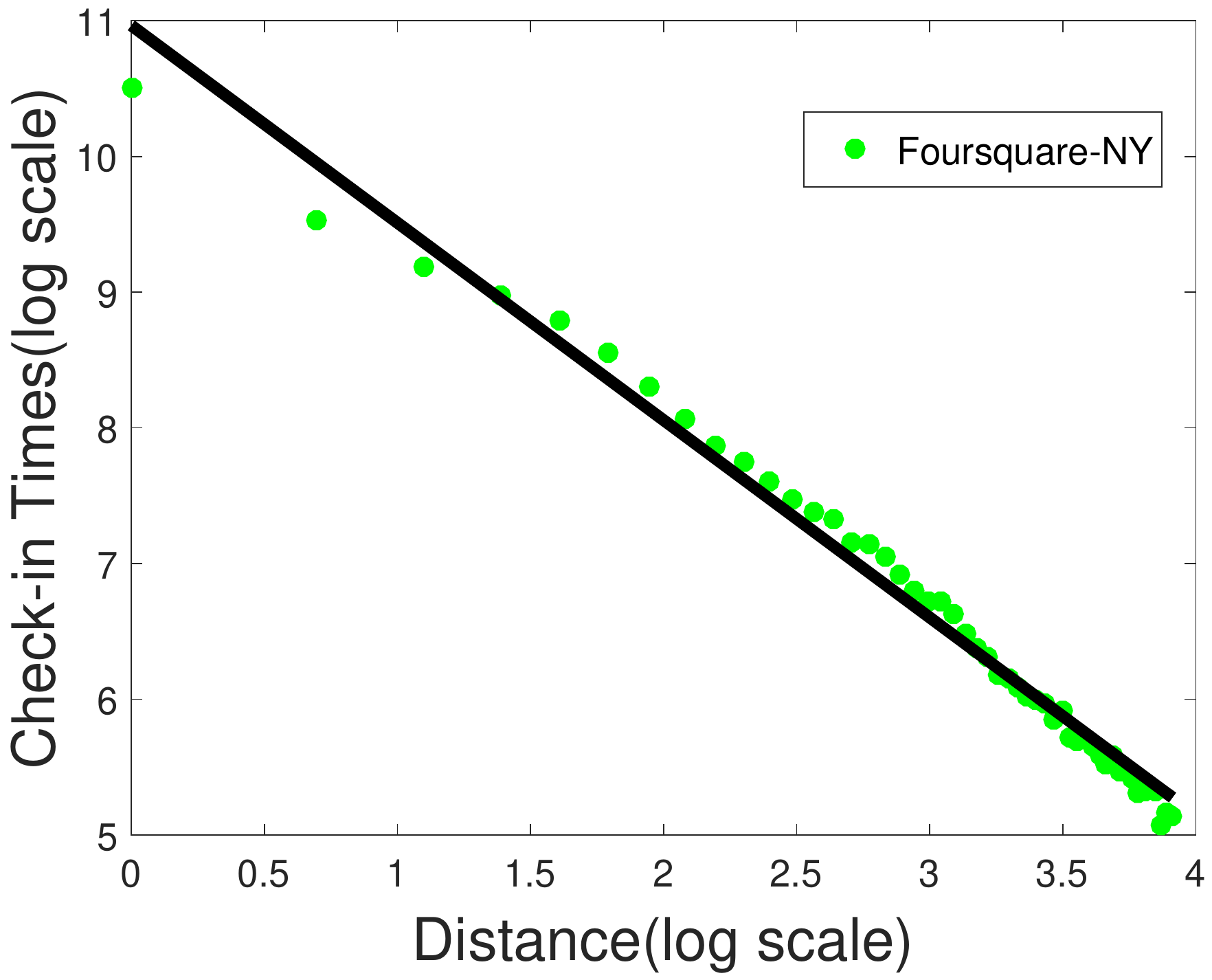} }
	\subfigure{\includegraphics[clip=true,width=2.3in]{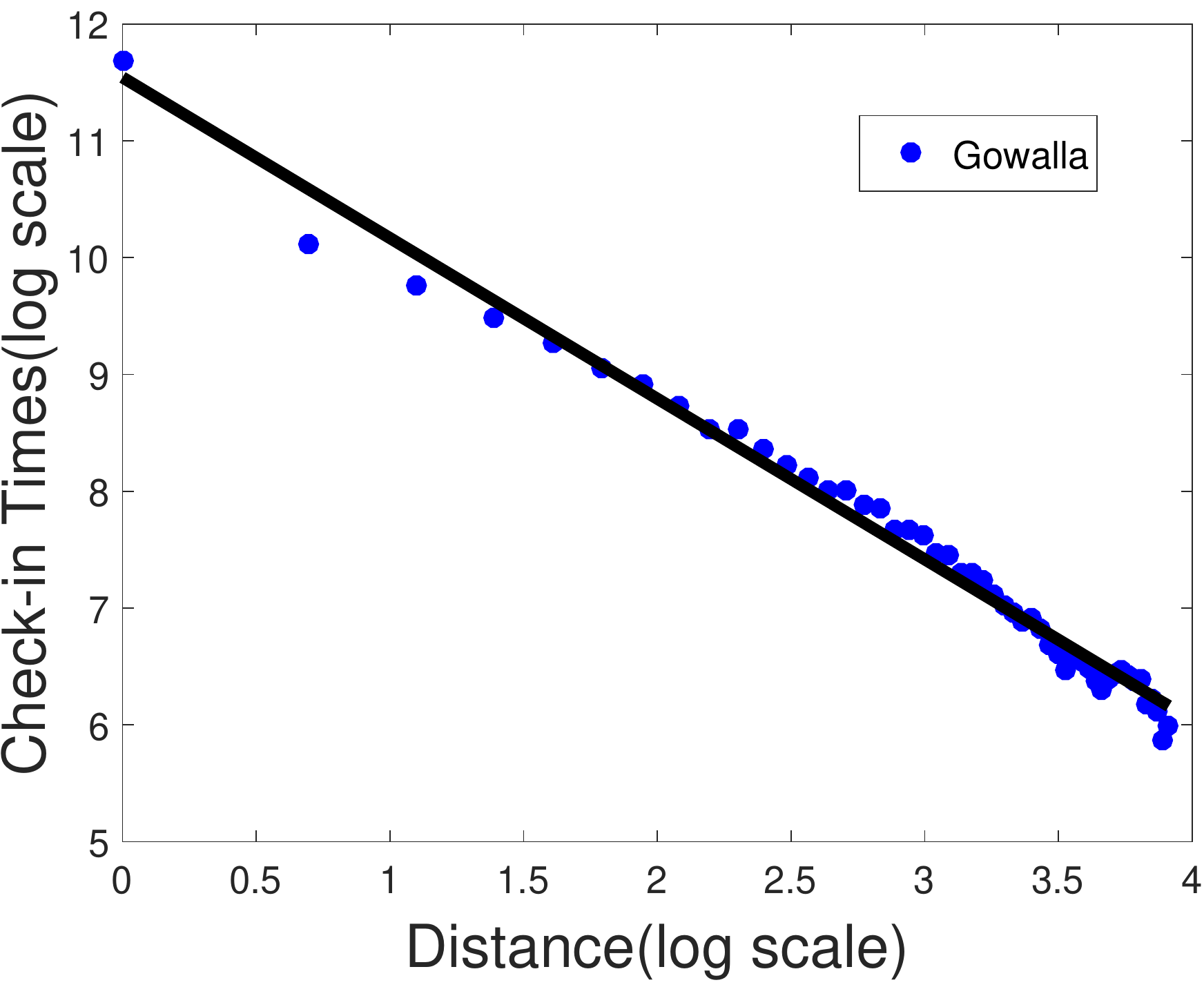} }
	\caption{Spatial behavior statistics}
	\label{fig:distance-log}
\end{figure*}

\emph{Spatial Preference}.
Inspired by \cite{Cho2011}, human mobility is constrained geographically by the distance one can travel within a day and their preference to visit a location decreases as the geographic distance increases. Moreover, most of POIs are likely explored near to users' residence, workplace, and frequently visited POIs.
Fig.\ref{fig:distance-log} shows the relation (in log scale) between the check-in counts and the distance between two successive checked-in locations for Foursquare data and Gowalla data respectively. It is obvious that the relation follows a power law distribution and the venues that a user has checked in are geographically dense.
Different from the existing works which simply remove locations out of the candidate list based on predefined distance threshold, we leverage on the distance constraint by defining power law distribution as the spatial preference $sp(d_{i,l})$ of user $u$ to visit a $d_{i,l}$km far away POI as follows:
\begin{eqnarray}\label{eq:power-law}
	sp(d_{i,l}) = a \times d_{i,l}^{k}
\end{eqnarray}
where $a$ and $k$ are parameters of the power law distribution.

After taking logarithmic on both side of Eq.(\ref{eq:power-law}), the linear function can be easily learned by the least-square regression.
\begin{eqnarray}
	\log sp(d_{i,l})= \log a + k \cdot \log d_{i,l}
\end{eqnarray}
Note that, in learning the two parameters, check-ins having distance larger than 50km is mot considered for these check-ins represent fewer than 29.5\%, 26\% and 9.8\% of the total number of check-ins in Foursquare-LA dataset, Foursquare-NY dataset and Gowalla dataset, respectively.
As the result, we learn the parameter of $a$ equals to 10.5 for Foursquare-LA dataset, 11.0 for Foursquare-NY dataset and 11.5 Gowalla dataset respectively and the parameter of $k$ equals to -1.25 for Foursquare-LA dataset, -1.45 for Foursquare-NY dataset and -1.37 Gowalla dataset respectively. The empirical settings of parameters of the power law distribution are 11 for $a$ and -1 for $k$, where the spatial preference is represented as:
\begin{eqnarray}\label{eq:power-law}
sp(d_{i,l}) = 11 \times d_{i,l}^{-1}
\end{eqnarray}

Combining these two types of preference linearly, we have an updated transition probability estimation, given as:
\begin{eqnarray}
	\hat{x}_{u,i,l} = u_{U,L} \cdot l_{L,U} + l_{L,I} \cdot i_{I,L} + \rho \cdot d_{i,l}^{-1}.
\end{eqnarray}
where $\rho$ is a tradeoff parameter used to fuse the two preferences and the parameter of a in power law distribution will be absolved into $\rho$ since the optimal setting of $\rho$ will be learned during model inference phase.
Thus, even the locations far away from the previously checked-in location have the chance to be recommended when personal preference dominates. And some occasional long journey could be predicted.

\subsection{Incorporating Pattern-Level Preference}
With assumption that user mobility can be classified into some latent behavior patterns, each pattern has distinct impact to user's transition preference, which indicates that users' transition probability is pattern-sensitive.
Here, we propose a novel model by introducing an intermediate latent patterns layer to capture the pattern-level preference in POI recommendation. $s$ is the latent variable to indicate the pattern-level influence. The joint probability of $x_{u,i,l}$ and $s$ is represented as:
\begin{eqnarray}\label{eq:jointdistribution}
	p(L_{u,l}, s|\textbf{c})=
	p(L_{u,l}|s, \textbf{c})p(s|\textbf{c})
\end{eqnarray}
where $p(s|\textbf{c})$ is the mixing coefficient, i.e. $\pi$. The pattern-level preference can be defined as:
\begin{eqnarray}
	\nonumber \hat{x}_{u,i,l}^s &=& p(L_{u,l}|s, \textbf{c})\\
	&=& u_{U,L}^s \cdot l_{L,U}^s + l_{L,I}^s \cdot i_{I,L}^s + \rho^s \cdot d_{i,l}^{-1}
\end{eqnarray}

\begin{figure}
	\centering
	\includegraphics[clip=true,width=2.6in]{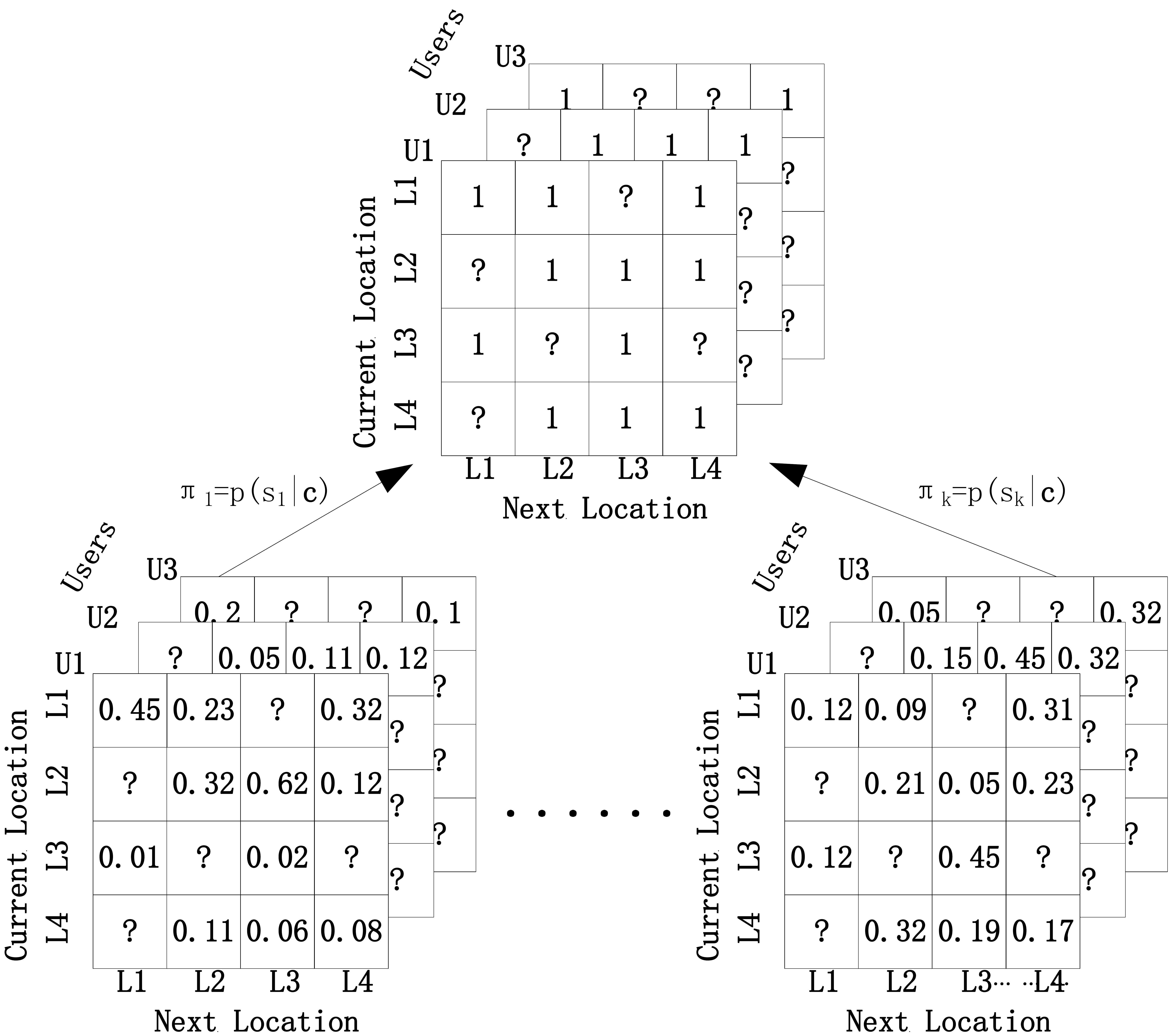}
	\caption{A graphical illustration of our proposed model}
	\label{fig:tensor1}
\end{figure}

By marginalizing out the latent variable $s$, the corresponding transition probability can be written as follows:
\begin{eqnarray}\label{nosoftmax}
	\hat{x}_{u,i,l}=\sum_s \hat{x}_{u,i,l}^s p(s|\textbf{c})
\end{eqnarray}
Fig.\ref{fig:tensor1} gives a graphical illustration of our proposed model. The upper tensor contains the historical check-in data which is in fact the transition tensor $\chi$, where the transition probability between two locations is labeled as ``1" if we observe that a transition happens between the two locations for a user, or ``?" otherwise. Each user, however, may have distinct pattern-level preference under different pattern. And each entry of lower tensors denotes the pattern-level transition probability. It is noted that transition tensor $\chi$ is a mixture of the pattern-level transition tensors, and $p(s|\textbf{c})$ is the mixing coefficient. Then, our goal is to infer the proper pattern-level transition probabilities and pattern distribution to recover the unobserved transition preference by fitting model.

We adopt a soft-max function $\frac{1}{S_c} \exp(\sum_{j=1}^F \alpha_j^{s} g_j(c))$ to infer the multi-patterns and $p(s|\textbf{c})$.
$\alpha_j^{s}$ is the weight associated with the $j_{th}$ feature for latent pattern $s$ and $S_c$ is the normalization factor that scaled the exponential function to be a proper probability distribution $\Pi$, i.e. $S_c = \sum_{k=1}^K \exp(\sum_{j=1}^F \alpha_j^{s_k}g_j(c))$. In this representation, contextual scenario $\textbf{c}$ is denoted by a bag of features $\{g_1(c),...,g_F(c)\}$ where F is the number of features. By plugging the soft-max function into Eq.(\ref{nosoftmax}), $\hat{x}_{u,i,l}$ is rewritten as:
\begin{eqnarray}
	\hat{x}_{u,i,l}= \frac{1}{S_c} \sum_s \hat{x}_{u,i,l}^s \exp(\sum_{j=1}^F \alpha_j^{s} g_j(c))
\end{eqnarray}
Because the learned pattern distribution is identical for all users and regardless of personalized difference, this model is also denoted as global pattern distribution model (GPDM) in this paper.

\subsection{Optimization Criterion}
The task of next POI recommendation is to recommend top-N POIs to users, and we make the parameter learning via learning the ranking order of successive check-in possibilities. We care more about the ranking order of the candidate POIs rather than the real values of check-in possibilities, thus we can model it as a ranking $>_{u,i}^s$ over locations, where $ \hat{x}_{u,i,l}^s$ denotes a personalized ranking score of transition from location $i$ to location $l$ for user $u$ under pattern $s$.
\begin{eqnarray}\label{equivalent}
	m >_{u,i}^s n \Leftrightarrow \hat{x}_{u,i,m}^s > \hat{x}_{u,i,n}^s
\end{eqnarray}
Eq.(\ref{equivalent}) indicates user $u$ prefers location $m$ to location $n$ under pattern $s$.

Next, we derive the sequential Bayesian Personalized Ranking (S-BPR) optimization criterion which is similar to the general BPR approach \cite{Rendle2009BPR}.
Then for user $u$ influenced by the pattern-level preference $s$, the best ranking can be modeled as:
\begin{eqnarray}
	p(\Theta| >_{u,i}^s) \propto p( >_{u,i}^s|\Theta)p(\Theta)
\end{eqnarray}
where $\Theta$ is the set of model parameters, i.e. $\Theta=\{ \bm{\alpha}^S, \bm{\rho}^S, U_{U,L}^S, V_{L,U}^S, V_{L,I}^S,  V_{I,L}^S  \} $.

Then we estimate the model by maximizing the posterior with assumption that users and their check-in history are independent:
\begin{eqnarray}
	\nonumber \mathop{argmax}_{\Theta} \prod_{u \in U} \prod_{i \in L_u}  \prod_{m \in L_u^t} \prod_{n \notin L_u^t} \sum_s p(m  >_{u,i}^s n|\Theta) \cdot \\
	p(s|\textbf{c})p(\Theta)
\end{eqnarray}

The ranking probability can be futher expressed by:
\begin{eqnarray}\label{nosigma}
	\nonumber p(m >_{u,i}^s n|\Theta) =& p(x_{u,i,m}^s > x_{u,i,n}^s|\Theta)\\
	=& p(x_{u,i,m}^s - x_{u,i,n}^s>0|\Theta)
\end{eqnarray}

Similar to \cite{rendle2010factorizing}, we use the logistic function $\sigma(z)=\frac{1}{1+e^{-z}}$ to approximate the likelihood of user's preference over location $m$ and $n$:
\begin{eqnarray}
	p(m  >_{u,i}^s n|\Theta) = \sigma(x_{u,i,m}^s - x_{u,i,n}^s)
\end{eqnarray}

By assuming the model parameters' prior follows a Guassian distribution $p(\Theta) \sim N(0,\frac{2}{\lambda_{\Theta}} I)$, the MAP estimation is now given as:
\begin{eqnarray}\label{objective}
	\nonumber \mathop{argmax}_{\Theta} \prod_{u \in U} \prod_{i \in L_u}  \prod_{m \in L_u^t} \prod_{n \notin L_u^t} \{ \frac{1}{S_c} \cdot\\
	\sum_s \sigma(x_{u,i,m}^s - x_{u,i,n}^s)
	\exp(\sum_{j=1}^F \alpha_j^{s} g_j(c))
	e^{- \frac{\lambda_{\Theta}}{2} ||\Theta||^2}\}
\end{eqnarray}

\subsection{Model Inference}
Furthermore, $\Theta$ can also be estimated by maximizing the following log-scale objective function:
\begin{eqnarray}\label{logfunction}
	\nonumber \mathop{argmax}_{\Theta} \sum_{u \in U} \sum_{i \in L_u}  \sum_{m \in L_u^t} \sum_{n \notin L_u^t} \ln \{ \frac{1}{S_c} \cdot \\
	\sum_s \sigma(x_{u,i,m}^s - x_{u,i,n}^s)
	\exp(\sum_{j=1}^F \alpha_j^{s} g_j(c))
	e^{- \frac{\lambda_{\Theta}}{2} ||\Theta||^2}\}
\end{eqnarray}
Here, we adopt Expectation Maximization(EM) algorithm \cite{dempster1977maximum} to estimate the model parameters.

In E-Step, $\gamma(s)$ the posterior distribution of $s$ is given as:
\begin{eqnarray}
	\nonumber \gamma(s)=& P(s| >_{u,i}^s, \Theta, \textbf{c}) \\
	=& \frac{\sigma(x_{u,i,m}^s - x_{u,i,n}^s)
		\exp(\sum_{j=1}^F \alpha_j^{s} g_j(c))}
	{\sum_s \sigma(x_{u,i,m}^s - x_{u,i,n}^s)
		\exp(\sum_{j=1}^F \alpha_j^{s} g_j(c))}
\end{eqnarray}

And in M-Step, $\bm{\alpha}^S$ and $\{ \Theta\setminus \bm{\alpha}^S \}$ can be derived by optimizing the Q-function of Eq.(\ref{Malpha}) and Eq.(\ref{Mrest}), respectively. The detailed algorithm and the parameter updating rules are shown in \textbf{Algorithm} 1.
\begin{eqnarray}\label{Malpha}
	\nonumber \bm{\alpha}^S=
	\mathop{argmax}_{\bm{\alpha}^S} \sum_{u \in U} \sum_{i \in L_u}  \sum_{m \in L_u^t} \sum_{n \notin L_u^t} \sum_s \gamma(s) \cdot \\
	\{ \ln (\frac{1}{S_c} \exp(\sum_{j=1}^F \alpha_j^{s} g_j(c))) -\frac{\lambda_{\Theta}}{2} ||\Theta||^2) \}
\end{eqnarray}
\begin{eqnarray}\label{Mrest}
	\nonumber \{ \Theta\setminus \bm{\alpha}^S \}
	=\mathop{argmax}_{\{\Theta \setminus \bm{\alpha}^S\}}
	\sum_{u \in U} \sum_{i \in L_u}  \sum_{m \in L_u^t} \sum_{n \notin L_u^t} \sum_s \gamma(s) \cdot \\
	\{ \ln \sigma(x_{u,i,m}^s - x_{u,i,n}^s) -\frac{\lambda_{\Theta}}{2} ||\Theta||^2) \}
\end{eqnarray}

\renewcommand{\algorithmicrequire}{\textbf{Input:}}
\renewcommand{\algorithmicensure}{\textbf{procedure:}}
\renewcommand{\algorithmicreturn}{\textbf{Return:}}
\begin{algorithm}[t]
	\caption{Our Proposed Methodology}
	\begin{algorithmic}[1]
		\STATE {\bfseries Input:} the number of patterns $K$, check-in data D		
		\STATE draw $\Theta$ from $N(0,\frac{2}{\lambda_{\Theta}} I)$
		\REPEAT
		\STATE {\bfseries E-Step:}
		
		\STATE $S_c \leftarrow \sum_{k=1}^K \exp(\sum_{j=1}^F \alpha_j^{s_k}g_j(c))$
		
		\STATE $p(s|\bm{c}) \leftarrow \frac{1}{S_c} \exp(\sum_{j=1}^F \alpha_j^{s} g_j(c))$
		
		\STATE $\gamma(s) \leftarrow \frac{\sigma(x_{u,i,m}^s - x_{u,i,n}^s)
			\exp(\sum_{j=1}^L \alpha_j^{s} g_j(c))}
		{\sum_s \sigma(x_{u,i,m}^s - x_{u,i,n}^s)
			\exp(\sum_{j=1}^L \alpha_j^{s} g_j(c))}$
		
		\STATE {\bfseries M-Step:}
		\STATE $\delta \leftarrow (1 - \sigma(x_{u,i,m}^s - x_{u,i,n}^s))$
		
		
		\STATE $u_{U,L}^s \leftarrow \frac{\sum_d \delta \cdot \gamma(s) \cdot (m_{L,U}^s - n_{L,U}^s) }
		{\lambda_{\Theta}  \sum_d \gamma(s)}$
		
		\STATE $i_{I,L}^s \leftarrow \frac{\sum_d \delta \cdot \gamma(s) \cdot (m_{L,I}^s - n_{L,I}^s) }
		{\lambda_{\Theta}  \sum_d \gamma(s)}$
		
		\STATE $m_{L,U}^s \leftarrow \frac{\sum_d \delta \cdot \gamma(s) \cdot u_{U,L}^s }
		{\lambda_{\Theta}  \sum_d \gamma(s)}$
		
		\STATE $n_{L,U}^s \leftarrow \frac{\sum_d \delta \cdot \gamma(s) \cdot (-u_{U,L}^s) }
		{\lambda_{\Theta}  \sum_d \gamma(s)}$
		
		\STATE $m_{L,I}^s \leftarrow \frac{\sum_d \delta \cdot \gamma(s) \cdot i_{I,L}^s }
		{\lambda_{\Theta}  \sum_d \gamma(s)}$
		
		\STATE $n_{L,I}^s \leftarrow \frac{\sum_d \delta \cdot \gamma(s) \cdot (-i_{I,L}^s) }
		{\lambda_{\Theta}  \sum_d \gamma(s)}$
		
		\STATE $\bm{\alpha}^s \leftarrow \frac{\sum_d \gamma(s) \cdot g(\bm{c}) \cdot (1-p(s|\bm{c})) }
		{\lambda_{\Theta}  \sum_d \gamma(s)}$
		
		\STATE $\bm{\rho}^s \leftarrow \frac{\sum_d \delta \cdot \gamma(s) \cdot (d_{i,m}^{-1} - d_{i,n}^{-1} ) }
		{\lambda_{\Theta}  \sum_d \gamma(s)}$
		
		\UNTIL {convergence}
		\RETURN $\Theta$
	\end{algorithmic}
\end{algorithm}

\subsection{Personalized Pattern Distribution Model}
\begin{figure}
	\centering
	\includegraphics[clip=true,width=2.6in]{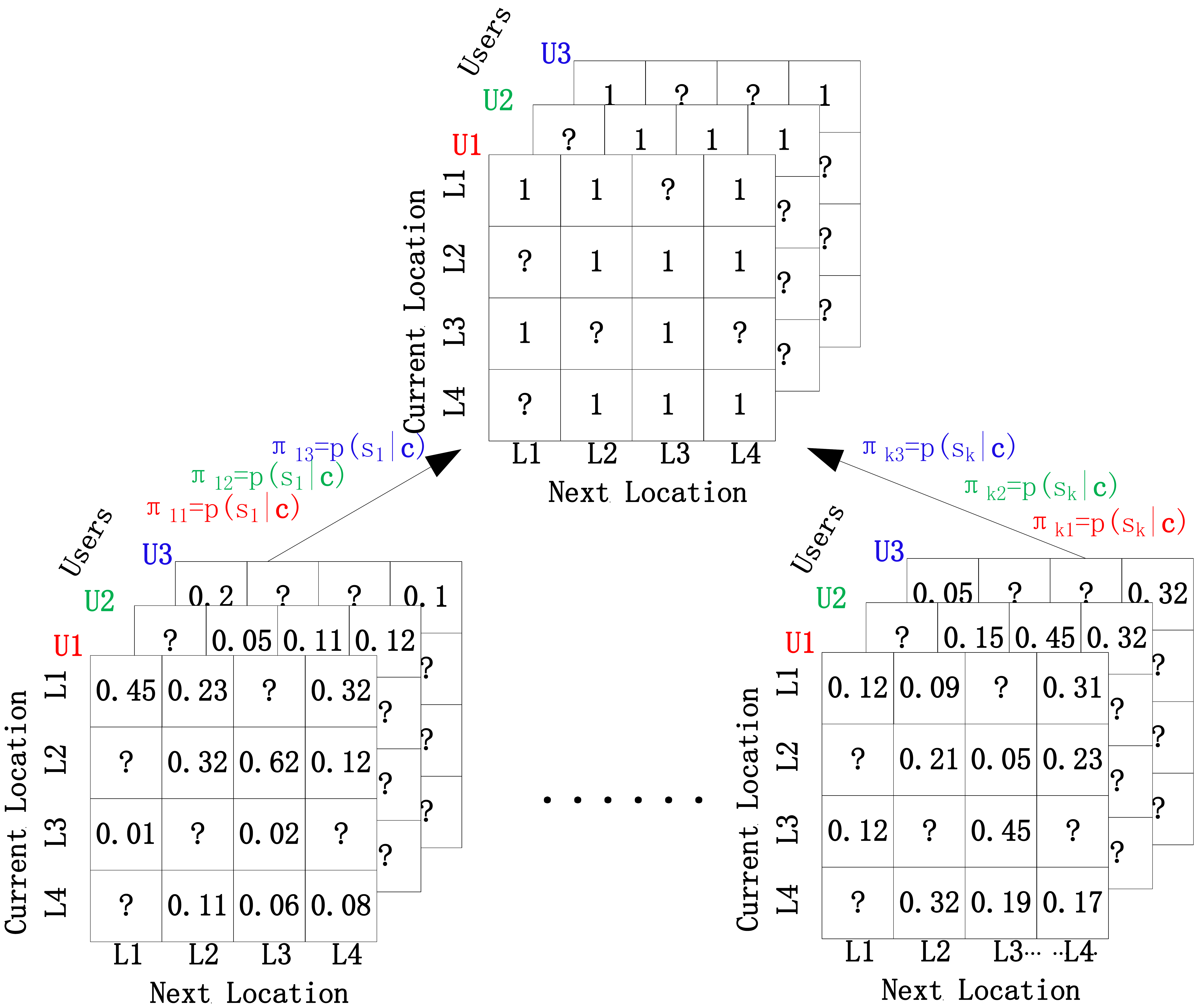}
	\caption{A graphical illustration of the personalized model}
	\label{fig:tensor2}
\end{figure}
In the global model introduced in Subsection A, a fixed pattern distribution is learned from the global perspective to optimize the overall performance for all users. However, the best pattern distribution for a given user is not always the best for others due to the personalization of individual users. The users in LBSNs are extremely diverse according to the various properties, including age, gender, home city, occupation, etc. Different users may have personalized pattern distribution under the same contextual scenario. 

Instead of inferring a fixed pattern distribution $\Pi$ for every user, personalized pattern distribution model (PPDM) infers pattern distributions for each user, i.e. $\Pi_{u}$, as shown in Fig.\ref{fig:tensor2}. It is noted that the mixing coefficients, i.e.  $p_{u}(s|\textbf{c})$, are personalized. By inferring $\alpha_{u,j}^{s}$ as the personalized weight associated with the $j_{th}$ feature for latent pattern $s$, the corresponding transition probability $\hat{x}_{u,i,l}$ for PPDM can be rewritten as follows: 
\begin{eqnarray}
\hat{x}_{u,i,l}= \frac{1}{S_{u,c}} \sum_s \hat{x}_{u,i,l}^s \exp(\sum_{j=1}^F \alpha_{u,j}^{s} g_{u,j}(c))
\end{eqnarray}
and the optimization function of $\bm{\alpha}^S$ in Eq.(\ref{Malpha}) is rewritten as:
\begin{eqnarray}
	\nonumber \bm{\alpha}^S_u=
	\mathop{argmax}_{\bm{\alpha}^S_u} \sum_{i \in L_u}  \sum_{m \in L_u^t} \sum_{n \notin L_u^t} \sum_s \gamma(s) \cdot \\
	\{ \ln (\frac{1}{S_c} \exp(\sum_{j=1}^F \alpha_j^{s} g_j(c))) -\frac{\lambda_{\Theta}}{2} ||\Theta||^2) \}
\end{eqnarray}
The formulations for updating the parameters $\{ \Theta\setminus \bm{\alpha}^S \}$ are the same as in Eq.(\ref{Mrest}) and the parameter updating rule of $\bm{\alpha}^S$ in Line 16 of Algorithm 1 is rewritten as:
\begin{eqnarray}
\bm{\alpha}^s_u \leftarrow \frac{\sum_{u,d} \gamma(s) \cdot g(\bm{c}) \cdot (1-p(s|\bm{c})) }{\lambda_{\Theta}  \sum_{u,d} \gamma(s)}
\end{eqnarray}


\section{Experiments}
In this section, we evaluate the following: (1) how is the proposed approaches in comparison with other state-of-the-art recommendation techniques? (2) how does the number of latent classes affect the model accuracy? (3) how does the features perform in the POI recommendation task? (4) how is the performance of proposed models in recommending new POIs? (5) how is the performance difference between GPMD and PPDM?



\subsection{Evaluation Metrics}
Given a top-N recommendation list $S_{N,u,rec}$ sorted in descending order of the prediction values to user $u$, we adapt a precision metric to evaluate the performance of our proposed next POI recommendation, given as:
\begin{eqnarray}\label{recall}
	Precision@N = \frac{1}{|U|} \sum_{u \in U} \frac{|S_{N,u,rec} \cap S_{visited}|}{|S_{visited}|}
\end{eqnarray}
where $S_{visited}$ are the visited locations of user $u$ and $|U|$ denotes the number of the users, N is the size of the next POI candidate list.

We evaluate the performance of next new POI recommendation by defining precision as:
\begin{eqnarray}\label{recall}
Precision@N_{new} = \frac{1}{|U|} \sum_{u \in U} \frac{|S_{N,u,rec} \cap S_{visited}^{new}|}{|S_{visited}^{new}|}
\end{eqnarray}
where $S_{visited}^{new}$ denotes locations that a user does not visit before and will be visited in the next time.
\subsection{Evaluated Methods and Parameter Settings}
We compare the proposed model with the following methods:
\begin{itemize}
	\item \textbf{MF:} matrix factorization \cite{5197422} is widely used in conventional recommender systems, which factorizes the user-item preference matrix.
	\item \textbf{PMF:} probabilistic matrix factorization is a well-known method for modeling time evolving relation data \cite{mnih2007probabilistic}. It is widely used in traditional recommender systems.
	\item \textbf{FPMC-LR:} this method is proposed in \cite{cheng2013you}, which is the state-of-the-art personalized successive POI recommendation method.
	\item \textbf{PRME-G:} this approach utilizes two Euclidean distances in the latent space: one is the distance between current location and next location, the other is the distance between user and next location, then takes the combination of two distances in predicting \cite{feng2015personalized}.
\end{itemize}
In the experiments, we use the three datasets introduced in Section\uppercase\expandafter{\romannumeral3} and Table \ref{tab:foursquare-LA}, Table \ref{tab:foursquare-NY} and Table \ref{tab:Gowalla} report the comparison results between our models and the baseline methods. We set $\lambda_{\Theta}$ to be 1 for both FPMC-LR and our proposed model, and the number of latent dimensions to 60 for all the compared models. The time window size is set to be 6 hours for both FPMC-LR and PRME-G. We set regularization term $\lambda$ = 0.03 and component weight $\alpha$ = 0.2 for PRME-G following \cite{feng2015personalized}. The empirical settings of the number of latent behavior patterns are 4 and 6 for \textit{Gowalla} dataset and \textit{Foursquare} datasets, respectively. For other parameters, we tune them in the training sets to find the optimal values, and subsequently use them in the test set.

\subsection{Comparison of Next POI Recommendation}
In the left of three Tables compares the recommendation accuracy of the evaluated methods on the next POI recommendation. The results show that:
\begin{itemize}
	\item Both FPMC-LR, PRME-G and the proposed models outperform MF and PMF significantly, which indicates that the conventional POI recommendation algorithms are not effective for the successive POI recommendation. One possible explanation could be that MF and PMF mainly exploit the user preference rather than making use of the sequential information. More specifically, our proposed models achieve a relative improvement of at least 91\% for MF and 81\% for PMF respectively, while FPMC-LR and PRME-G also achieve an improvement compared with MF and PMF. This demonstrates that spatial influence plays an important role in next POI recommendation.
	
	\item Both GPDM and PPDM consistently outperform FPMC-LR, improving around 35\% and 45\% over FPMC-LR for Foursquare dataset and Gowalla dataset, respectively. We make similar observations by comparing the proposed models with PRME-G. Again, PPDM does around 35\% for Foursquare dataset and 45\% for Gowalla dataset better than PRME-G, respectively. It illustrates that inferring user latent behavior patterns can better capture user mobility preference in LBSNs, and therefore, help us recommend POIs to users more accurately.
\end{itemize}

\subsection{Comparison of Next New POI Recommendation}
In the right of three Tables contrast the recommendation accuracy of the evaluated methods on the next new POI recommendation, which indicates that our model is capable of predicting the periodicity of user mobility as well as providing new POIs to users. According to the accuracy results, we have the following three observations:
\begin{itemize}
	\item MF and PMF aim to tune the latent factor vectors of users and locations to explain observed check-ins and recover unobserved check-ins. However, it's challenging for conventional methods to recommend the new POIs to users without any extra information, because new POIs have received few preference from users and assigned low credit by conventional methods compared with the visited POIs. As a result, they report the lowest recommendation precision.
	
	\item Both FPMC-LR and PRME-G show the increasing precision by taking geographical influence into account. Moreover, PRME-G is better than FPMC-LR, since PRME-G has been customized to predict next new POI by representing each POI as one point in latent space rather than two independent vectors.
	
	\item The proposed models always achieve the highest recommendation precision by a large margin, which implies that inferring user latent behavior patterns plays an important role when performing next new POI recommendation.
\end{itemize}

Another interesting observation is that the proposed models reach lower precision than FPMC-LR and PRME-G in terms of P@1 on Gowalla dataset. Note that the categorical information of POIs are not included in Gowalla dataset and the proposed models fail to integrate the categorical information to infer the latent behavior patterns. Intuitively we expect that categorical information is useful for modeling the specific preference of a user for recommending new POIs. 

\begin{figure}[htb]
	\centering
	\subfigure[Foursquare-LA]{\includegraphics[clip=true,width=2.6in]{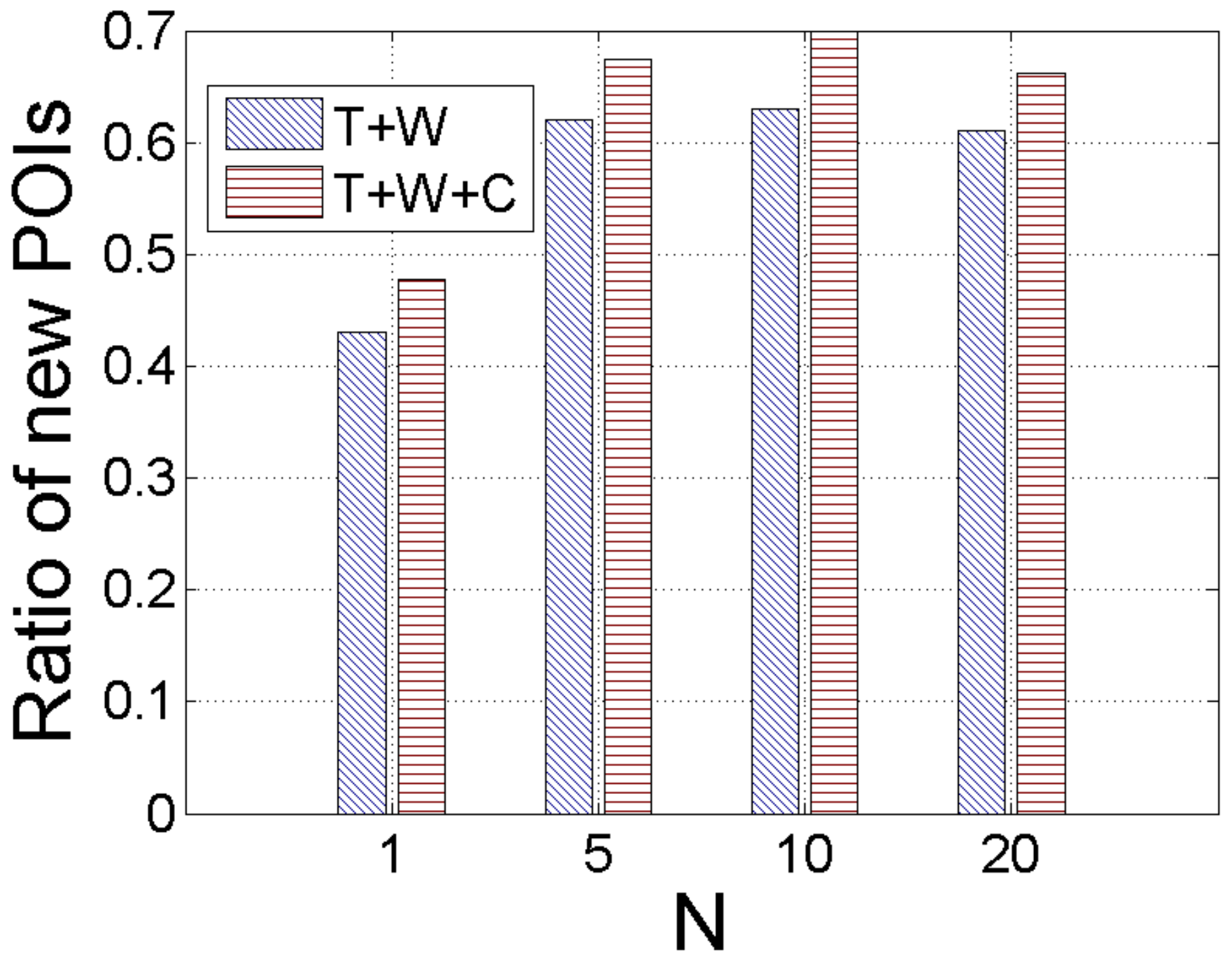}}
	\subfigure[Foursquare-NYC]{\includegraphics[clip=true,width=2.6 in]{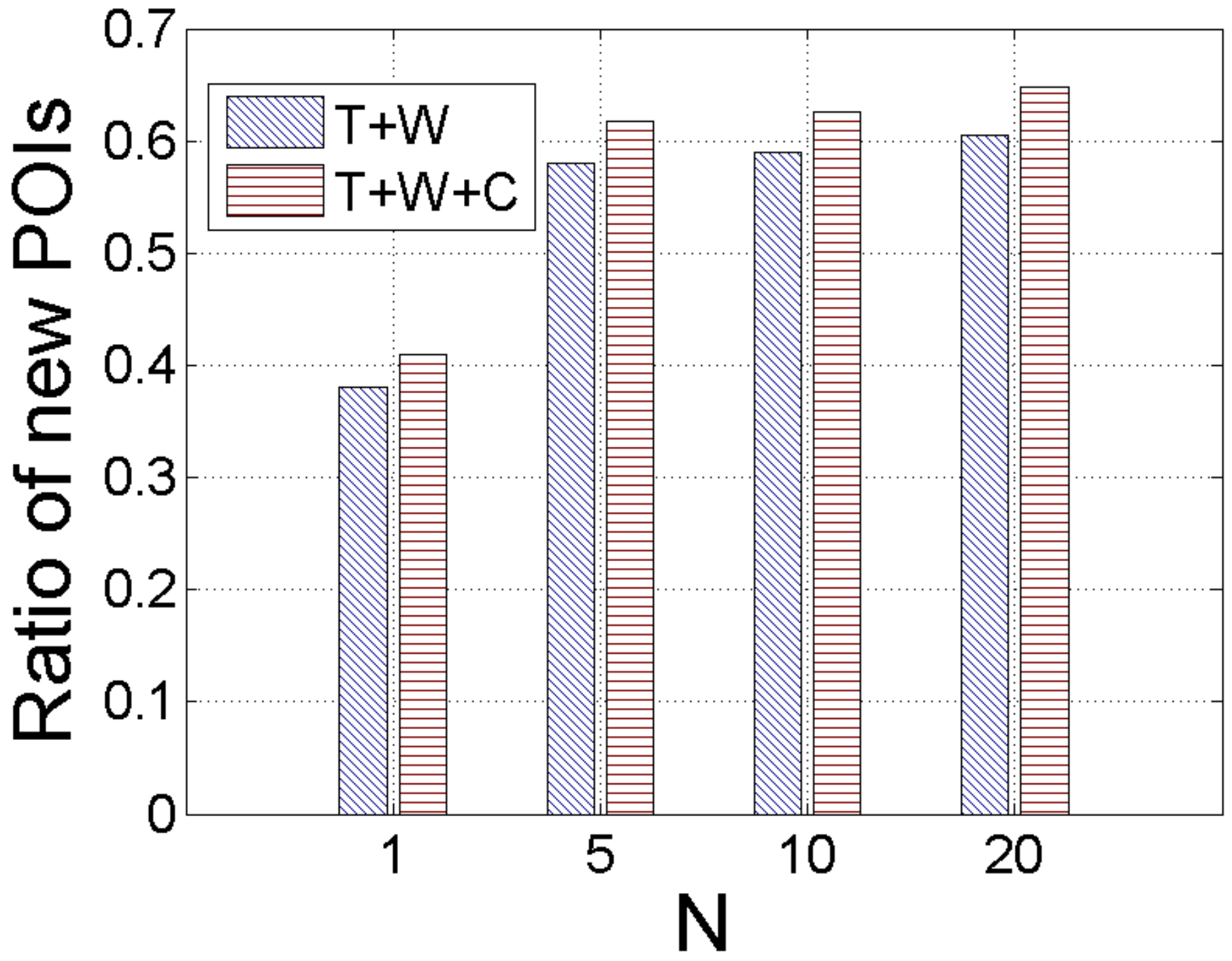}}
	\caption{The ratio of new POIs with Precision@N \label{fig:next_new}}
\end{figure}

To quantify this importance, Fig.\ref{fig:next_new} further depicts the fraction of new POIs over all accurately predicted POIs, which is $\frac{|S_{N,u,rec} \cap S_{visited}^{new}|}{|S_{N,u,rec} \cap S_{visited}|}$. We observe that GPDM with the contextual feature of previous location's category achieve much better performance in recommending new POIs than GPDM without the contextual feature of previous location's category, which suggests that the predictive ability of proposed models on next new POI recommendation can be uplifted through incorporating categorical influence into inferring latent behavior patterns. Our explanation is that: the categories of POIs visited by a user implicitly indicate the activities of the user int the POIs. In reality people have different biases on the categories of POIs: a foodie often visits restaurants to taste a variety of food, and a tourism enthusiast usually travels on tourism attractions all over the world. Accordingly, the proposed model can deduce the relevance score of a user to an unvisited POI based on the categorical information in the categories of the user's visited POIs and the unvisited POIs.

\begin{table*}[!t]
	\newcommand{\tabincell}[2]{\begin{tabular}{@{}#1@{}}#2\end{tabular}}
	\centering
	\caption{Performance Comparison on Foursquare-LA}\label{tab:foursquare-LA}
	\begin{tabular*}{18cm}{@{\extracolsep{\fill}}c||c|c|c|c|c|c|c|c|c|c|c|c}
		\hline
		\hline
		\multirow{2}{*}{Metrics}&\multicolumn{6}{c|}{Next POI Recommendation}&\multicolumn{6}{c}{Next New POI Recommendation}\\
		\cline{2-13}
		& MF & PMF & FPMC-LR & PRME-G & GPDM & PPDM & MF & PMF & FPMC-LR & PRME-G & GPDM & PPDM \\
		\hline\hline
		
		\tabincell{c}{P@1\\Improve}&\tabincell{c}{0.023\\91.30\%}&\tabincell{c}{0.024\\83.33\%}
		&\tabincell{c}{0.032\\37.50\%}&\tabincell{c}{0.034\\29.41\%}&\tabincell{c}{\textbf{0.044}}&\tabincell{c}{\textbf{0.044}}&\tabincell{c}{0.007\\425.0\%}&\tabincell{c}{0.007\\425.0\%}&\tabincell{c}{0.027\\31.25\%}&\tabincell{c}{0.034\\5.00\%}&\tabincell{c}{\textbf{0.036}}&\tabincell{c}{\textbf{0.036}} \\
		\hline
		
		\tabincell{c}{P@5\\Improve}&\tabincell{c}{0.067\\92.54\%}&\tabincell{c}{0.071\\81.69\%}
		&\tabincell{c}{0.097\\32.99\%}&\tabincell{c}{0.097\\32.99\%}&\tabincell{c}{\textbf{0.129}}&\tabincell{c}{\textbf{0.125}}&\tabincell{c}{0.039\\278.3\%}&\tabincell{c}{0.039\\278.3\%}&\tabincell{c}{0.093\\58.18\%}&\tabincell{c}{0.114\\29.85\%}&\tabincell{c}{\textbf{0.147}}&\tabincell{c}{\textbf{0.141}} \\
		\hline
		
		\tabincell{c}{P@10\\Improve}&\tabincell{c}{0.089\\91.01\%}&\tabincell{c}{0.093\\82.80\%}
		&\tabincell{c}{0.128\\32.81\%}&\tabincell{c}{0.124\\37.10\%}&\tabincell{c}{\textbf{0.170}}&\tabincell{c}{\textbf{0.163}}&\tabincell{c}{0.066\\205.1\%}&\tabincell{c}{0.069\\190.2\%}&\tabincell{c}{0.124\\63.01\%}&\tabincell{c}{0.144\\40.00\%}&\tabincell{c}{\textbf{0.202}}&\tabincell{c}{\textbf{0.185}}\\
		\hline
		
		\tabincell{c}{P@20\\Improve}&\tabincell{c}{0.108\\97.22\%}&\tabincell{c}{0.116\\83.62\%}&
		\tabincell{c}{0.155\\37.42\%}
		&\tabincell{c}{0.150\\42.00\%}&\tabincell{c}{\textbf{0.213}}&\tabincell{c}{\textbf{0.199}}&\tabincell{c}{0.107\\123.8\%}&\tabincell{c}{0.110\\116.9\%}&\tabincell{c}{0.153\\56.67\%}&\tabincell{c}{0.173\\38.24\%}&\tabincell{c}{\textbf{0.239}}&\tabincell{c}{\textbf{0.229}}\\
		\hline
		
	\end{tabular*}
	*improved by GPDM
\end{table*}

\begin{table*}[!t]
	\newcommand{\tabincell}[2]{\begin{tabular}{@{}#1@{}}#2\end{tabular}}
	\centering
	\caption{Performance Comparison on Foursquare-NY}\label{tab:foursquare-NY}
	\begin{tabular*}{18cm}{@{\extracolsep{\fill}}c||c|c|c|c|c|c|c|c|c|c|c|c}
		\hline
		\hline
		\multirow{2}{*}{Metrics}&\multicolumn{6}{c|}{Next POI Recommendation}&\multicolumn{6}{c}{Next New POI Recommendation}\\
		\cline{2-13}
		& MF & PMF & FPMC-LR & PRME-G & GPDM & PPDM & MF & PMF & FPMC-LR & PRME-G & GPDM & PPDM \\
		\hline\hline
		
		\tabincell{c}{P@1\\Improve}&\tabincell{c}{0.011\\300.0\%}&\tabincell{c}{0.015\\193.3\%}
		&\tabincell{c}{0.022\\100.0\%}&\tabincell{c}{0.034\\29.41\%}&\tabincell{c}{\textbf{0.044}}&\tabincell{c}{\textbf{0.045}}&\tabincell{c}{0.004\\800.0\%}&\tabincell{c}{0.006\\500.0\%}
		&\tabincell{c}{0.030\\12.50\%}&\tabincell{c}{0.032\\5.88\%}&\tabincell{c}{\textbf{0.034}}&\tabincell{c}{\textbf{0.036}} \\
		\hline
		
		\tabincell{c}{P@5\\Improve}&\tabincell{c}{0.034\\261.8\%}&\tabincell{c}{0.041\\200.0\%}
		&\tabincell{c}{0.068\\80.88\%}&\tabincell{c}{0.092\\33.70\%}&\tabincell{c}{\textbf{0.123}}&\tabincell{c}{\textbf{0.129}}&\tabincell{c}{0.021\\590.9\%}&\tabincell{c}{0.030\\375.0\%}
		&\tabincell{c}{0.095\\52.00\%}&\tabincell{c}{0.109\\33.33\%}&\tabincell{c}{\textbf{0.145}}&\tabincell{c}{\textbf{0.154}} \\
		\hline
		
		\tabincell{c}{P@10\\Improve}&\tabincell{c}{0.051\\219.6\%}&\tabincell{c}{0.062\\162.9\%}
		&\tabincell{c}{0.096\\69.79\%}&\tabincell{c}{0.115\\41.74\%}&\tabincell{c}{\textbf{0.163}}&\tabincell{c}{\textbf{0.169}}
		&\tabincell{c}{0.040\\385.7\%}&\tabincell{c}{0.057\\240.0\%}&\tabincell{c}{0.131\\47.83\%}&\tabincell{c}{0.139\\39.73\%}&\tabincell{c}{\textbf{0.194}}&\tabincell{c}{\textbf{0.210}}\\
		\hline
		
		\tabincell{c}{P@20\\Improve}&\tabincell{c}{0.072\\180.6\%}&\tabincell{c}{0.083\\143.4\%}&\tabincell{c}{0.124\\62.90\%}
		&\tabincell{c}{0.137\\47.45\%}&\tabincell{c}{\textbf{0.202}}&\tabincell{c}{\textbf{0.212}}&\tabincell{c}{0.076\\227.5\%}&\tabincell{c}{0.095\\162.0\%}&\tabincell{c}{0.156\\59.76\%}
		&\tabincell{c}{0.164\\52.33\%}&\tabincell{c}{\textbf{0.250}}&\tabincell{c}{\textbf{0.265}}\\
		\hline
		
	\end{tabular*}
	*improved by GPDM
\end{table*}

\begin{table*}[!t]
	\newcommand{\tabincell}[2]{\begin{tabular}{@{}#1@{}}#2\end{tabular}}
	\centering
	\caption{Performance Comparison on Gowalla}\label{tab:Gowalla}
	\begin{tabular*}{18cm}{@{\extracolsep{\fill}}c||c|c|c|c|c|c|c|c|c|c|c|c}
		\hline
		\hline
		\multirow{2}{*}{Metrics}&\multicolumn{6}{c|}{Next POI Recommendation}&\multicolumn{6}{c}{Next New POI Recommendation}\\
		\cline{2-13}
		& MF & PMF & FPMC-LR & PRME-G & GPDM & PPDM & MF & PMF & FPMC-LR & PRME-G & GPDM & PPDM \\
		\hline\hline
		
				\tabincell{c}{P@1\\Improve}&
               \tabincell{c}{0.022\\100.00\%}&\tabincell{c}{0.024\\83.33\%}
               &\tabincell{c}{0.029\\51.72\%}&\tabincell{c}{0.040\\10.00\%}&\tabincell{c}{\textbf{0.044}}&\tabincell{c}{\textbf{0.041}}&\tabincell{c}{0.006\\300.00\%}&\tabincell{c}{0.006\\300.0\%}&\tabincell{c}{0.028\\-20.0\%}&\tabincell{c}{0.045\\-50.0\%}&\tabincell{c}{\textbf{0.022}}&\tabincell{c}{\textbf{0.020}} \\
               \hline

               \tabincell{c}{P@5\\Improve}&
              \tabincell{c}{0.086\\96.51\%}&\tabincell{c}{0.093\\81.72\%}
              &\tabincell{c}{0.116\\45.69\%}&\tabincell{c}{0.142\\19.01\%}&\tabincell{c}{\textbf{0.169}}&\tabincell{c}{\textbf{0.160}}&\tabincell{c}{0.031\\781.8\%}&\tabincell{c}{0.036\\646.2\%}&\tabincell{c}{0.118\\130.9\%}&\tabincell{c}{0.182\\49.23\%}&\tabincell{c}{\textbf{0.271}}&\tabincell{c}{\textbf{0.238}} \\
                \hline

               \tabincell{c}{P@10\\Improve}&\tabincell{c}{0.147\\99.32\%}&\tabincell{c}{0.158\\85.44\%}&\tabincell{c}{0.198\\47.98\%}&
               \tabincell{c}{0.195\\50.26\%}&\tabincell{c}{\textbf{0.293}}&\tabincell{c}{\textbf{0.259}}&\tabincell{c}{0.059\\533.3\%}&\tabincell{c}{0.067\\454.2\%}&\tabincell{c}{0.182\\104.6\%}&\tabincell{c}{0.249\\49.44\%}&\tabincell{c}{\textbf{0.372}}&\tabincell{c}{\textbf{0.322}}\\
               \hline

              \tabincell{c}{P@20\\Improve}&\tabincell{c}{0.188\\100.53\%}&
               \tabincell{c}{0.202\\86.63\%}&\tabincell{c}{0.247\\52.63\%}
               &\tabincell{c}{0.246\\53.25\%}&\tabincell{c}{\textbf{0.377}}&\tabincell{c}{\textbf{0.333}}&\tabincell{c}{0.109\\323.1\%}&\tabincell{c}{0.118\\292.9\%}&\tabincell{c}{0.252\\83.33\%}&\tabincell{c}{0.311\\48.65\%}&\tabincell{c}{\textbf{0.462}}&\tabincell{c}{\textbf{0.395}}\\
                \hline
		
		
		
		
	\end{tabular*}
	*improved by GPDM
\end{table*}

\begin{table*}
	\newcommand{\tabincell}[2]{\begin{tabular}{@{}#1@{}}#2\end{tabular}}
	\centering
	\caption{Performance Comparison}\label{tab:recall}
	\begin{tabular*}{18cm}{@{\extracolsep{\fill}}c|c|c|c|c|c|c}
		\hline
		\hline
		User ID	&\ current Venue&\ Check-in T. &\ next Venue &\ Check-in T. &\ Dist.(km) &\ T. interval(h) \\
		\hline
		
		2282 &\ Arches National Park Visitor Center &\ 14:24, Fri &\ Patagonia Outlet &\ 10:26, Sun &\ 304 &\ 44 \\
		\hline
		
		1598 &\ Silvia's Hair Design &\ 13:09, Fri &\ Don Carlos &\ 12:19, Sat &\  0.5577 &\  23.16   \\
		\hline
		
		192 &\tabincell{c} holiday Inn Express  &\ 06:01, Sun &\tabincell{c}  {Hartsfield-Jackson Atlanta \\ International Airport} &\ 08:50, Sun &\ 82.82 &\ 2.81  \\
		\hline
		
		1121 &\tabincell{c}  {John Ascuaga's Nugget Casino Resort}  &\ 22:22, Fri &\ John Ascuaga's Nugget Casino Resort &\ 22:49, Sat &\ 0 &\ 24.45  \\
		\hline
		
		2446 &\ Blue Bayou Restaurant  &\ 18:11, Sun &\ Pirates of the Caribbean &\ 20:15, Sun &\ 0.0268 &\ 2.06  \\
		\hline

	\end{tabular*}
\end{table*}


\begin{figure*}
	\centering
	\subfigure[Foursquare-LA]{\includegraphics[clip=true,width=2.3in]{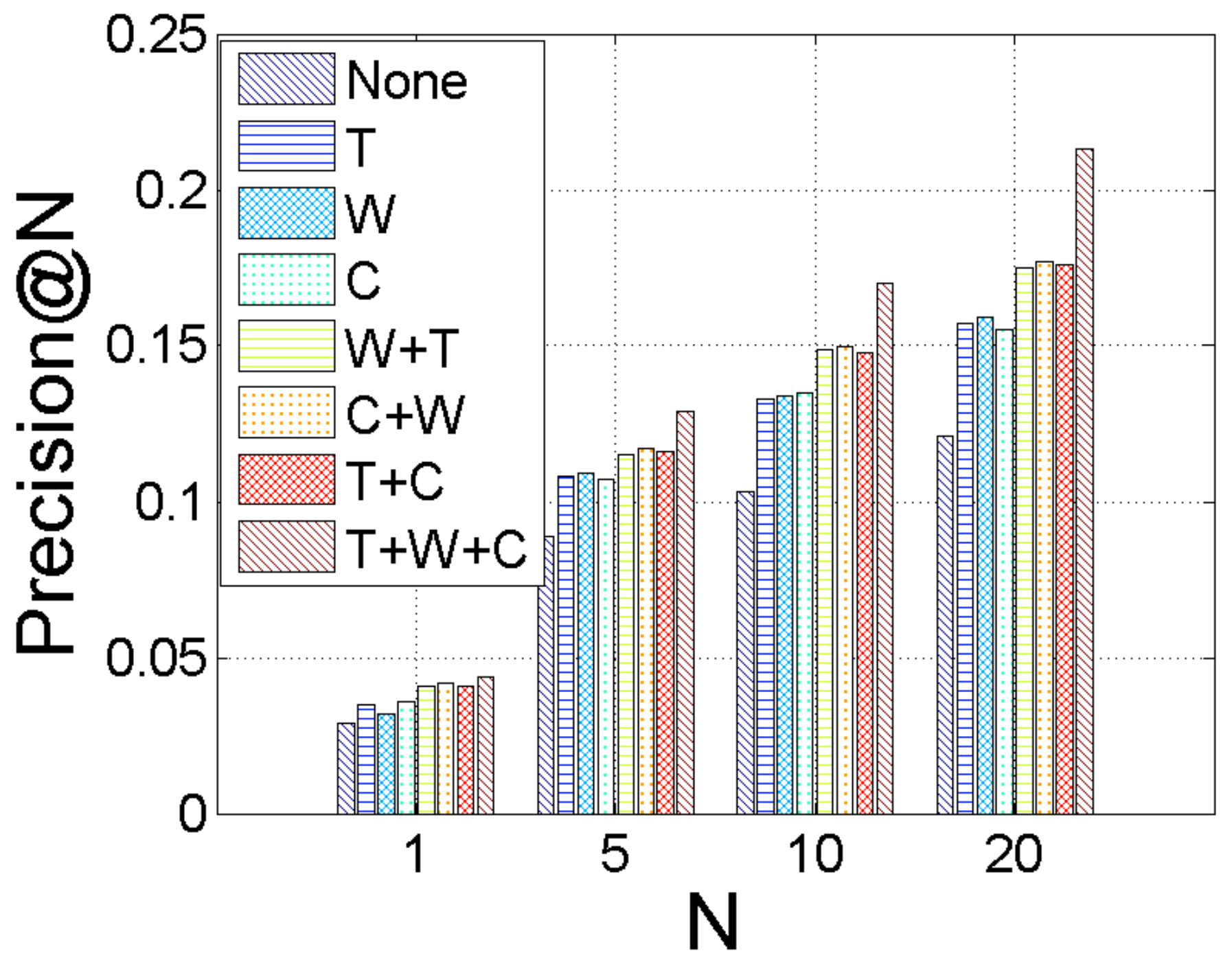} }
	\subfigure[Foursquare-NY]{\includegraphics[clip=true,width=2.3in]{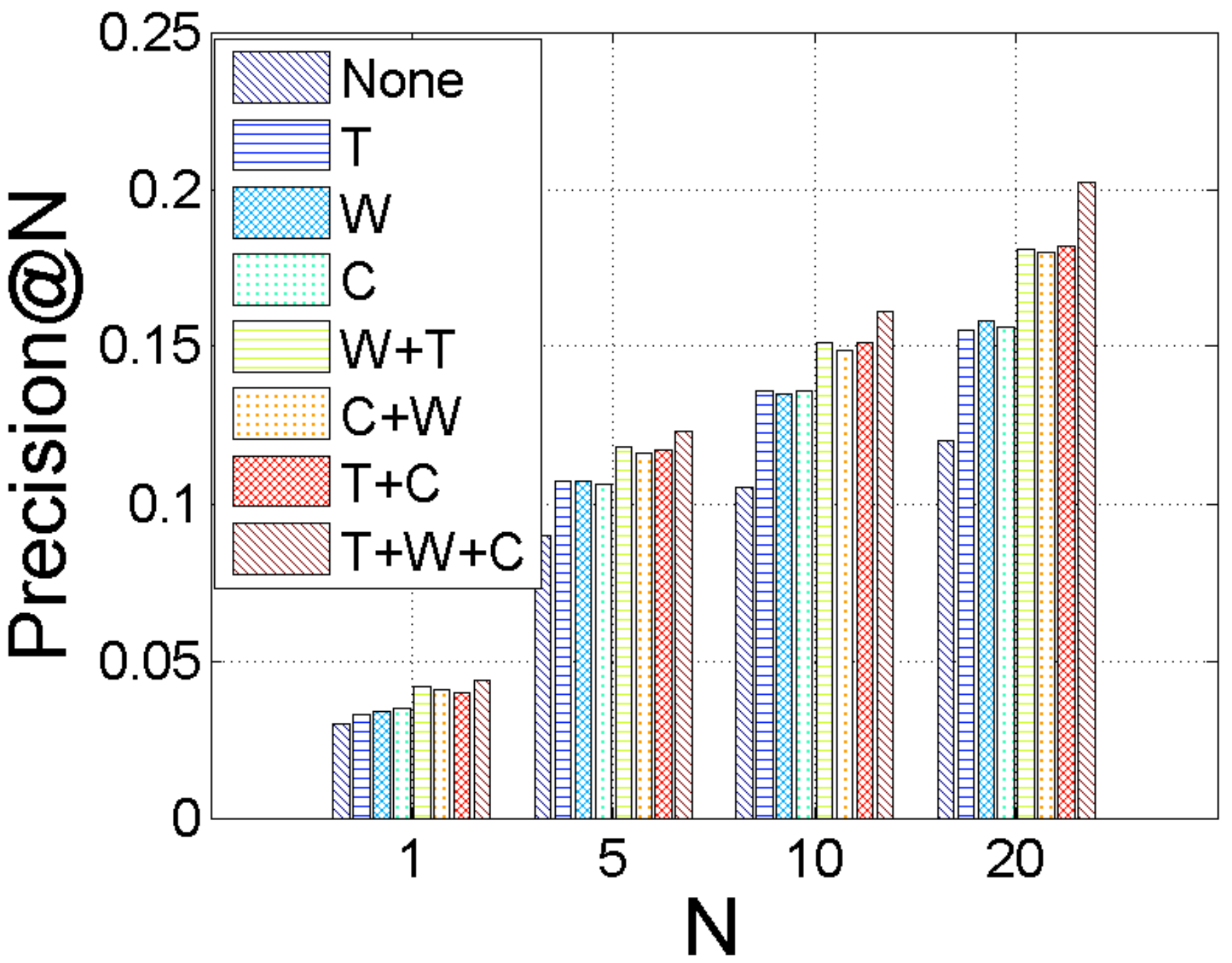} }
	\subfigure[Gowalla]{\includegraphics[clip=true,width=2.3in]{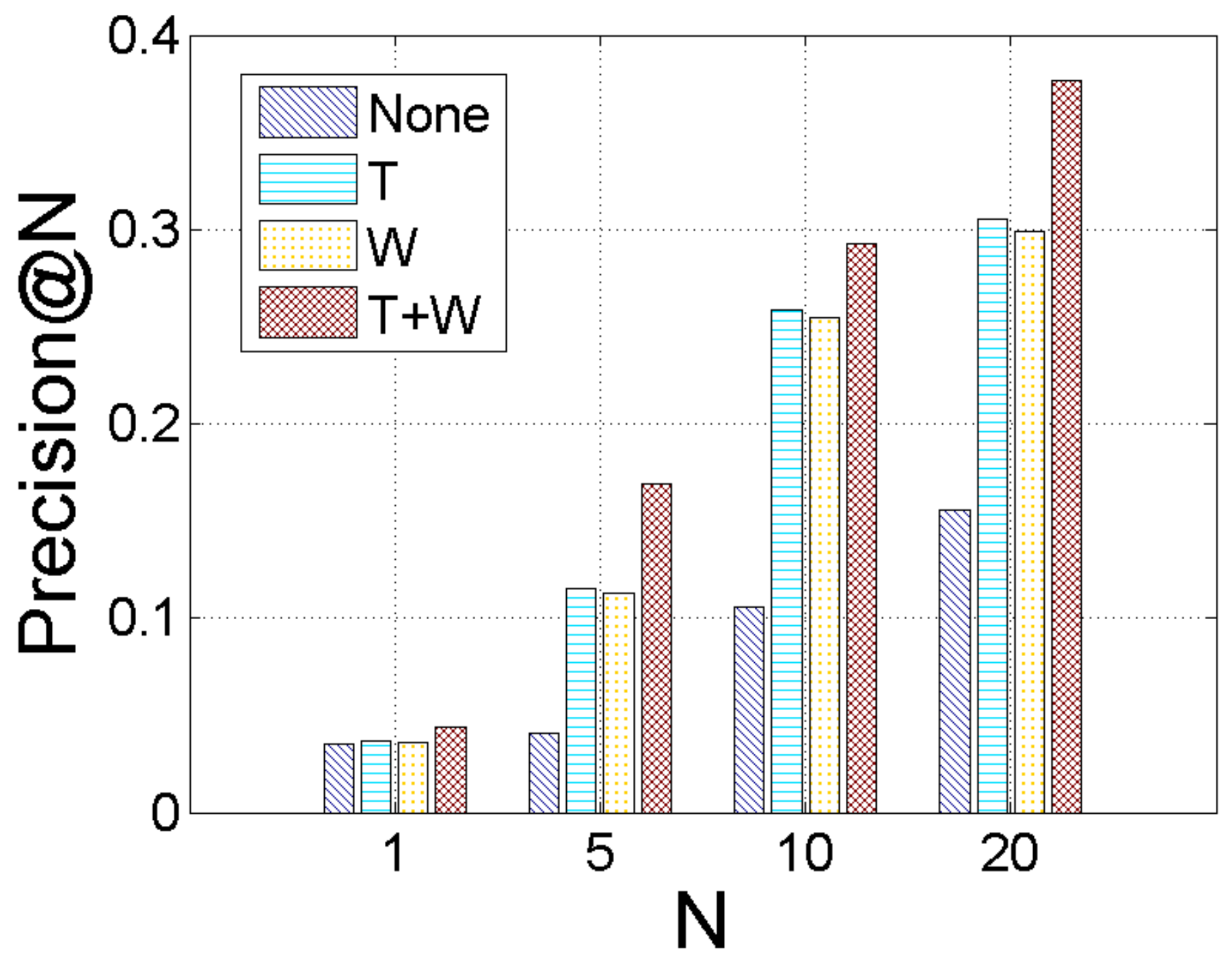}}
	\caption{Impact of Feature Combos(\emph{T stands for time, W stands for week and C stands for }) \label{fig:feature}}
\end{figure*}

\begin{figure*}
	\centering
	\subfigure[Foursquare-LA]{\includegraphics[clip=true,width=2.3in]{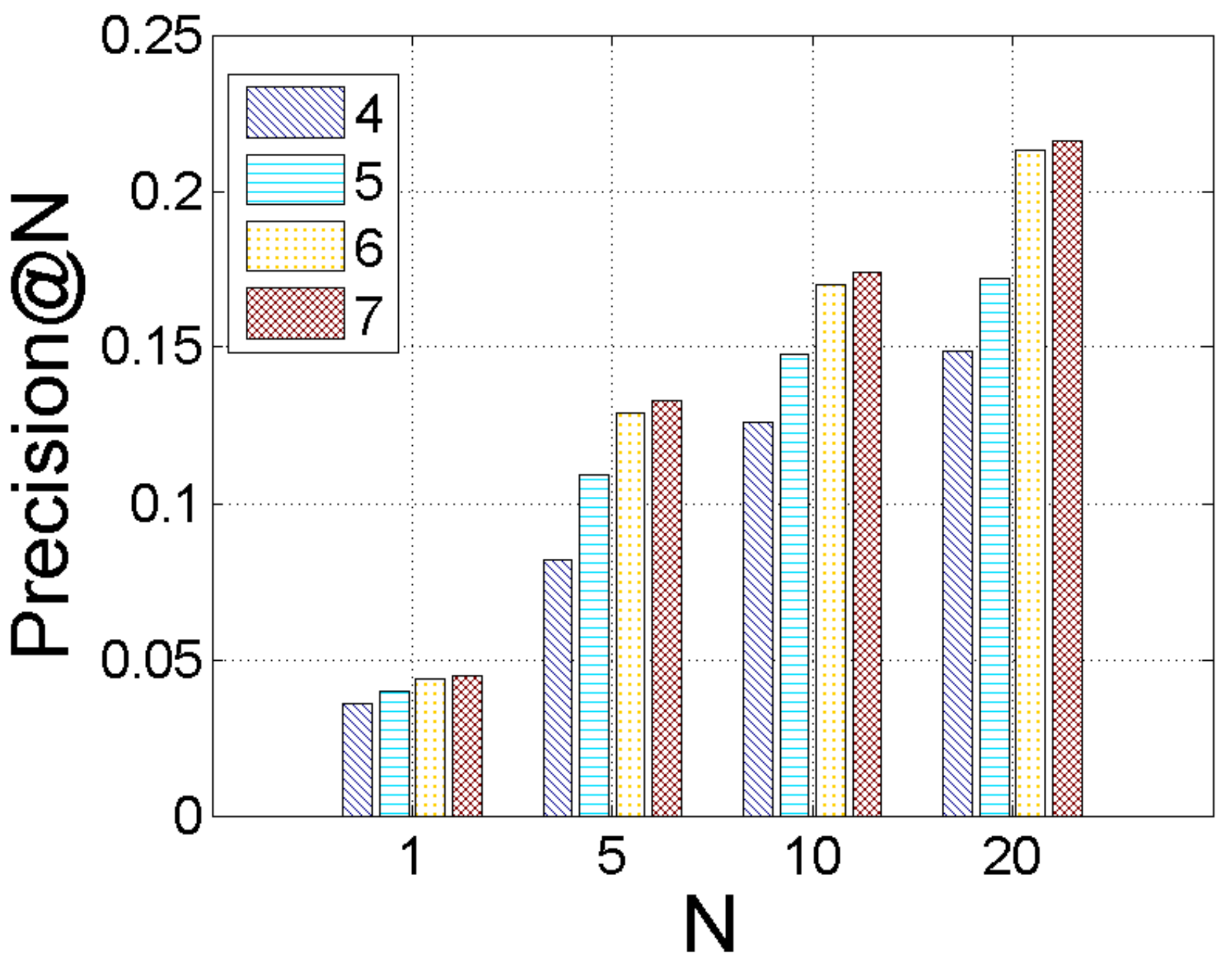}}
	\subfigure[Foursquare-NY]{\includegraphics[clip=true,width=2.3in]{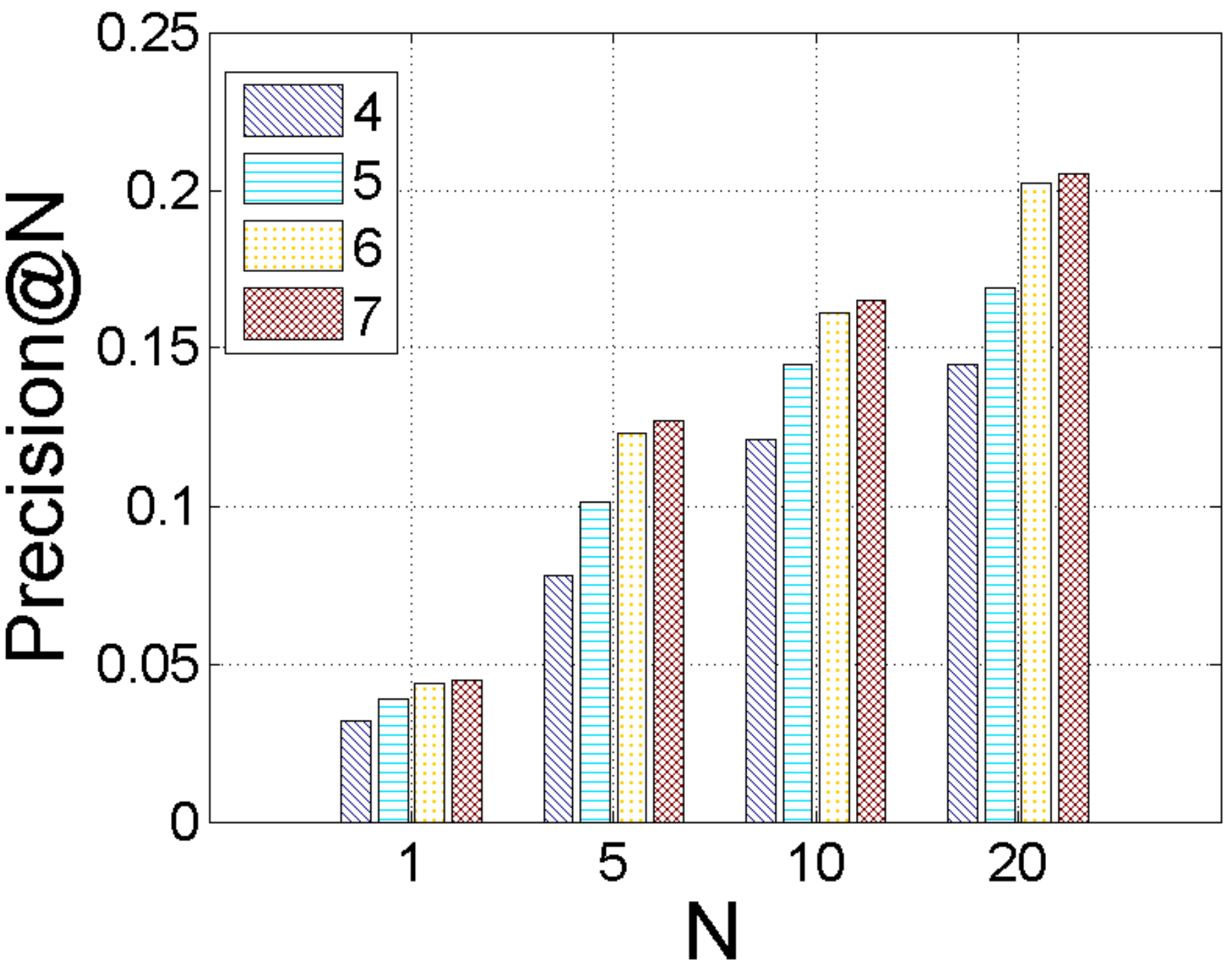}}
	\subfigure[Gowalla]{\includegraphics[clip=true,width=2.3in]{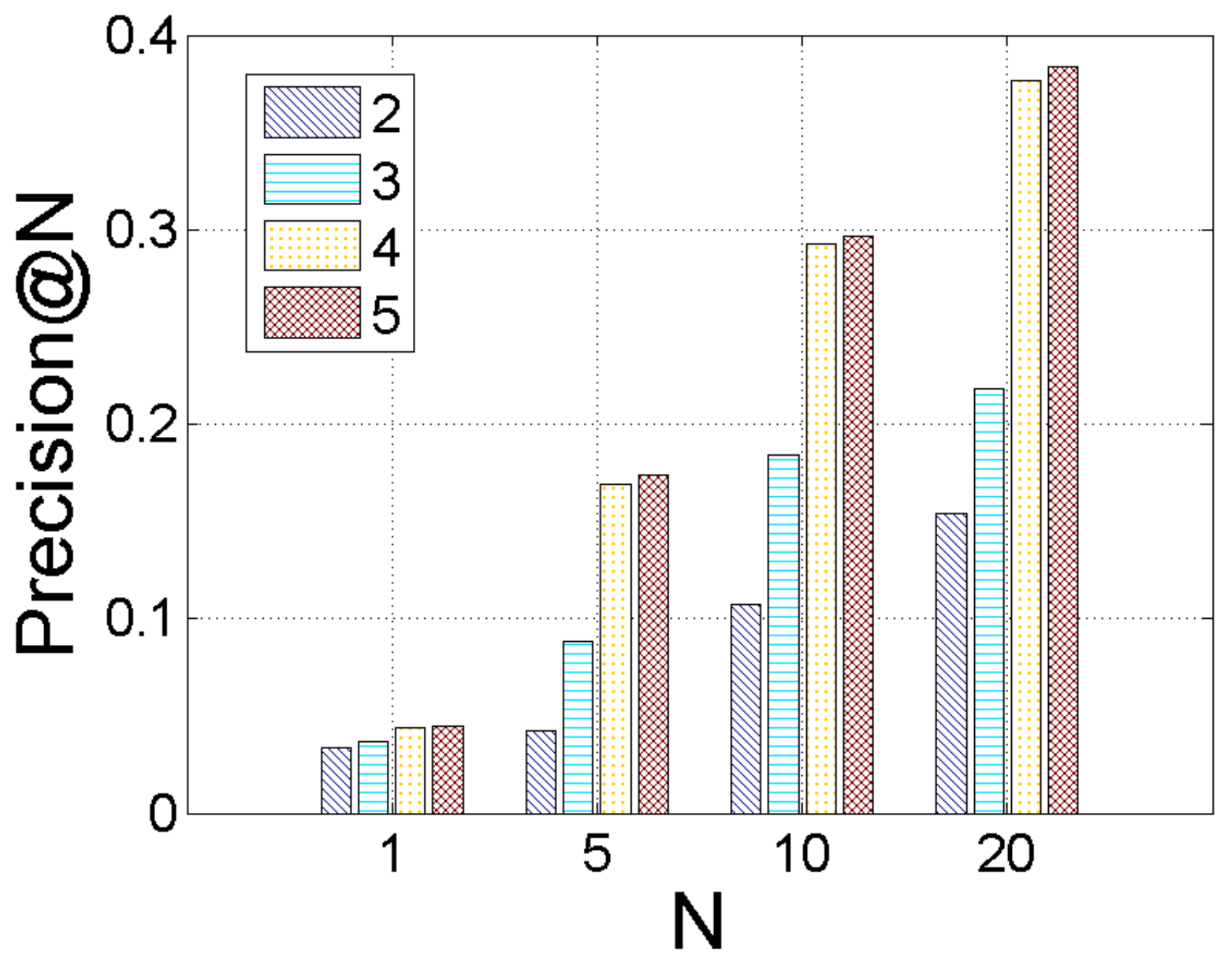}}
	\caption{Number of Latent Pattern \label{fig:class}}
\end{figure*}

\subsection{GPDM vs. PPDM}
\begin{figure*}
	\centering
	\subfigure[Distribution of user groups]{\includegraphics[clip=true,width=2.3in]{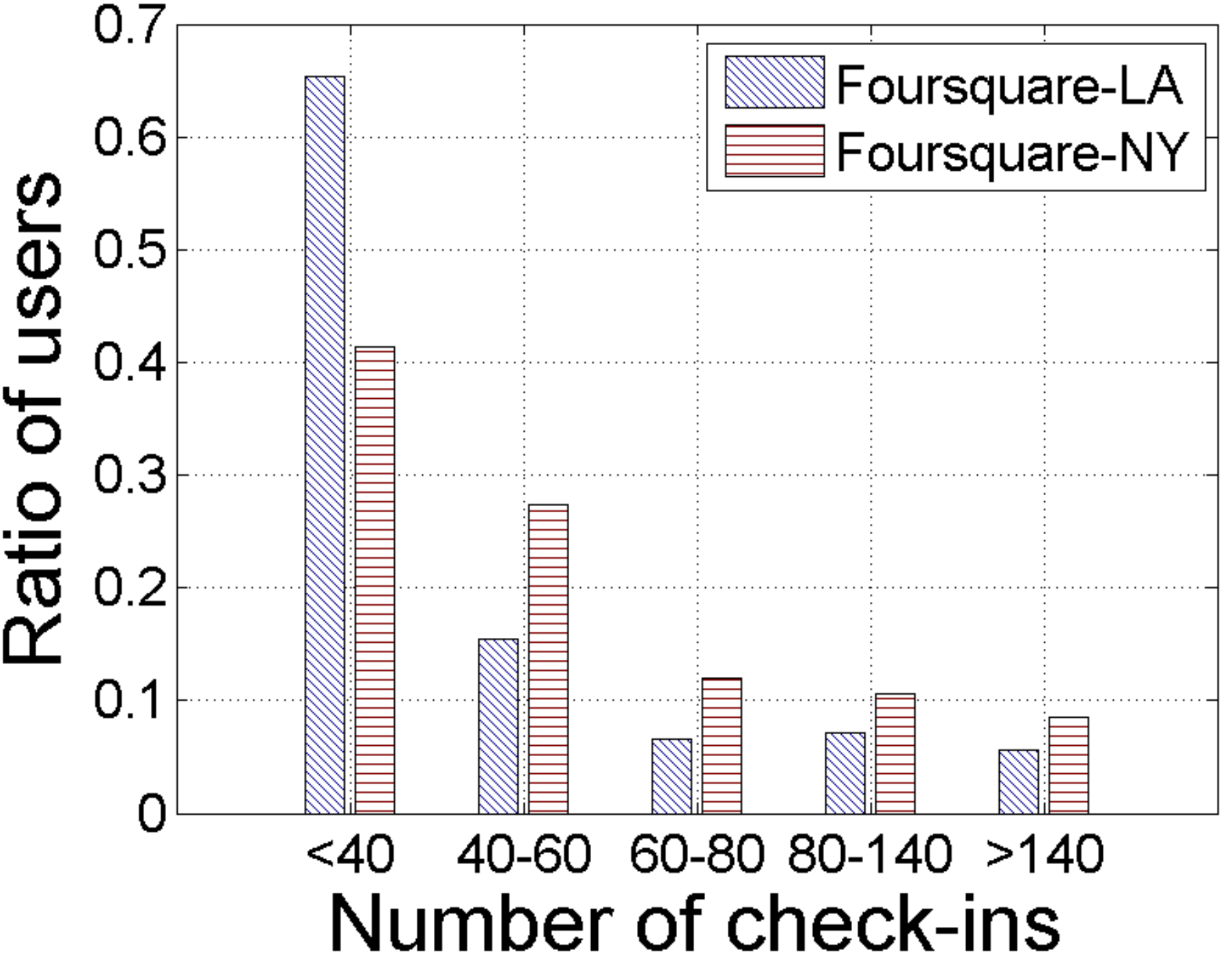}
		\label{fig:user_distribution} }
	\subfigure[User group of $<$40]{\includegraphics[clip=true,width=2.3in]{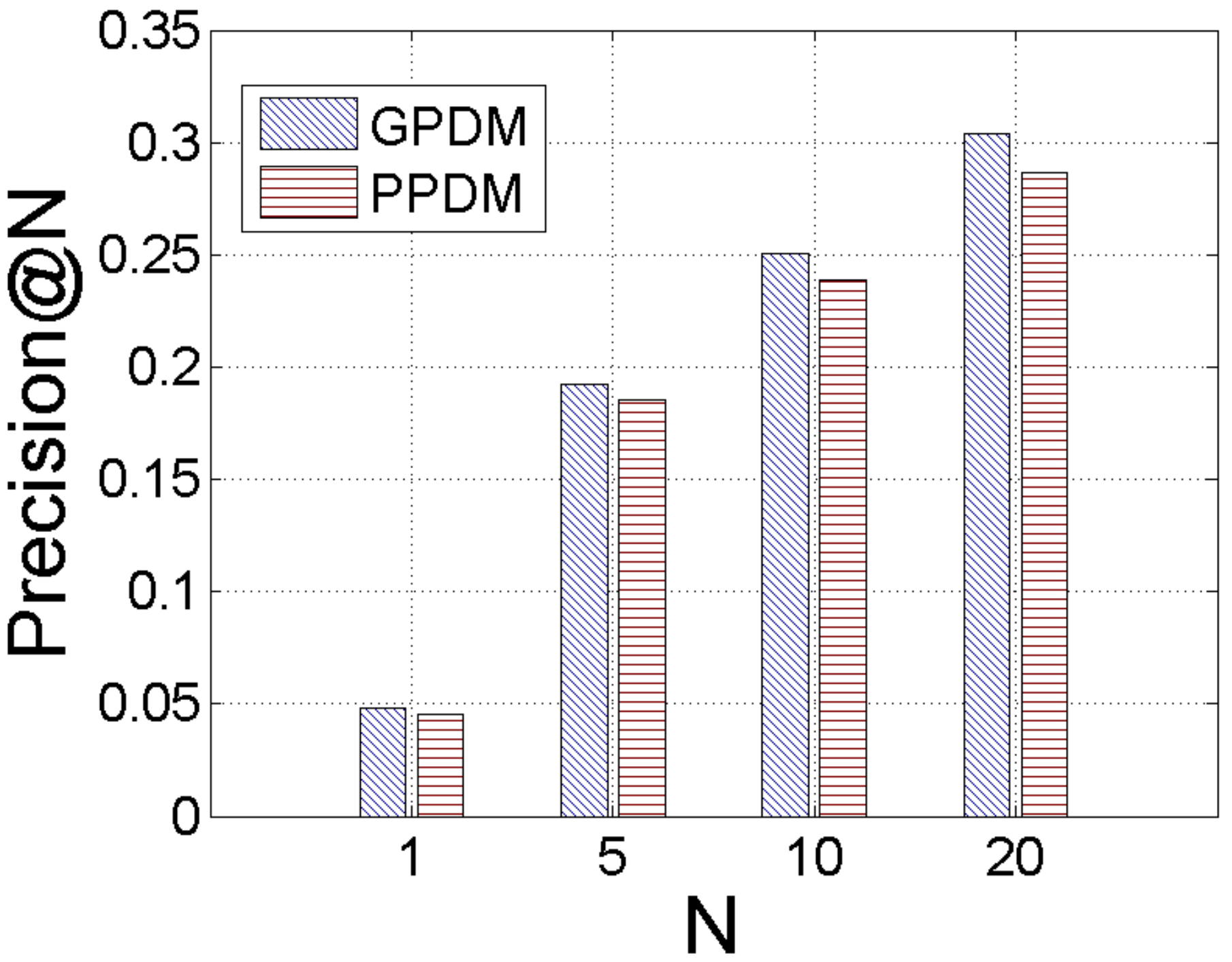}
		\label{fig:group1} }
	\subfigure[User group of 40-60]{\includegraphics[clip=true,width=2.3in]{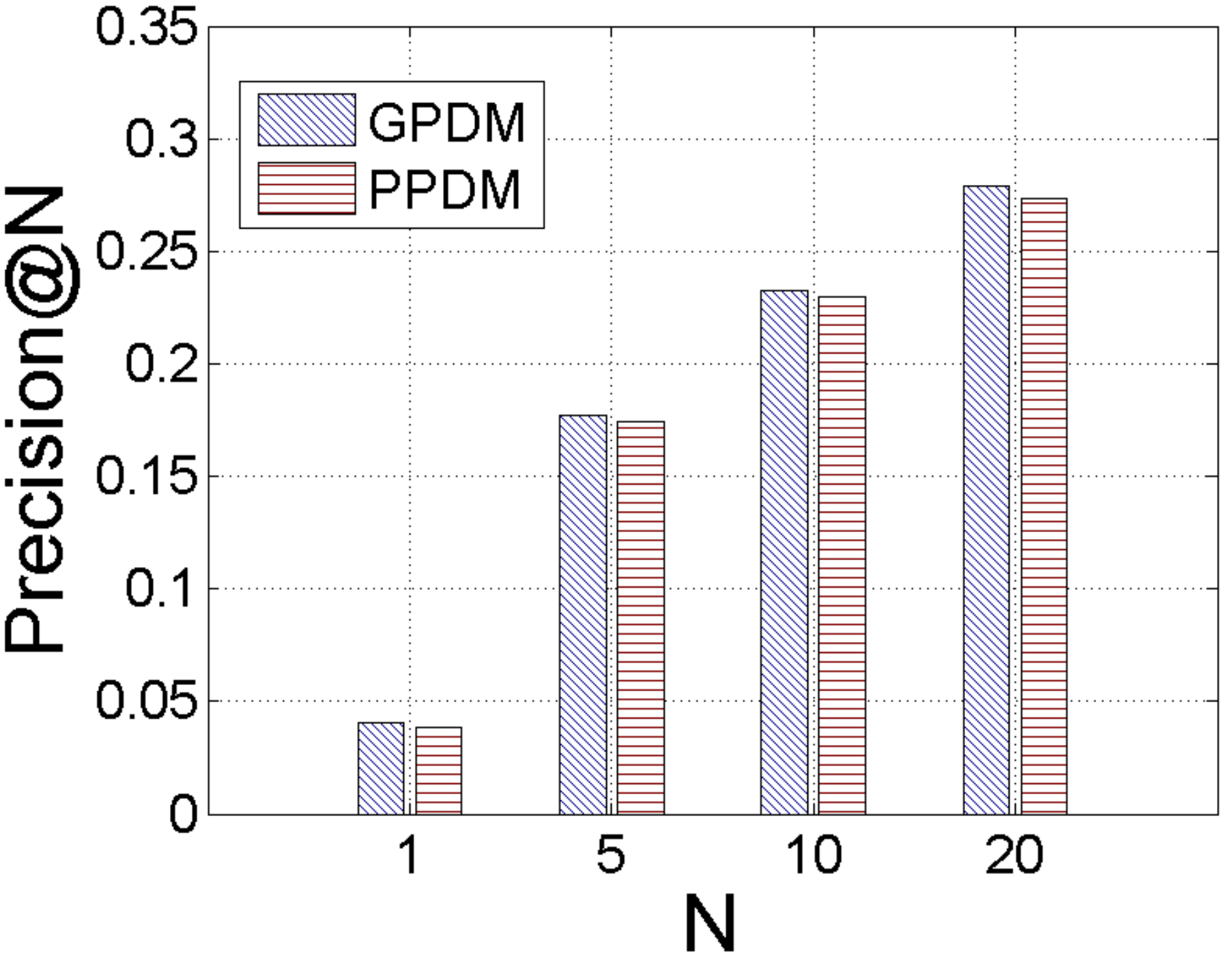}
		\label{fig:group2} }
	\subfigure[User group of 60-80]{\includegraphics[clip=true,width=2.3in]{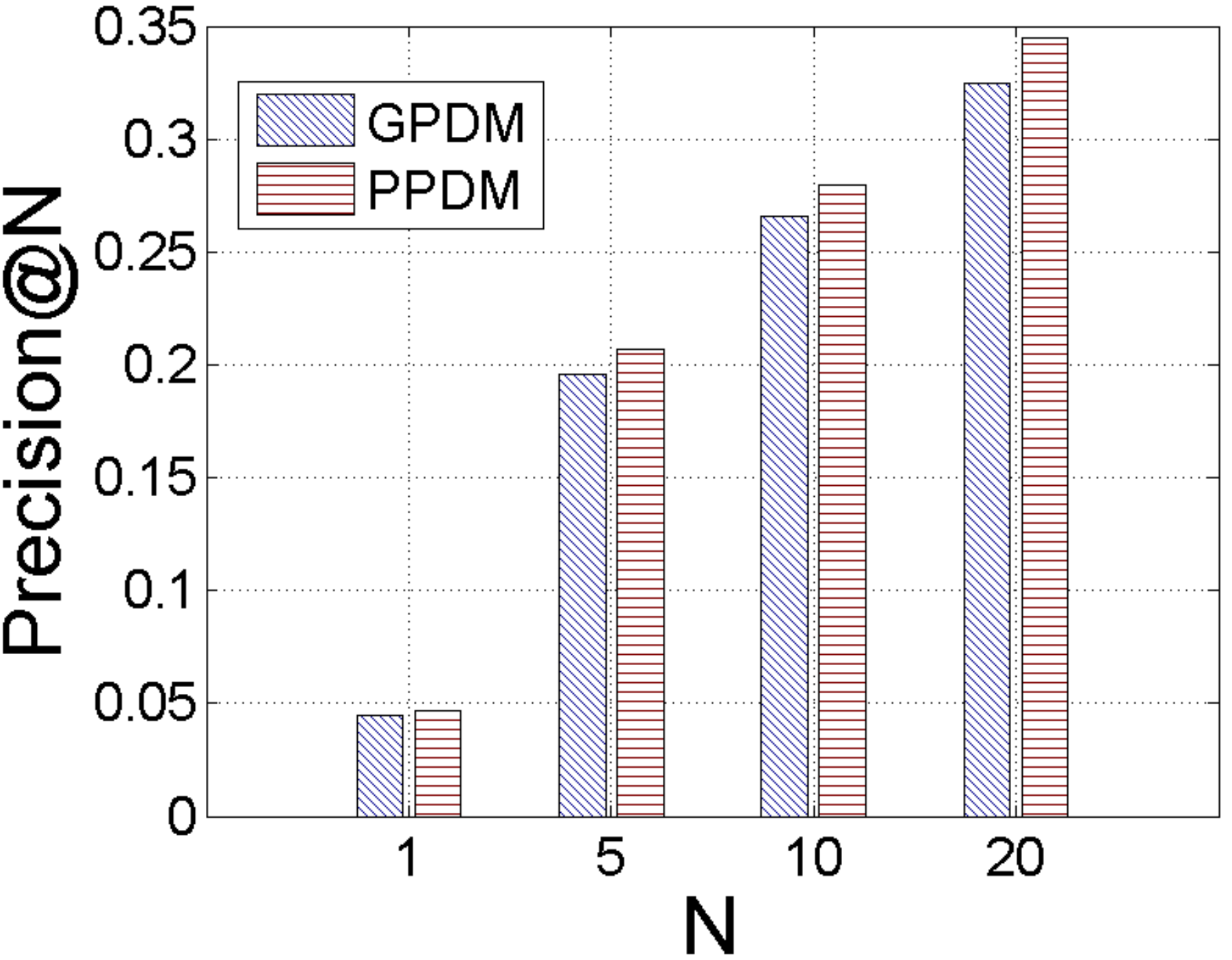}
		\label{fig:group3} }
	\subfigure[User group of 80-140]{\includegraphics[clip=true,width=2.3in]{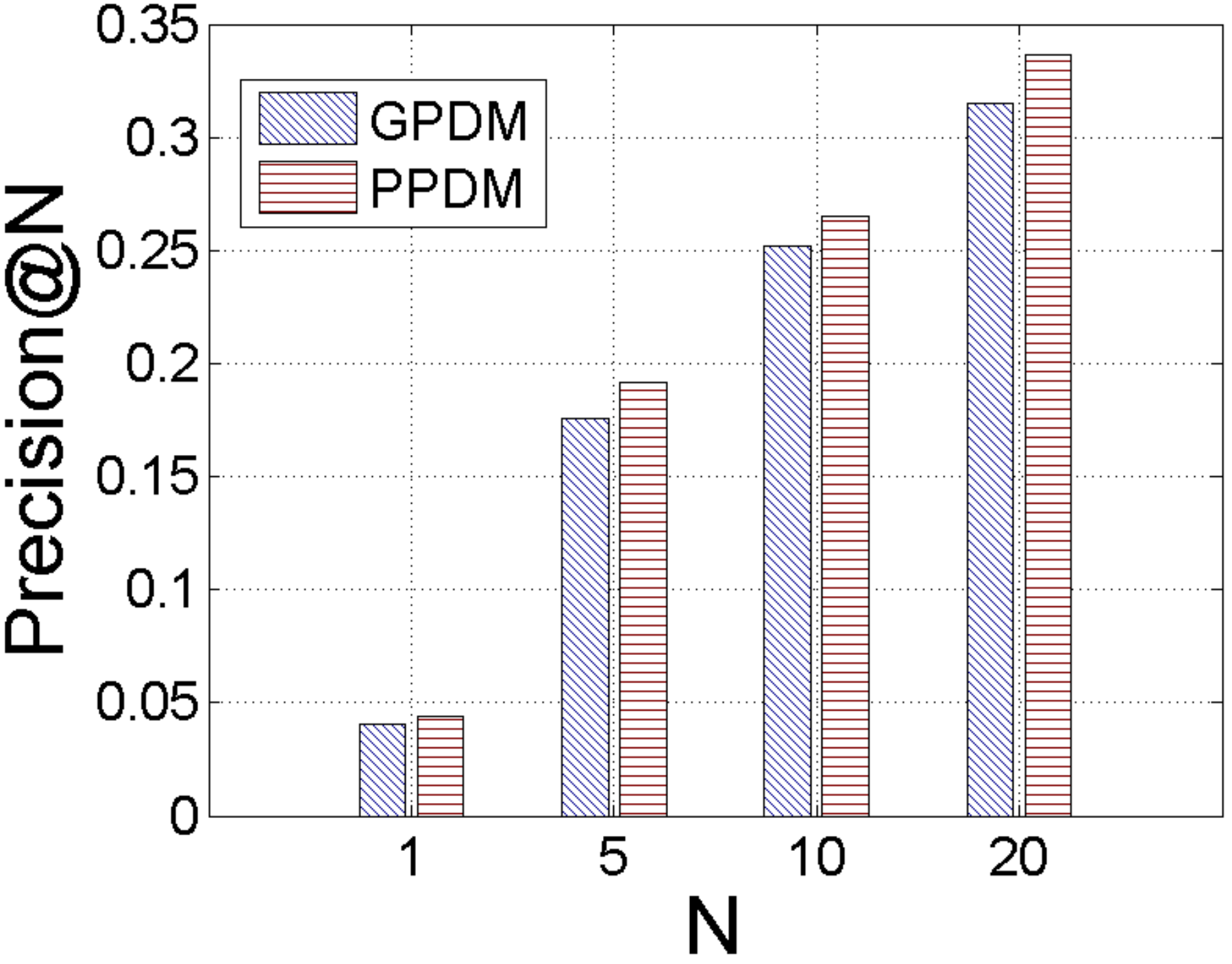}
		\label{fig:group4} }
	\subfigure[User group of $>$140]{\includegraphics[clip=true,width=2.3in]{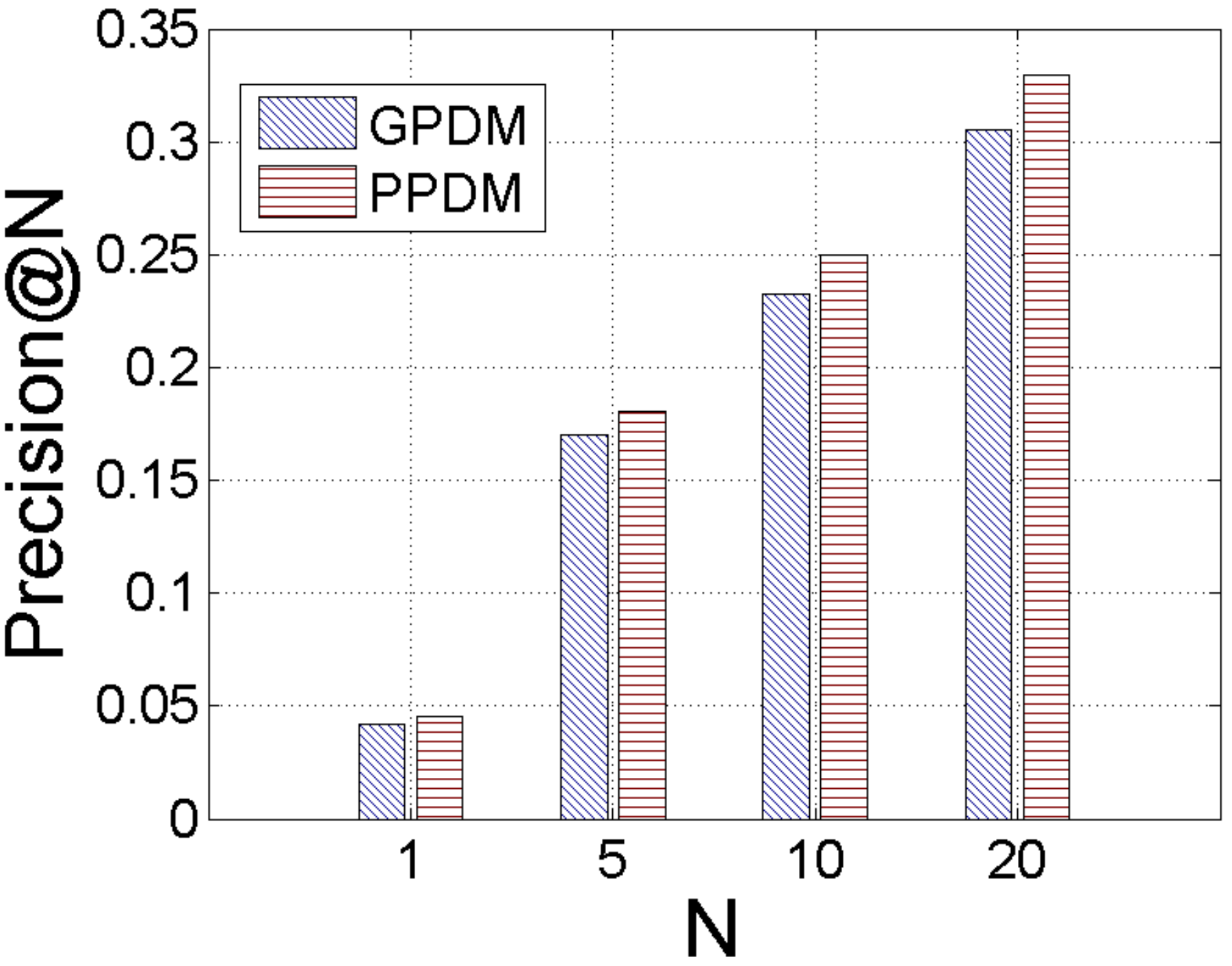}
		\label{fig:group5} }
	\caption{Performance comparison on different user groups }
\end{figure*}
Both GPDM and PPDM are much better than other algorithms, which demonstrates the effectiveness of latent behavior patterns assumptions. GPDM further improves the performance on all evaluation metrics on Foursquare-LA datasets and Gowalla dataset, while PPDM outperforms GPDM on Foursquare-NY dataset. The reason is twofold: (1) Fig.\ref{fig:user_distribution} summarizes the distribution in different ranges of users' check-in frequency in Foursquare datasets. From Fig.\ref{fig:group1} to Fig.\ref{fig:group5}, we observe that GPDM outperforms PPDM when users' check-in frequency is small. However, PPDM performs better than GPDM when users' check-in frequency becomes larger. It is reasonable since when users' check-in frequency is small, it is difficult to learn users' latent behavior patterns, and GPDM is able to recommend POI to the users based on the shared distribution of latent behavior patterns. Furthermore, one challenge of the POI recommendation is that it is difficult to recommend POI to those users who have very few check-in history, and the latent behavior patterns for those users cannot be accurately obtained, which leads to the low performance of PPDM. (2) For Gowalla dataset, the contextual features only include time of day and day of week, which indicates that most users share the similar latent behavior patters and the low contextual scenario diversity can be learned even for users with few check-ins.



\subsection{Quantitative Evaluation of Accumulated Precision}
\begin{figure}[htb]
	\centering
	\subfigure[Quantitative Evaluation of accumulated precision along with time]{\includegraphics[clip=true,width=2.6in]{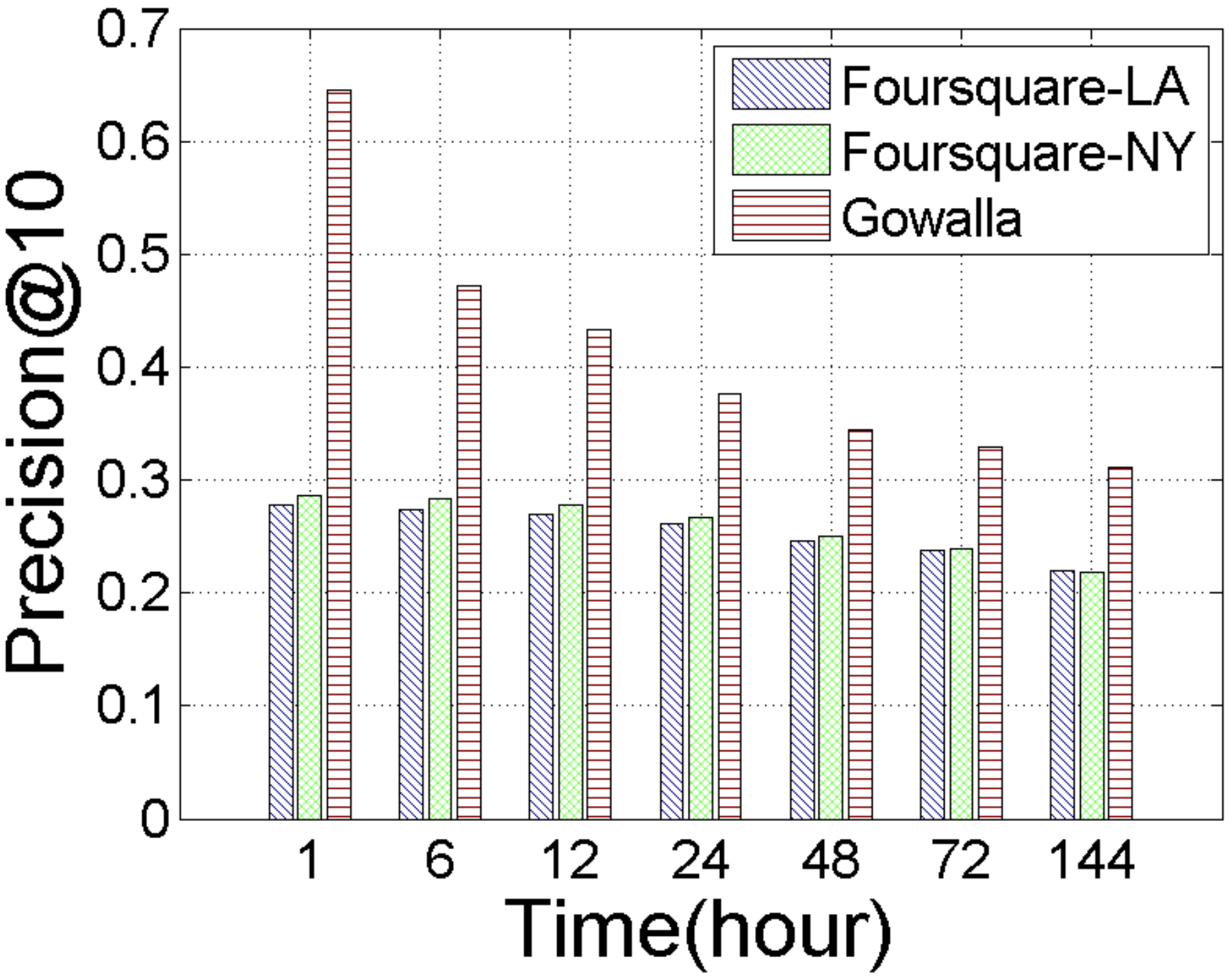}}
	\subfigure[Quantitative Evaluation of accumulated precision along with distance]{\includegraphics[clip=true,width=2.6 in]{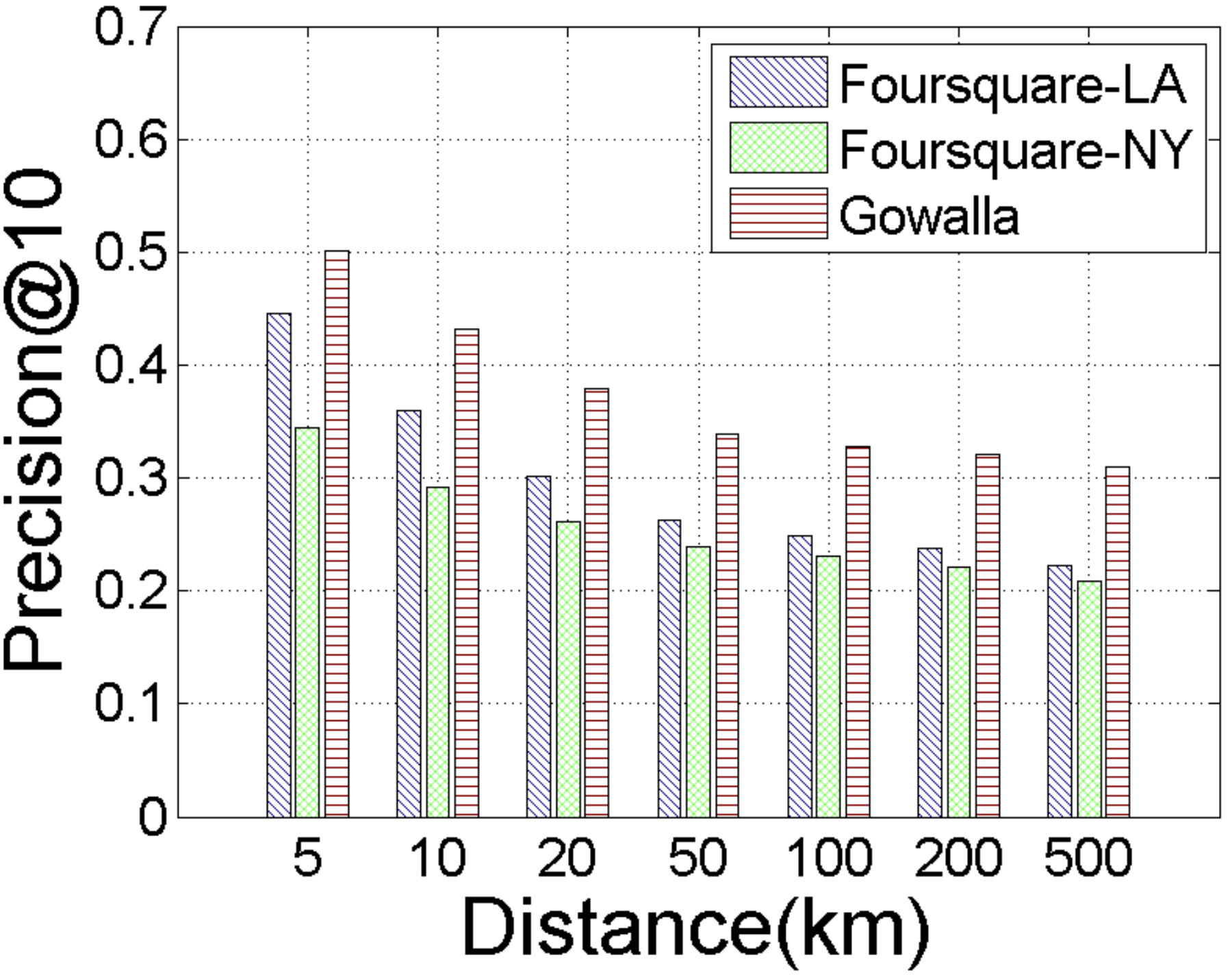}}
	\caption{Prediction Ability vs. Distance and Time \label{fig:caseDis}}
\end{figure}
Fig.\ref{fig:caseDis} shows the prediction ability vs. distance and time. 
The quantitative results along with distance (see in Fig. \ref{fig:caseDis}(b)) manifest that our model is capable of predicting transitions within a localized region as well as an occasional journey with long distance and the quantitative results along with time (see in Fig. \ref{fig:caseDis}(a)) imply that our model is suited to periodicity of mobility within big time interval as well as random movement within small interval.

\subsection{Impact of the Contextual Features}
Here, we discuss the recommendation efforts of different types of contextual information, i.e. Previous Location's Category, Time of Day, Day of Week. Figure.\ref{fig:feature} depicts the experimental results with variants of combinations of contextual information incorporated. In general, the model accuracy increases with more contextual information added in. Most importantly, the proposed model uplifts the performance significantly by integrating all contextual information to infer latent behavior patterns in recommendation. It indicates that finer latent patterns are obtained to better capture user preference.

\subsection{Impact of the Number of Latent Patterns}
Figure.\ref{fig:class} shows the experimental results with different settings of the number of latent patterns. We can see that for both datasets, the model accuracy increases with the increasing of the number of patterns. When the number of latent patterns reaches 6 for Foursquare and 4 for Gowalla, the returns diminish largely. Even the performance gained by adding one more latent pattern is minor compared to the difference between the number of patterns less than 6 for Foursquare and 4 for Gowalla. For example, P@10 on Gowalla dataset is 0.184 using 3 latent patterns, whereas the four-latent-pattern model has a P@10 of 0.293, which is a 59.2\% relative improvement. Using a five-latent-pattern model only increase performance by another 1.4\%. Besides considering the additional computation cost of inferring preference for each pattern, we conclude that the 6 latent patterns for Foursquare and 4 latent patterns for Gowalla is rich enough to complete the task of next personalized POI recommendation.

\subsection{Case Study}
Table tabulates 5 representative successful predictions for Foursquare data.For each user, we show the user ID, the current venue, check-in time of current venue, the next venue, check-in time of next venue, the distance and time interval between successive check-ins. The distance between two successive POIs varies from 0.5km to 304km and the time interval varies from half an hour to 44 hour, which manifests that our model is capable of predicting transitions within a localized region as well as an occasional journey with long distance.



\section{Conclusion and Future Work}
To address the personalized next POI recommendation problem, in this paper we propose a unified tensor-based latent model to capture the successive check-in behavior by exploring the latent pattern-level preference for each user. We derive a BPR-like optimization criterion accordingly and then use Expectation Maximization (EM) to estimate the model parameters. Performance evaluation conducted on two large-scale real-world LBSNs datasets shows that our proposed approach improves the recommendation accuracy significantly compared against other state-of-the-art methods. More specifically, our proposed method is capable of predicting journey of long distance and the consecutive check-ins which span a long period of time. For future work, we will soon evaluate our proposed model's ability for next new POI recommendation by redefining the transition tensor in a categorical dimension.

\bibliographystyle{IEEEtran}
\bibliography{extendaaai}

\end{document}